# Lattice Boltzmann methods for multiphase flow and phase-change heat transfer

Q. Li [a, b], K. H. Luo [c, *], Q. J. Kang [a], Y. L. He [d], Q. Chen [e], and Q. Liu [d]

[a] Computational Earth Science Group, Los Alamos National Laboratory, Los Alamos, NM 87545, USA

[b] School of Energy Science and Engineering, Central South University, Changsha 410083, China

[c] Department of Mechanical Engineering, University College London, Torrington Place, London WC1E 7JE, United Kingdom

[d] Key Laboratory of Thermo-Fluid Science and Engineering of Ministry of Education, School of Energy and Power Engineering, Xi'an Jiaotong University, Xi'an, Shaanxi 710049, China

[e] School of Mechanical and Electronic Engineering, Nanjing Forestry University, Nanjing, Jiangsu 210037, China

*Corresponding author at: Department of Mechanical Engineering, University College London, Torrington Place, London WC1E 7JE, United Kingdom. Tel.: +44 (0)20 7679 3916. E-mail address: k.luo@ucl.ac.uk (K. H. Luo).

**Abstract**

Over the past few decades, tremendous progress has been made in the development of particle-based discrete simulation methods versus the conventional continuum-based methods. In particular, the lattice Boltzmann (LB) method has evolved from a theoretical novelty to a ubiquitous, versatile and powerful computational methodology for both fundamental research and engineering applications. It is a kinetic-based mesoscopic approach that bridges the microscales and macroscales, which offers distinctive advantages in simulation fidelity and computational efficiency. Applications of the LB method are now found in a wide range of disciplines including physics, chemistry, materials, biomedicine and various branches of engineering. The present work provides a comprehensive review of the LB method for thermofluids and energy applications, focusing on multiphase flows, thermal flows and thermal multiphase flows with phase change. The review first covers the theoretical framework of the LB method, revealing certain inconsistencies and defects as well as common features of multiphase and thermal LB models. Recent developments in improving the thermodynamic and hydrodynamic consistency, reducing spurious currents, enhancing the numerical stability, etc., are highlighted. These efforts have put the LB method on a firmer theoretical foundation with enhanced LB models that can achieve larger liquid-gas density ratio, higher Reynolds number and flexible surface tension. Examples of applications are provided in fuel cells and batteries, droplet collision, boiling heat transfer and evaporation, and energy storage. Finally, further developments and future prospect of the LB method are outlined for thermofluids and energy applications.

# Contents

## 1. Introduction

Energy and combustion systems typically involve fluid dynamics, chemical reactions, heat transfer, multiphase flows and phase change that occur over scales ranging from macroscale via mesoscale to microscale. Extensive examples are found in batteries, fuel cells, gas turbines, fluidized beds, coal-fired power plants, solar thermal power plants and nuclear power plants. The performance, reliability and safety of these technologies depend crucially on how to organize the fundamental thermal-fluids processes, which in turn requires accurate and reliable predictive and diagnostic methods. Since the 1970s, general-purpose computational fluid dynamics (CFD) based on solving the Reynolds-averaged Navier-Stokes (RANS) equations, pioneered by D. B. Spalding and others, has been developed to compute fluid flow, heat transfer and combustion with considerable success. With the emergence of supercomputers in the 1990s, more accurate but computationally demanding methods such as large eddy simulation (LES) and direct numerical simulation (DNS) have been in increasing use. These macroscopic methods, however, are all based on the assumption of continuum, which makes it difficult or even impossible to treat certain physical phenomena, especially at micro- and meso-scales.

More fundamental approaches are particle-based (i.e. molecular cluster-based) discrete methods, such as molecular dynamics (MD), direct simulation Monte Carlo (DSMC), and dissipative particle dynamics (DPD). These methods are capable of simulating phenomena where the continuum assumption breaks down. On the other hand, the high computational cost renders these methods impractical for the majority of problems of practical concern in energy and combustion. The lattice Boltzmann (LB) method, sitting in the middle of the hierarchy of modeling and simulation methods (see Fig. 1), is a mesoscopic approach based on the kinetic theory expressed by the original Boltzmann equation. The LB equation can be either viewed as a special discrete solver for the Boltzmann equation or a minimal form of the Boltzmann equation in which the microscopic kinetic principles are preserved to recover the hydrodynamic behavior at the macroscopic scale [1]. Therefore, the LB method is based on a particle picture, but principally aims to predict macroscopic properties [2]. This scale-bridging nature of the LB method is a fundamental advantage, which allows it to incorporate the essential microscopic or mesoscopic physics while recovering the macroscopic laws and properties at affordable computational cost.

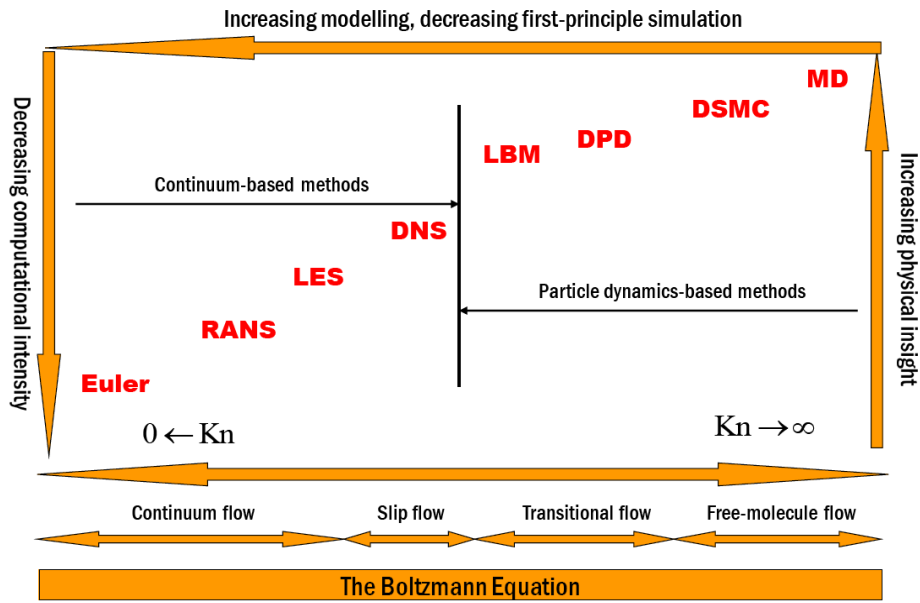

**Fig. 1**. A hierarchy of modeling and simulation approaches [3].

In the last 25 years, the LB method has been developed into an efficient and powerful simulation method for a wide range of phenomena and processes [1-16], such as single-phase flows, multiphase flows, turbulence, heat transfer, and phase change, as well as a numerical tool for nonlinear partial differential equations [17-22]. It exhibits many distinctive advantages over conventional numerical methods [2, 23]. First, in the LB equation the convective operator (the streaming process) is completely linear, whereas the convective terms of the Navier-Stokes equations are nonlinear. Second, in conventional numerical methods it is usually necessary and costly to solve a Poisson equation for the pressure field of incompressible flows, while in the LB method the fluid pressure can be simply calculated with an equation of state [2] (such an advantage can also be found in the artificial compressibility method, but only for steady-state flows). Third, complex boundary conditions in the LB method can be formulated with elementary mechanical rules such as bounce-back and reflection according to the interactions of the LB "molecules" with solid walls [23]. Moreover, the LB method is ideal for parallel computing because of its explicit scheme, local interactions, and consequently very low communication/computation ratio. It is ideally situated to exploit the massively parallel supercomputers based on either CPUs or GPUs or heterogeneous architectures. Meanwhile, it should be noted that, as a natural-born dynamic scheme, the LB method is not a method of choice for steady-state computations [16]. In addition, the standard LB method is not well suited to body-fitted coordinates and adaptive time-stepping [16].

Since the emergence of the LB method, its application in multiphase flows has always been a very important theme of the method. With the development in the past two decades, many multiphase LB models have been proposed. These models mostly fall into one of the following categories: the color-gradient LB method [24-30], the pseudopotential LB method [31-39], the free-energy LB method [40-44], and the phase-field LB method [45-53]. In addition, several multiphase LB models were recently developed based on the entropic LB method [54] and the discrete Boltzmann equation [55]. The *color-gradient LB method* was introduced by Gunstensen *et al.* [24], who employed red and blue particle distribution functions to represent two different fluids. Besides the standard collision operator, an additional collision operator was also utilized in this method, which can be regarded as a source term for generating the surface tension. Furthermore, to separate different phases and maintain interfaces, a recoloring process is required in the color-gradient LB models [24-30, 56].

*The free-energy LB method* was proposed by Swift *et al.* [40, 41] based on thermodynamics considerations. The second-order moment of the equilibrium density distribution function was modified to include a non-ideal thermodynamic pressure tensor. The phase separation was therefore described by a non-ideal equation of state in the thermodynamic theory such as the van der Waals equation of state. However, the original free-energy LB model suffered from the lack of Galilean invariance owing to some non-Navier-Stokes terms [57], which resulted from the incorporation of pressure tensor using the equilibrium distribution function. To restore the Galilean invariance, some correction terms should be added to the equilibrium distribution function [41-44]. A similar problem also exists in the color-gradient LB method, in which the pressure is changed by modifying the equilibrium distribution function. Therefore the color-gradient multiphase LB models also need some correction terms to eliminate the non-Navier-Stokes terms in the recovered macroscopic equations [58-60].

*The pseudopotential LB method*, which is the simplest multiphase LB method, was devised by Shan and Chen [31, 32]. In this method, the fluid interactions are mimicked by an interparticle potential, through which the separation of fluid phases or components can be achieved automatically, without resorting to any techniques to track or capture interfaces [61]. In fact, the interparticle potential will lead to a non-ideal pressure tensor, although it is different from that in the free-energy LB method. The pseudopotential LB method has become very popular in the multiphase LB community due to its conceptual simplicity and computational efficiency [2, 12, 62-64] and has been applied with great success to many problems.

The fourth category, *the phase-field LB method*, represents the multiphase LB models that are based on the phase-field theory, in which the interface dynamics is described by an order parameter that obeys the Cahn-Hilliard equation or a Cahn-Hilliard-like equation [65]. In 1999, an incompressible multiphase LB model was proposed by He *et al*. [45]. In this model, the liquid-gas interface was captured with the evolution of an index function (order parameter). Meanwhile, the recovered interface-capturing equation was found to be a Cahn-Hilliard-like equation. In this regard, He *et al.*'s model is a phase-field LB model, even though the model was not directly built on the phase-field theory. Similarly, the multiphase LB model devised by Lee and Lin [47] also belongs to this category.

Much progress has been made in the above four categories of multiphase LB methods since the aforementioned early studies. However, these multiphase LB methods exhibit different performances in simulating *dynamic* multiphase flows at *large* liquid-gas density ratios ( $\rho_l / \rho_g \sim 10^3$ in real world), which may be related to the following issues. First, it can be found that in these multiphase LB methods the physical quantities that need to be evaluated across the liquid-gas interface are different. For instance, in the free-energy and the pseudopotential LB methods, the density and the pseudopotential are used, respectively. Second, as just mentioned, the free-energy and the color-gradient LB methods need some correction terms to remove the non-Navier-Stokes terms in the macroscopic equations. These correction terms will introduce additional sources of numerical instability as they involve many density-gradient terms such as $\mathbf{v}\nabla\rho$ and $\mathbf{v}\cdot\nabla\rho$, where $\mathbf{v}$ is the velocity [41-44, 58-60]. This is one of the reasons why the free-energy and the color-gradient multiphase LB methods usually suffer from severe numerical instability in simulating dynamic multiphase flows at large density ratios and high Reynolds numbers, although they are successful in static or quasi-static cases with large density ratios.

In comparison with the free-energy and the color-gradient LB methods, the pseudopotential LB method and the phase-field LB method have been successfully applied to dynamic multiphase flows at large density ratios ( $\rho_l/\rho_g \sim 10^3$ ) and relatively high Reynolds numbers (e.g., droplet splashing and droplet collision) [39, 46, 47, 49, 66-72]. Moreover, the pseudopotential and the phase-field multiphase LB methods have been widely employed to simulate multiphase flows in fuel cells (water-gas two-phase transport) and batteries (the electrolyte transport dynamics) [73-93] as well as phase-change heat transfer (boiling, evaporation, etc.) [94-116]. It is noticed that these two multiphase LB methods play an increasingly important role in modeling multiphase flow and phase-change heat transfer that are involved in energy science and technologies from the viewpoint of a mesoscopic numerical approach.

With the increase of practical applications of the LB method, it is very necessary to review the related theories and clarify some theoretical issues that are crucial to applications. The purpose of this article is therefore to present a comprehensive review of the advances in the pseudopotential and the phase-field multiphase LB methods. Various theoretical aspects will be addressed, such as the fundamental theory and basic models, the elimination of thermodynamic/hydrodynamic inconsistency, the surface tension treatment, the adjustment of interface thickness, and the implementation of contact angles. Meanwhile, the thermal LB models based on these two multiphase LB methods for simulating liquid-vapor phase change will also be critically reviewed. Furthermore, we will summarize several forcing schemes that are widely used in the LB method, which also play a crucial role in the multiphase LB methods. In addition, the thermal LB approaches on standard lattices, which are extensively involved in the LB simulations of phase-change heat transfer, will be discussed in detail.

It is worth mentioning that there have been some excellent books on the LB method, covering different aspects of theories and applications, e.g., by Succi [1], Wolf-Gladrow [11], Sukop and Thorne [12], Mohamad [14], and Guo and Shu [15]. There have also been several comprehensive reviews in the field, e.g., by Benzi *et al.* [10], Chen and Doolen [2], and Aidun and Clausen [13]. The rest of the present paper is organized as follows. Section 2 describes the fundamentals of the LB method, the forcing schemes, and the thermal LB approaches on standard lattices. The pseudopotential and the phase-field multiphase LB methods are comprehensively reviewed in Sections 3 and 4, respectively. Section 5 reviews some applications of the multiphase and thermal LB methods, including the applications in fuel cells and batteries, droplet collision, boiling heat transfer and evaporation, and energy storage with phase change materials. Finally, Section 6 summarizes the key points of the present review and gives a brief discussion about the further developments and future prospects of the multiphase and thermal LB methods for thermofluids and energy applications.

## 2. The basic LB formulations

### 2.1 The LB-BGK formulation

Historically, the LB method [4-9] originated from the lattice gas automata method [6, 117], which can be considered as a simplified fictitious molecular dynamics model in which the space, time, and particle velocities are all discrete. Later it was demonstrated that [118, 119] the LB equation can be rigorously derived from the Boltzmann equation in the kinetic theory, which not only establishes a direct connection between the LB method and the kinetic theory but also greatly solidifies the physics base of the LB method. Nevertheless, it should be noted that the LB method is not limited to dilute gases (a limitation of the Boltzmann equation) because it can be extended to incorporate non-ideal interactions through the effective interactions in the spirit of density functional theory. In this paper we start with the Boltzmann equation [1, 11], which can be written as (without external forces)

$$\frac{\partial f}{\partial t} + \boldsymbol{\xi} \cdot \nabla f = \Omega_f ,\qquad(1)$$

where $f = f(\mathbf{x}, \boldsymbol{\xi}, t)$ is the single particle distribution function, $\boldsymbol{\xi}$ is the microscopic velocity, and $\Omega_f$ is the collision term. Using the Bhatnagar-Gross-Krook (BGK) collision operator [8, 9], the collision term is given by $\Omega_f = -(f - f^{eq})/\tau_f$, in which $\tau_f$ is the relaxation time and $f^{eq}$ is the continuous Maxwell-Boltzmann distribution function [1, 11]

$$f^{eq} = \frac{\rho}{(2\pi RT)^{D/2}} \exp\left[-\frac{(\boldsymbol{\xi} - \mathbf{v})^2}{2RT}\right],\qquad(2)$$

where $R$ is the gas constant, $D$ is the spatial dimension, $\rho$ is the density, $T$ is the temperature, and $\mathbf{v}$ is the macroscopic velocity. By discretizing the velocity $\boldsymbol{\xi}$ into a set of lattice velocities: $\{\mathbf{e}_\alpha\}$, where $\alpha = 0, 1, \cdots, n-1$, the following discrete Boltzmann-BGK equation can be obtained:

$$\frac{\partial f_\alpha}{\partial t} + \mathbf{e}_\alpha \cdot \nabla f_\alpha = -\frac{f_\alpha - f_\alpha^{eq}}{\tau_f},\qquad(3)$$

where $f_\alpha$ is the discrete density distribution function and $f_\alpha^{eq}$ is its equilibrium distribution. Integrating Eq. (3) over a time interval $\delta_t$, we can obtain [118]

$$f_\alpha\left(\mathbf{x} + \mathbf{e}_\alpha \delta_t, t + \delta_t\right) - f_\alpha(\mathbf{x}, t) = -\int_t^{t+\delta_t} \frac{f_\alpha - f_\alpha^{eq}}{\tau_f} \mathrm{dt} .\qquad(4)$$

If the integrand on the right-hand side of Eq. (4) is assumed to be constant over the time interval, then the following equation can be attained:

$$f_\alpha\left(\mathbf{x} + \mathbf{e}_\alpha \delta_t, t + \delta_t\right) - f_\alpha(\mathbf{x}, t) = -\frac{1}{\tau}\left(f_\alpha - f_\alpha^{eq}\right),\qquad(5)$$

where $\tau = \tau_f/\delta_t$ is the non-dimensional relaxation time. The above equation is just the *standard LB equation*. The assumption that the collision term is constant in the time interval $[t,\ t + \delta_t]$ will yield an artificial viscosity at the Navier-Stokes level. Fortunately, the artificial viscosity can be absorbed into the real viscosity of the fluid [2, 9].

The space $\mathbf{x}$ is usually discretized in such a way that $\mathbf{e}_\alpha \delta_t$ is the distance between two neighboring grid points. Then after one time step $\delta_t$, $f_\alpha(\mathbf{x},t)$ will arrive at its neighboring grid site along the lattice velocity direction $\mathbf{e}_\alpha$. Hence the LB equation can be split into two processes: the "collision" process

$$f_\alpha^*(\mathbf{x},t) = f_\alpha(\mathbf{x},t) - \frac{1}{\tau}\left[f_\alpha(\mathbf{x},t) - f_\alpha^{eq}(\mathbf{x},t)\right], \tag{6}$$

and the "streaming" process

$$f_\alpha(\mathbf{x} + \mathbf{e}_\alpha \delta_t, t + \delta_t) = f_\alpha^*(\mathbf{x},t). \tag{7}$$

From Eqs. (6) and (7) we can see that the collision process is completely local and the streaming process is completely linear. Actually, many advantages of the LB method arise from such a feature. For instance, it can be seen that most of the computations take place locally at the collision process, which makes the LB method highly amenable to parallel computing [1, 23].

In the literature, the set of the lattice velocities $\{\mathbf{e}_\alpha\}$ is often denoted as the D$d$Q$n$ lattice model [9], where $d$ and $n$ represent the spatial dimension and the total number of the lattice velocities, respectively. The widely used two-dimensional nine-velocity lattice model is named "D2Q9" and its lattice velocities $\{\mathbf{e}_\alpha\}$ are given by

$$\mathbf{e}_\alpha = \begin{cases} (0,\ 0), & \alpha = 0, \\ c\left(\cos\left[(\alpha-1)\pi/2\right],\ \sin\left[(\alpha-1)\pi/2\right]\right), & \alpha = 1-4, \\ \sqrt{2}c\left(\cos\left[(2\alpha-9)\pi/4\right],\ \sin\left[(2\alpha-9)\pi/4\right]\right), & \alpha = 5-8, \end{cases} \tag{8}$$

where $c = \delta_x/\delta_t$ is the lattice constant and $\delta_x$ is the lattice spacing. The equilibrium distribution function $f_\alpha^{eq}$ is given by [1, 2, 9]

$$f_\alpha^{eq} = \omega_\alpha \rho \left[1 + \frac{\mathbf{e}_\alpha \cdot \mathbf{v}}{c_s^2} + \frac{\mathbf{v}\mathbf{v} : \left(\mathbf{e}_\alpha \mathbf{e}_\alpha - c_s^2 \mathbf{I}\right)}{2c_s^4}\right], \tag{9}$$

where $\mathbf{I}$ is the unit tensor, $c_s = c/\sqrt{3}$ is the lattice sound speed, and $\omega_\alpha$ are the weights, which are given as follows: $\omega_0 = 4/9$, $\omega_{1-4} = 1/9$, and $\omega_{5-8} = 1/36$.

## 2.2 The LB-MRT formulation

Owing to its extreme simplicity, the BGK collision operator is still the most frequently used operator in the LB community. Nevertheless, the LB-BGK equation usually suffers from severe numerical instability at high Reynolds numbers (the relaxation time $\tau$ is close to 0.5). Hence several alternative collision operators, such as the multiple-relaxation-time (MRT) collision operator [120-123] and the two-relaxation-time collision operator [124-126], have been proposed. The entropic LB method [127-132] and the cascaded LB method [133-136] have also attracted significant attention.

The MRT collision operator is an important extension of the relaxation LB method proposed by Higuera $et$ $al.$ [5, 6] and the standard LB-MRT equation is given by [14, 120-123]:

$$f_\alpha\left(\mathbf{x}+\mathbf{e}_\alpha\delta_t,t+\delta_t\right)=f_\alpha\left(\mathbf{x},t\right)-\bar{\Lambda}_{\alpha\beta}\left[f_\beta\left(\mathbf{x},t\right)-f_\beta^{eq}\left(\mathbf{x},t\right)\right], \tag{10}$$

where $\bar{\Lambda}=\mathbf{M}^{-1}\Lambda\mathbf{M}$ is the collision matrix, in which $\Lambda$ is a diagonal matrix, $\mathbf{M}$ is an orthogonal transformation matrix, and $\mathbf{M}^{-1}$ is the inverse matrix of $\mathbf{M}$. For the D2Q9 model, the transformation matrix $\mathbf{M}$ can be given by (the lattice constant $c=1$) [121]

$$\mathbf{M}=\begin{bmatrix} 1 & 1 & 1 & 1 & 1 & 1 & 1 & 1 & 1 \\ -4 & -1 & -1 & -1 & -1 & 2 & 2 & 2 & 2 \\ 4 & -2 & -2 & -2 & -2 & 1 & 1 & 1 & 1 \\ 0 & 1 & 0 & -1 & 0 & 1 & -1 & -1 & 1 \\ 0 & -2 & 0 & 2 & 0 & 1 & -1 & -1 & 1 \\ 0 & 0 & 1 & 0 & -1 & 1 & 1 & -1 & -1 \\ 0 & 0 & -2 & 0 & 2 & 1 & 1 & -1 & -1 \\ 0 & 1 & -1 & 1 & -1 & 0 & 0 & 0 & 0 \\ 0 & 0 & 0 & 0 & 0 & 1 & -1 & 1 & -1 \end{bmatrix}. \tag{11}$$

Using the transformation matrix, $f_\alpha$ and its equilibrium distribution function $f_\alpha^{eq}$ can be projected onto the moment space. For the D2Q9 model, the following results can be obtained [121, 137]:

$$\mathbf{m}=\mathbf{Mf}=\mathrm{M}_{\alpha\beta}f_\beta=\left(\rho,e,\varsigma,j_x,q_x,j_y,q_y,p_{xx},p_{xy}\right)^{\mathrm{T}}, \tag{12}$$

$$\begin{aligned}\mathbf{m}^{eq}=\mathbf{Mf}^{eq}=\mathrm{M}_{\alpha\beta}f_\beta^{eq}&=\left(\rho,e^{eq},\varsigma^{eq},j_x,q_x^{eq},j_y,q_y^{eq},p_{xx}^{eq},p_{xy}^{eq}\right)^{\mathrm{T}}\\ &=\rho\left(1,-2+3\left|\mathbf{v}\right|^2,1-3\left|\mathbf{v}\right|^2,v_x,-v_x,v_y,-v_y,v_x^2-v_y^2,v_xv_y\right)^{\mathrm{T}},\end{aligned} \tag{13}$$

where $\mathbf{f}=\left(f_0,f_1,\cdots,f_8\right)^{\mathrm{T}}$, $\mathbf{f}^{eq}=\left(f_0^{eq},\cdots,f_8^{eq}\right)^{\mathrm{T}}$, $\rho$ is the density, $\left|\mathbf{v}\right|^2=\mathbf{v}\cdot\mathbf{v}$, $e$ is the energy mode, $\varsigma$ is related to energy square, $\left(j_x,j_y\right)$ are the momentum components, $\left(q_x,q_y\right)$ correspond to energy flux, $\left(p_{xx},p_{xy}\right)$ are related to the diagonal and off-diagonal components of the stress tensor, $v_x$ and $v_y$ are the $x$- and $y$-components of the macroscopic velocity $\mathbf{v}$, respectively, and the superscript "T" denotes the transpose operator.

For the D2Q9 model, the diagonal matrix $\mathbf{\Lambda}$, which consists of the relaxation times, is given by

$$\mathbf{\Lambda} = \mathrm{diag}\left( \tau_\rho^{-1}, \tau_e^{-1}, \tau_\varsigma^{-1}, \tau_j^{-1}, \tau_q^{-1}, \tau_j^{-1}, \tau_q^{-1}, \tau_v^{-1}, \tau_v^{-1} \right), \tag{14}$$

where $\tau_\rho$ and $\tau_j$ are the relaxation times of conserved moments and can be set to 1.0, $\tau_v$ determines the dynamic viscosity $\mu = \rho c_s^2 \left( \tau_v - 0.5 \right) \delta_t$, and $\tau_e$ is related to the bulk viscosity. The transformation matrix $\mathbf{M}$, the equilibria $\mathbf{m}^{eq}$, and the diagonal matrix $\mathbf{\Lambda}$ of the D3Q15 and D3Q19 models can be found in Refs. [122, 138]. The collision process of the LB-MRT equation (10) can be carried out in the moment space by multiplying through the transformation matrix to obtain

$$\mathbf{m}^* = \mathbf{m} - \mathbf{\Lambda}\left( \mathbf{m} - \mathbf{m}^{eq} \right), \tag{15}$$

where $\mathbf{m}^* = \left( m_0^*, m_1^*, \cdots, m_8^* \right)^{\mathrm{T}}$. Then the streaming process can be formulated as follows:

$$f_\alpha \left( \mathbf{x} + \mathbf{e}_\alpha \delta_t, t + \delta_t \right) = f_\alpha^* \left( \mathbf{x}, t \right), \tag{16}$$

where $f_\alpha^* = \mathbf{M}^{-1}\mathbf{m}^* = M_{\alpha\beta}^{-1} m_\beta^*$.

The Chapman-Enskog analysis can be applied to the LB-MRT equation to derive the macroscopic equations. For details about this procedure, readers are referred to Refs. [137, 139-141]. In the literature, it has been demonstrated [123, 140, 142-148] that the MRT collision model is superior over the BGK collision model in terms of numerical stability because the relaxation times in the MRT collision operator can be individually tuned to achieve "optimal" stability [122].

## 2.3 The forcing schemes

The forcing scheme, which is used to incorporate an external force into the LB equation, plays an important role in the LB method. It determines whether the force of the system is correctly implemented and therefore affects the numerical accuracy of the LB model. Here we focus on introducing several forcing schemes that have been demonstrated to be accurate in recovering the *unsteady* macroscopic equations at the Navier-Stokes level in the low Mach number limit. In addition, a forcing scheme proposed by Kupershtokh *et al*. [149] will also be reviewed. For *steady* problems, readers are referred to the studies of Mohamad and Kuzmin [150] and Silva and Semiao [151].

### 2.3.1 He et al.'s scheme and Guo et al.'s scheme

In 1998, He *et al*. [152] proposed a forcing scheme based on the complete Boltzmann-BGK equation with an external body force, which is given by

$$\frac{\partial f}{\partial t} + \boldsymbol{\xi} \cdot \nabla f + \boldsymbol{a} \cdot \nabla_{\xi} f = -\frac{f - f^{eq}}{\tau_f}, \tag{17}$$

where $\boldsymbol{a}$ is the acceleration due to the body force $\mathbf{F} = \rho \boldsymbol{a}$ [139] and $\nabla_{\xi} f = \partial f / \partial \boldsymbol{\xi}$. Considering that the equilibrium distribution function $f^{eq}$ is the leading part of the distribution function $f$ and the gradient of $f^{eq}$ has the most important contribution to the gradient of $f$, He *et al*. assumed [152] that $\nabla_{\xi} f \approx \nabla_{\xi} f^{eq}$. With the aid of Eq. (2), they obtained $\nabla_{\xi} f^{eq} = -(\boldsymbol{\xi} - \mathbf{v}) f^{eq} / RT$. Then the discrete Boltzmann-BGK equation with an external force can be written as follows:

$$\frac{\partial f_{\alpha}}{\partial t} + \mathbf{e}_{\alpha} \cdot \nabla f_{\alpha} = -\frac{f_{\alpha} - f_{\alpha}^{eq}}{\tau_f} + \frac{(\mathbf{e}_{\alpha} - \mathbf{v}) \cdot \mathbf{F}}{\rho c_s^2} f_{\alpha}^{eq}, \tag{18}$$

where $c_s = \sqrt{RT}$. As mentioned earlier, the collision term can be integrated constantly over the integral interval. For the second term on the right-hand side of Eq. (18), He *et al*. [152] stressed that a trapezoidal rule is required in order to achieve second-order accuracy in time, which gives

$$f_{\alpha}\left(\mathbf{x} + \mathbf{e}_{\alpha}\delta_t, t + \delta_t\right) - f_{\alpha}\left(\mathbf{x}, t\right) = -\frac{f_{\alpha} - f_{\alpha}^{eq}}{\tau} + \frac{\delta_t}{2}\left[\frac{(\mathbf{e}_{\alpha} - \mathbf{v}) \cdot \mathbf{F} f_{\alpha}^{eq}}{\rho c_s^2}\bigg|_{t+\delta_t} + \frac{(\mathbf{e}_{\alpha} - \mathbf{v}) \cdot \mathbf{F} f_{\alpha}^{eq}}{\rho c_s^2}\bigg|_t\right], \tag{19}$$

where $f_{\alpha}^{eq} = f_{\alpha}^{eq}\left(\rho, \mathbf{v}\right)$ and $\tau = \tau_f / \delta_t$. The implicitness of Eq. (19) can be eliminated with [152]

$$\overline{f}_{\alpha} = f_{\alpha} - \frac{\delta_t}{2} \frac{(\mathbf{e}_{\alpha} - \mathbf{v}) \cdot \mathbf{F}}{\rho c_s^2} f_{\alpha}^{eq}. \tag{20}$$

Using $\overline{f}_{\alpha}$, Eq. (19) will become

$$\overline{f}_{\alpha}\left(\mathbf{x} + \mathbf{e}_{\alpha}\delta_t, t + \delta_t\right) - \overline{f}_{\alpha}\left(\mathbf{x}, t\right) = -\frac{\overline{f}_{\alpha} - f_{\alpha}^{eq}}{\tau} + \delta_t\left(1 - \frac{1}{2\tau}\right)\frac{(\mathbf{e}_{\alpha} - \mathbf{v}) \cdot \mathbf{F}}{\rho c_s^2} f_{\alpha}^{eq}. \tag{21}$$

The second term on the right-hand side of Eq. (21) is just the forcing term. Meanwhile, according to Eq.

(20), the macroscopic density and velocity should be calculated by [152]

$$\rho = \sum_{\alpha} \bar{f}_{\alpha} , \quad \rho \mathbf{v} = \sum_{\alpha} \bar{f}_{\alpha} \mathbf{e}_{\alpha} + \frac{\delta_t}{2} \mathbf{F} . \tag{22}$$

To sum up, Eq. (21) together with Eq. (22) constitutes He *et al.*'s forcing scheme. In fact, it can be seen that Eqs. (21) and (22) are self-consistent; hence the hat "$-$" of $\bar{f}_{\alpha}$ in Eqs. (21) and (22) can be dropped in practical applications. However, the transformation given by Eq. (20) should be kept in mind in some cases, which will be mentioned below.

A similar forcing scheme was later devised by Guo *et al.* [153] in 2002. The major difference is that Guo *et al.*'s forcing term is written in a power series of $\mathbf{e}_{\alpha}$:

$$F_{\alpha} = \omega_{\alpha} \delta_t \left[ \frac{\mathbf{B} \cdot \mathbf{e}_{\alpha}}{c_s^2} + \frac{\mathbf{C} : \left( \mathbf{e}_{\alpha} \mathbf{e}_{\alpha} - c_s^2 \mathbf{I} \right)}{2 c_s^4} \right] , \tag{23}$$

where $\mathbf{B}$ and $\mathbf{C}$ are functions of the force $\mathbf{F}$ and determined by requiring that the moments of the forcing term $F_{\alpha}$ are consistent with the target macroscopic equations. Using the Chapman-Enskog analysis, Guo *et al.* [153] found that $\mathbf{B}$ and $\mathbf{C}$ should be chosen as follows so as to recover the correct unsteady macroscopic equations at the Navier-Stokes level:

$$\mathbf{B} = \left( 1 - \frac{1}{2\tau} \right) \mathbf{F} , \quad \mathbf{C} = \left( 1 - \frac{1}{2\tau} \right) \left( \mathbf{v}\mathbf{F} + \mathbf{F}\mathbf{v} \right) . \tag{24}$$

The LB equation with Guo *et al.*'s forcing term is then given by [153]

$$f_{\alpha} \left( \mathbf{x} + \mathbf{e}_{\alpha} \delta_t , t + \delta_t \right) - f_{\alpha} \left( \mathbf{x}, t \right) = - \frac{f_{\alpha} - f_{\alpha}^{eq}}{\tau} + \delta_t \left( 1 - \frac{1}{2\tau} \right) \omega_{\alpha} \left[ \frac{\mathbf{e}_{\alpha} - \mathbf{v}}{c_s^2} + \frac{\left( \mathbf{e}_{\alpha} \cdot \mathbf{v} \right)}{c_s^4} \mathbf{e}_{\alpha} \right] \cdot \mathbf{F} , \tag{25}$$

where $f_{\alpha}^{eq} = f_{\alpha}^{eq} \left( \rho, \mathbf{v} \right)$. Furthermore, Guo *et al.* [153] pointed out that the macroscopic variables should be defined as follows:

$$\rho = \sum_{\alpha} f_{\alpha} , \quad \rho \mathbf{v} = \sum_{\alpha} f_{\alpha} \mathbf{e}_{\alpha} + \frac{\delta_t}{2} \mathbf{F} . \tag{26}$$

Comparing Eqs. (25) and (26) with Eqs. (21) and (22), we can find that He *et al.*'s scheme and Guo *et al.*'s scheme are basically the same except that the form of the forcing term is different. Actually, Guo and Zheng [137] have shown that the difference between these two forcing schemes lies in the second-order moment of the forcing term:

$$\sum_\alpha \mathbf{e}_\alpha \mathbf{e}_\alpha F_{\alpha,\,\mathrm{Guo}} = \delta_t \left(1 - \frac{1}{2\tau}\right)\left(\mathbf{vF} + \mathbf{Fv}\right),$$ (27)

$$\sum_\alpha \mathbf{e}_\alpha \mathbf{e}_\alpha F_{\alpha,\,\mathrm{He}} = \delta_t \left(1 - \frac{1}{2\tau}\right)\left(\mathbf{vF} + \mathbf{Fv} - \frac{\mathbf{F}\cdot\mathbf{v}}{c_s^2}\mathbf{vv}\right).$$ (28)

The third-order velocity term in Eq. (28) is an error term resulting from the deviation between the third-order moment of $f_\alpha^{eq}$ and that of the Maxwell-Boltzmann distribution function $f^{eq}$. However, for low Mach number flows, the third-order velocity term can be neglected and then the two forcing schemes will be equivalent.

Previously we have mentioned that the hat "$-$" of $\bar{f}_\alpha$ in Eqs. (21) and (22) can be dropped in practical applications. Nevertheless, we should keep Eq. (20) in mind in some cases, e.g., in the case of calculating the strain rate tensor using the second-order moment of the non-equilibrium distribution function, which is often required in the LB simulations of non-Newtonian flows [154-157] and in the LB-based simulations of turbulent flows [158, 159]. According to the Chapman-Enskog analysis of the standard LB-BGK equation, namely Eq. (5), we can find that the strain rate tensor $\boldsymbol{\phi} = \left[\nabla\mathbf{v} + (\nabla\mathbf{v})^{\mathrm{T}}\right]\big/2$ can be computed locally as follows [154, 158]:

$$\boldsymbol{\phi} \approx -\frac{1}{2\rho c_s^2 \tau \delta_t} \sum_\alpha \mathbf{e}_\alpha \mathbf{e}_\alpha \left(f_\alpha - f_\alpha^{eq}\right).$$ (29)

The above formulation is also valid for Eq. (19). Hence, when Eqs. (21) and (22) are used, $f_\alpha$ in Eq. (29) should be evaluated according to Eq. (20), which yields

$$\boldsymbol{\phi} \approx -\frac{1}{2\rho c_s^2 \tau \delta_t}\left[\sum_\alpha \mathbf{e}_\alpha \mathbf{e}_\alpha \left(\bar{f}_\alpha - f_\alpha^{eq}\right) + 0.5\delta_t\left(\mathbf{vF} + \mathbf{Fv} - \frac{\mathbf{F}\cdot\mathbf{vvv}}{c_s^2}\right)\right].$$ (30)

Moreover, it can be found that $f_\alpha$ in Eqs. (25) and (26) is actually equivalent to $\bar{f}_\alpha$ in Eqs. (21) and (22). Hence, for Guo *et al.*'s forcing scheme, one can replace $\bar{f}_\alpha$ in Eq. (30) with $f_\alpha$ and remove the third-order velocity term [160].

### 2.3.2 Wagner's scheme and the exact-difference-method scheme

In 2006, Wagner [161] constructed a forcing scheme as follows:

$$f_\alpha \left( \mathbf{x} + \mathbf{e}_\alpha \delta_t, t + \delta_t \right) - f_\alpha \left( \mathbf{x}, t \right) = -\frac{f_\alpha - f_\alpha^{eq} \left( \rho, \mathbf{u} \right)}{\tau} + F_{\alpha, \mathrm{Wa}} \,, \tag{31}$$

where the forcing term $F_{\alpha, \mathrm{Wa}}$ satisfies the following constraints (see Eqs. (20), (21), and (59) in Ref. [161] and neglect the high-order term beyond the Navier-Stokes level):

$$\sum_\alpha F_{\alpha, \mathrm{Wa}} = 0 \,, \quad \sum_\alpha \mathbf{e}_\alpha F_{\alpha, \mathrm{Wa}} = \delta_t \mathbf{F} \,, \tag{32}$$

$$\sum_\alpha \mathbf{e}_\alpha \mathbf{e}_\alpha F_{\alpha, \mathrm{Wa}} = \delta_t \left[ \left( \mathbf{u}\mathbf{F} + \mathbf{F}\mathbf{u} \right) + \delta_t \left( 1 - \frac{1}{4\tau} \right) \frac{\mathbf{F}\mathbf{F}}{\rho} \right] . \tag{33}$$

By conducting the Taylor expansion analysis of Eq. (31), Wagner stressed that the forcing term must satisfy Eqs. (32) and (33) in order to recover the correct unsteady macroscopic equations at the Navier-Stokes level. It can be seen that, in Wagner's forcing scheme, the velocity in the equilibrium density distribution function and the forcing term is given by $\mathbf{u} = \sum_\alpha f_\alpha \mathbf{e}_\alpha / \rho$, while the actual fluid velocity is defined as $\mathbf{v} = \mathbf{u} + \delta_t \mathbf{F} / (2\rho)$ [161]. Hence the equilibrium distribution function in Eq. (31) can be rewritten as

$$f_\alpha^{eq} \left( \rho, \mathbf{u} \right) \equiv f_\alpha^{eq} \left( \rho, \mathbf{v} \right) + \left[ f_\alpha^{eq} \left( \rho, \mathbf{u} \right) - f_\alpha^{eq} \left( \rho, \mathbf{v} \right) \right], \tag{34}$$

which means that the actual forcing term of Wagner's formulation is given by

$$\widehat{F}_{\alpha, \mathrm{Wa}} = F_{\alpha, \mathrm{Wa}} + \frac{f_\alpha^{eq} \left( \rho, \mathbf{u} \right) - f_\alpha^{eq} \left( \rho, \mathbf{v} \right)}{\tau} . \tag{35}$$

According to Eqs. (32), (33), and (35), it can be found that

$$\sum_\alpha \mathbf{e}_\alpha \widehat{F}_{\alpha, \mathrm{Wa}} = \delta_t \left( 1 - \frac{1}{2\tau} \right) \mathbf{F} \,, \tag{36}$$

$$\sum_\alpha \mathbf{e}_\alpha \mathbf{e}_\alpha \widehat{F}_{\alpha, \mathrm{Wa}} = \delta_t \left( 1 - \frac{1}{2\tau} \right) \left[ \left( \mathbf{u}\mathbf{F} + \mathbf{F}\mathbf{u} \right) + \delta_t \frac{\mathbf{F}\mathbf{F}}{\rho} \right] . \tag{37}$$

Substituting $\mathbf{v} = \mathbf{u} + \delta_t \mathbf{F} / (2\rho)$ into Eq. (27), we can find that Eq. (27) and Eq. (37) are identical. Similarly, it can be found that Eq. (36) is the same as the result of Guo *et al.*'s forcing scheme, which means that Wagner's forcing scheme is identical to Guo *et al.*'s forcing scheme, although they were given in different forms.

A forcing scheme, which was named "exact-difference-method (EDM) scheme" by Kupershtokh *et al.* [149, 162], is now reviewed. On the basis of He *et al.*'s work [152], Kupershtokh *et al.* found that [162] $\nabla_\xi f^{eq} = -\nabla_{\mathbf{u}} f^{eq}$. Meanwhile, the acceleration $\boldsymbol{a}$ in Eq. (17) was expressed as $\boldsymbol{a} = \mathrm{d}\mathbf{u}/\mathrm{d}t$. For isothermal flows, $f^{eq}$ is a function of $\rho$ and $\mathbf{u}$. With the chain rule, we can obtain

$$\frac{\mathrm{d}f^{eq}(\rho,\mathbf{u})}{\mathrm{d}t} = \frac{\partial f^{eq}}{\partial \mathbf{u}} \cdot \frac{\mathrm{d}\mathbf{u}}{\mathrm{d}t} + \frac{\partial f^{eq}}{\partial \rho}\frac{\mathrm{d}\rho}{\mathrm{d}t}, \tag{38}$$

where $\partial f^{eq}/\partial \mathbf{u} = \nabla_{\mathbf{u}} f^{eq}$. Kupershtokh *et al.* [162] assumed that $f^{eq}$ is only functional of $\mathbf{u}$ and then obtained $\mathrm{d}f^{eq}(\mathbf{u})/\mathrm{d}t = -\boldsymbol{a}\cdot\nabla_{\xi}f^{eq}$, which yields the following discrete Boltzmann-BGK equation:

$$\frac{\partial f_{\alpha}}{\partial t} + \mathbf{e}_{\alpha}\cdot\nabla f_{\alpha} = -\frac{f_{\alpha}-f_{\alpha}^{eq}}{\tau_f} + \frac{\mathrm{d}f_{\alpha}^{eq}(\mathbf{u})}{\mathrm{d}t}. \tag{39}$$

Integrating Eq. (39) over the time interval $[t,\ t+\delta_t]$ gives

$$f_{\alpha}(\mathbf{x}+\mathbf{e}_{\alpha}\delta_t, t+\delta_t) = f_{\alpha}(\mathbf{x},t) - \frac{f_{\alpha}-f_{\alpha}^{eq}}{\tau} + f_{\alpha}^{eq}(\mathbf{u}^{t+\delta_t}) - f_{\alpha}^{eq}(\mathbf{u}^t). \tag{40}$$

The above equation is implicit because the velocity at the time level $t+\delta_t$ is unknown. Kupershtokh *et al.* therefore simplified Eq. (40) as

$$f_{\alpha}(\mathbf{x}+\mathbf{e}_{\alpha}\delta_t, t+\delta_t) - f_{\alpha}(\mathbf{x},t) = -\frac{1}{\tau}\Big[f_{\alpha}-f_{\alpha}^{eq}(\rho,\mathbf{u})\Big] + f_{\alpha}^{eq}(\rho,\mathbf{u}+\Delta\mathbf{u}) - f_{\alpha}^{eq}(\rho,\mathbf{u}), \tag{41}$$

where $\rho$ and $\mathbf{u} = \sum_{\alpha} f_{\alpha}\mathbf{e}_{\alpha}/\rho$ are the density and the velocity at the $t$ time level, respectively, and $\Delta\mathbf{u} = \mathbf{F}\delta_t/\rho$. Equation (41) is the EDM forcing scheme defined by Kupershtokh *et al.* [149]. By comparing Eq. (41) with Eq. (40), it can be found that the velocity $\mathbf{u}^{t+\delta_t}$ was evaluated as follows:

$$\mathbf{u}^{t+\delta_t} \equiv \mathbf{u}^t + \int_t^{t+\delta_t}\frac{\mathrm{d}\mathbf{u}}{\mathrm{d}t}\mathrm{d}t = \mathbf{u}^t + \int_t^{t+\delta_t}\frac{\mathbf{F}}{\rho}\mathrm{d}t \approx \mathbf{u}^t + \frac{\mathbf{F}\delta_t}{\rho}, \tag{42}$$

which indicates that the acceleration $\boldsymbol{a} = \mathbf{F}/\rho$ is assumed to be constant over the time.

From the above analysis, we can see that several assumptions were used in deriving the EDM forcing scheme. Furthermore, Eq. (41) can also be analyzed by substituting the expression of $f_{\alpha}^{eq}$ into the forcing term. According to Eq. (9), $f_{\alpha}^{eq}(\rho,\mathbf{u}+\Delta\mathbf{u})$ can be written as follows [163]:

$$f_{\alpha}^{eq}(\rho,\mathbf{u}+\Delta\mathbf{u}) \equiv f_{\alpha}^{eq}(\rho,\mathbf{u}) + \rho\omega_{\alpha}\left[\frac{\Delta\mathbf{u}\cdot\mathbf{e}_{\alpha}}{c_s^2} + \frac{(\Delta\mathbf{u}\Delta\mathbf{u}+2\mathbf{u}\Delta\mathbf{u}):(\mathbf{e}_{\alpha}\mathbf{e}_{\alpha}-c_s^2\mathbf{I})}{2c_s^4}\right]. \tag{43}$$

With the aid of Eq. (43) and $\Delta\mathbf{u} = \mathbf{F}\delta_t/\rho$, Li *et al.* [163] obtained

$$F_{\alpha,\text{EDM}} = \omega_{\alpha}\delta_t\left[\frac{\mathbf{F}\cdot\mathbf{e}_{\alpha}}{c_s^2} + \frac{(\mathbf{v}_{\text{EDM}}\mathbf{F}+\mathbf{F}\mathbf{v}_{\text{EDM}}):(\mathbf{e}_{\alpha}\mathbf{e}_{\alpha}-c_s^2\mathbf{I})}{2c_s^4}\right], \tag{44}$$

where $\mathbf{v}_{EDM} = \mathbf{u} + \Delta\mathbf{u}/2 = \mathbf{u} + \delta_t\mathbf{F}/(2\rho)$. The above equation reveals that the forcing term of the EDM scheme can be written in the general form of forcing schemes and the velocity in the EDM forcing term is given by $\mathbf{v}_{EDM} = \mathbf{u} + \delta_t\mathbf{F}/(2\rho)$ [163].

Actually, Wagner's forcing scheme and the EDM forcing scheme belong to the same class of forcing schemes, in which the velocity in $f_\alpha^{eq}$ and the actual fluid velocity are given by $\mathbf{u} = \sum_\alpha f_\alpha\mathbf{e}_\alpha/\rho$ and $\mathbf{v} = \mathbf{u} + \delta_t\mathbf{F}/(2\rho)$, respectively. For this class of forcing schemes, the usual Chapman-Enskog analysis in the literature should be revised so as to derive the actual macroscopic equations recovered from these forcing schemes. For the EDM forcing scheme, the following macroscopic momentum equation can be obtained at the Navier-Stokes level:

$$\partial_t(\rho\mathbf{v}) + \nabla\cdot(\rho\mathbf{v}\mathbf{v}) = -\nabla p + \nabla\cdot\boldsymbol{\Pi} + \mathbf{F} - \delta_t^2\nabla\cdot\left(\frac{\mathbf{F}\mathbf{F}}{4\rho}\right), \tag{45}$$

where $\boldsymbol{\Pi}$ is the viscous stress tensor. The last term on the right-hand side of Eq. (45) is an additional (error) term. When the force $\mathbf{F}$ is spatially uniform and the variation of density in space is very small, the error term in Eq. (45) can be neglected.

### 2.3.3 The MRT forcing scheme

In 2005, McCracken and Abraham [139] proposed a forcing scheme for the LB-MRT equation. Based on He *et al.*'s work [152], McCracken and Abraham started with the following equation:

$$f_\alpha(\mathbf{x}+\mathbf{e}_\alpha\delta_t, t+\delta_t) = f_\alpha(\mathbf{x},t) - \widehat{\Lambda}_{\alpha\beta}\left(f_\beta - f_\beta^{eq}\right)\Big|_{(\mathbf{x},t)} + \frac{\delta_t}{2}\left[G_\alpha|_{(\mathbf{x}+\mathbf{e}_\alpha\delta_t, t+\delta_t)} + G_\alpha|_{(\mathbf{x},t)}\right], \tag{46}$$

where $G_\alpha = (\mathbf{e}_\alpha - \mathbf{v})\cdot\mathbf{F}f_\alpha^{eq}/(\rho c_s^2)$. Similarly, the implicitness of Eq. (46) can be eliminated using

$$\overline{f}_\alpha = f_\alpha - \frac{\delta_t}{2}G_\alpha, \tag{47}$$

which yields [139]

$$\overline{f}_\alpha(\mathbf{x}+\mathbf{e}_\alpha\delta_t, t+\delta_t) = \overline{f}_\alpha(\mathbf{x},t) - \widehat{\Lambda}_{\alpha\beta}\left(\overline{f}_\beta - f_\beta^{eq}\right)\Big|_{(\mathbf{x},t)} + \delta_t\left(G_\alpha - 0.5\widehat{\Lambda}_{\alpha\beta}G_\beta\right)\Big|_{(\mathbf{x},t)}. \tag{48}$$

As previously mentioned, the collision process of the LB-MRT equation can be implemented in the moment space. From Eq. (48), the following equation can be obtained [164]

$$\mathbf{m}^* = \bar{\mathbf{m}} - \Lambda\left(\bar{\mathbf{m}} - \mathbf{m}^{eq}\right) + \delta_t\left(\mathbf{I} - \frac{\Lambda}{2}\right)\mathbf{S}, \tag{49}$$

where $\bar{\mathbf{m}} = \mathbf{M}\bar{\mathbf{f}}$ with $\bar{\mathbf{f}} = \left(\bar{f}_0, \bar{f}_1, \cdots, \bar{f}_8\right)^{\mathrm{T}}$ for the D2Q9 model and $\delta_t\left(\mathbf{I} - 0.5\Lambda\right)\mathbf{S}$ is the forcing

term in the moment space with $\mathbf{S} = \mathbf{M}\mathbf{G}$. The streaming process is given by

$$\bar{f}_\alpha\left(\mathbf{x} + \mathbf{e}_\alpha\delta_t, t + \delta_t\right) = f_\alpha^*\left(\mathbf{x}, t\right), \tag{50}$$

where $\mathbf{f}^* = \mathbf{M}^{-1}\mathbf{m}^*$. According to the formulation $\mathbf{S} = \mathbf{M}\mathbf{G}$ and the expression of $G_\alpha$, McCracken and Abraham [139] attained the following $\mathbf{S} = \left(S_0, S_1, \cdots, S_8\right)^{\mathrm{T}}$ for the D2Q9 model:

$$\mathbf{S} = \begin{bmatrix} 0 \\ 6\left(v_x F_x + v_y F_y\right) \\ -6\left(v_x F_x + v_y F_y\right) \\ F_x \\ -F_x \\ F_y \\ -F_y \\ 2\left(v_x F_x - v_y F_y\right) \\ \left(v_x F_y + v_y F_x\right) \end{bmatrix}, \tag{51}$$

where $F_x$ and $F_y$ are the $x$- and $y$-components of the force $\mathbf{F}$, respectively. Note that the

third-order velocity terms have been omitted in Eq. (51). When $G_\alpha$ in Eq. (46) is taken following Guo

*et al*.'s approach [153], the same $\mathbf{S}$ will be obtained.

The macroscopic density and velocity are also calculated via Eq. (22). Through the

Chapman-Enskog analysis, McCracken and Abraham [139] demonstrated that the macroscopic

equations at the Navier-Stokes level with an external force can be correctly recovered using the above

MRT forcing scheme. Moreover, according to the Chapman-Enskog analysis [137, 139], we can find that, due to the freedom of the MRT collision model, the MRT forcing term or any other source terms in the LB-MRT equation can be directly constructed in the moment space [15, 137] without resorting to their BGK forms.

In the LB-MRT method, the strain rate tensor $\boldsymbol{\phi} = \left[\nabla\mathbf{v} + \left(\nabla\mathbf{v}\right)^{\mathrm{T}}\right]\big/2$ can also be calculated from the non-equilibrium part of the distribution functions, i.e., $\left(\mathbf{m} - \mathbf{m}^{eq}\right)$. For the D2Q9 model, the

Chapman-Enskog analysis of the standard LB-MRT Eq. (10) will yield the following results [137]:

$$-m_1^{(1)} \approx 2\rho\tau_e\delta_t\left(\partial_x v_x + \partial_y v_y\right),\tag{52}$$

$$-m_7^{(1)} \approx \frac{2}{3}\rho\tau_v\delta_t\left(\partial_x v_x - \partial_y v_y\right),\tag{53}$$

$$-m_8^{(1)} \approx \frac{1}{3}\rho\tau_v\delta_t\left(\partial_x v_y + \partial_y v_x\right),\tag{54}$$

where $m_1$, $m_7$, and $m_8$ correspond to $e$, $p_{xx}$, and $p_{xy}$ in Eq. (12), respectively, and $m_\alpha^{(1)} \approx m_\alpha - m_\alpha^{eq}$. The components of the strain rate tensor can be expressed using $m_1^{(1)}$, $m_7^{(1)}$ and $m_8^{(1)}$, as given in Eqs. (52)-(54). It can be found that the above results are also valid for Eq. (46). However, with the transformation given by Eq. (47), we should use $\mathbf{m} = \bar{\mathbf{m}} + 0.5\delta_t\mathbf{S}$ [143]. Hence, when Eqs. (49)-(51) are employed, $m_1^{(1)}$, $m_7^{(1)}$, and $m_8^{(1)}$ in the above equations should be evaluated as follows:

$$m_1^{(1)} = \bar{m}_1^{(1)} + 0.5\delta_t S_1, \quad m_7^{(1)} = \bar{m}_7^{(1)} + 0.5\delta_t S_7, \quad m_8^{(1)} = \bar{m}_8^{(1)} + 0.5\delta_t S_8,\tag{55}$$

where $\bar{m}_\alpha^{(1)} \approx \bar{m}_\alpha - m_\alpha^{eq}$. For the results of the D3Q19 model, readers are referred to Ref. [165].

### 2.4 The thermal LB method on standard lattices

In this section, the thermal LB method for simulating thermal flows on standard lattices is introduced. The standard lattices [166] represent the lattice models that are commonly used in the LB method, such as the D2Q9, D3Q15, and D3Q19 lattice models [9]. The earliest attempts to construct thermal LB models were made by Massaioli *et al.* [167], Alexander *et al.* [168], and Qian [169] in 1993. Since then many thermal LB models have been developed and most of these models can be classified into the following categories [15, 137, 170]: the multispeed approach [145, 168, 169, 171-188], the double-distribution-function approach [189-211], and the hybrid approach [51, 104, 111, 212-216].

The multispeed approach is a straightforward extension of the isothermal LB method. In this approach, high-order lattices are utilized and the equilibrium distribution function usually includes the higher-order velocity terms so as to recover the energy equation at the macroscopic level [174]. Numerous studies have been conducted within the framework of the multispeed LB approach and deserve a comprehensive review, which is beyond the scope of the present work. Readers can refer to Refs. [1, 15, 179] and the references therein. Here we briefly introduce the DDF approach and the hybrid approach, which are often constructed on the standard lattices and widely encountered in the simulations of thermal multiphase flows with the pseudopotential and the phase-field LB methods.

### 2.4.1 The double-distribution-function (DDF) approach

Unlike the multispeed LB approach, in which only the density distribution function is involved, the DDF LB approach utilizes two different distribution functions, one (the density distribution function) for the flow field and the other for the energy or temperature field. In the literature, various DDF LB models have been devised from different points of view [189-195, 197-202]. For convenience, these models are classified according to their target macroscopic energy equation.

Firstly we give the macroscopic energy equation of ideal gases in terms of the internal energy and the total energy, respectively, which can be found elsewhere:

$$\partial_t (\rho e) + \nabla \cdot (\rho e \mathbf{v}) = \nabla \cdot (\lambda \nabla T) - p \nabla \cdot \mathbf{v} + \mathbf{\Pi} : \nabla \mathbf{v} , \qquad (56)$$

$$\partial_t (\rho E) + \nabla \cdot (\rho E \mathbf{v}) = \nabla \cdot (\lambda \nabla T) - \nabla \cdot (p \mathbf{v}) + \nabla \cdot (\mathbf{\Pi} \cdot \mathbf{v}) , \qquad (57)$$

where $e = c_v T$ is the internal energy, $E = e + 0.5|\mathbf{v}|^2$ is the total energy, $\lambda$ is the thermal conductivity, $p$ is the pressure, $\mathbf{\Pi}$ is the viscous stress tensor, $p \nabla \cdot \mathbf{v}$ denotes the compression work, and $\mathbf{\Pi} : \nabla \mathbf{v}$ is the viscous heat dissipation.

The above two equations are equivalent for an ideal gas system without external forces. In the presence of an external force $\mathbf{F}$, a term $\mathbf{F} \cdot \mathbf{v}$ should be added to the right-hand side of Eq. (57). When the compression work and the viscous heat dissipation can be neglected, the internal energy equation (56) will reduce to

$$\partial_t (\rho c_v T) + \nabla \cdot (\rho c_v T \mathbf{v}) = \nabla \cdot (\lambda \nabla T) . \qquad (58)$$

Furthermore, if the specific heat $c_v$ is constant, then the following temperature equation can be derived from Eq. (58):

$$\partial_t T + \mathbf{v} \cdot \nabla T = \frac{1}{\rho c_v} \nabla \cdot (\lambda \nabla T) . \qquad (59)$$

According to the target macroscopic equations, most of the existing DDF LB models fall into the following three types: the internal-energy-based model, the total-energy-based model, and the temperature-based model.

### 2.4.1.1 The internal-energy- and total-energy-based DDF models

The internal-energy-based DDF LB model, which is also the most well-known DDF model in the LB community, was proposed by He $et$ $al.$ [189], who introduced an internal energy distribution function $g_\alpha$ to simulate the internal energy field, while the density and velocity fields are still modeled with the density distribution function $f_\alpha$. The two distribution functions $f_\alpha$ and $g_\alpha$ satisfy the following discrete Boltzmann equations [189]:

$$\frac{\partial f_\alpha}{\partial t} + \left(\mathbf{e}_\alpha \cdot \boldsymbol{\nabla}\right) f_\alpha = -\frac{1}{\tau_f}\left(f_\alpha - f_\alpha^{eq}\right) + G_\alpha ,  \qquad (60)$$

$$\frac{\partial g_\alpha}{\partial t} + \left(\mathbf{e}_\alpha \cdot \boldsymbol{\nabla}\right) g_\alpha = -\frac{1}{\tau_g}\left(g_\alpha - g_\alpha^{eq}\right) - f_\alpha q_\alpha ,  \qquad (61)$$

where $\tau_g$ is the relaxation time for the internal energy, which corresponds to the thermal conductivity $\lambda = \tau_g \rho c_s^2 \left(D+2\right) R / 2$, $G_\alpha$ is the forcing term, $g_\alpha^{eq}$ is the equilibrium internal energy distribution function, and $q_\alpha = \left(\mathbf{e}_\alpha - \mathbf{v}\right) \cdot \left[\partial_t \mathbf{v} + \left(\mathbf{e}_\alpha \cdot \boldsymbol{\nabla}\right)\mathbf{v}\right]$. The LB equations of $f_\alpha$ and $g_\alpha$ can be obtained by adopting a second-order integration for Eqs. (60) and (61) [189]. He $et$ $al.$ showed [189] that their model has excellent numerical stability and the viscous heat dissipation together with the compression work done by the pressure can be taken into account. The internal-energy-based DDF model has received much attention since its emergence and has been applied in many applications.

From Eq. (61) we can see that a complicated term $q_\alpha$, which involves the temporal and spatial derivatives of macroscopic variables, exists in the evolution equation of the internal energy distribution function. Guo $et$ $al.$ [190] pointed out that such a complicated term may introduce some additional errors and affect the numerical stability of the model. Based on He $et$ $al.$'s work, Shi $et$ $al.$ [192] found that $f_\alpha q_\alpha$ in Eq. (61) can be simplified as $f_\alpha q_\alpha = \left(f_\alpha - f_\alpha^{eq}\right)\mathbf{e}_\alpha \mathbf{e}_\alpha : \boldsymbol{\nabla}\mathbf{v}$ and demonstrated that such a change does not influence the recovered macroscopic equations at the Navier-Stokes level.

Nevertheless, the calculation of the velocity gradient term $\nabla \mathbf{v}$ is still required.

To solve this problem, Guo *et al*. [190] proposed a total-energy-based DDF LB model by introducing a total energy distribution function $h_\alpha$ to replace the internal energy distribution function $g_\alpha$. In Guo *et al*.'s model, the density distribution function $f_\alpha$ still obeys Eq. (60) but $G_\alpha$ is different. The evolution equation of the total energy distribution function was also derived from the Boltzmann equation and is given by [190]

$$\frac{\partial h_\alpha}{\partial t} + (\mathbf{e}_\alpha \cdot \nabla) h_\alpha = -\frac{1}{\tau_h}\left(h_\alpha - h_\alpha^{eq}\right) + \frac{Z_\alpha}{\tau_{hf}}\left(f_\alpha - f_\alpha^{eq}\right) + I_\alpha, \tag{62}$$

where $\tau_h$ is the relaxation time for the total energy, which corresponds to $\lambda = \tau_h \rho c_s^2 (D+2) R / 2$, $h_\alpha^{eq}$ is the equilibrium total energy distribution function, $\tau_{hf} = (\tau_f - \tau_h) / \tau_h \tau_f$, $Z_\alpha = \mathbf{e}_\alpha \cdot \mathbf{v} - |\mathbf{v}|^2 / 2$, and $I_\alpha$ is related to the external force $\mathbf{F}$. Similarly, the LB equations of $f_\alpha$ and $h_\alpha$ can be attained by integrating Eqs. (60) and (62) with a second-order integration [190]. From Eq. (62) it can be found that there are no complicated terms that consist of the gradients of macroscopic quantities. Hence in the total-energy-based DDF model, the inclusion of the compression work and the viscous heat dissipation is simpler and easier than that in the internal-energy-based DDF model. Recently, a similar total-energy-based DDF model has been developed by Karlin *et al*. [193]. In addition, based on Guo *et al*.'s work, Li *et al*. [144] have constructed a coupling DDF model on the standard D2Q9 lattice, which can recover the equation of state of ideal gases, $p = \rho R T$.

In some applications, the compression work and the viscous heat dissipation are negligible; thus Eq. (58) can be taken as the target macroscopic energy equation. On the basis of this consideration, several simplified internal-energy-based DDF models have been proposed by Peng *et al*. [191], Shi *et al*. [192], and Li *et al*. [194], respectively. The simplified thermal LB equation for $g_\alpha$ is given by

$$g_\alpha\left(\mathbf{x} + \mathbf{e}_\alpha \delta_t, t + \delta_t\right) - g_\alpha\left(\mathbf{x}, t\right) = -\frac{1}{\tau_g}\left(g_\alpha - g_\alpha^{eq}\right), \tag{63}$$

where the equilibrium internal energy distribution function can be chosen as $g_\alpha^{eq} = e f_\alpha^{eq}$. In the presence of an external force $\mathbf{F}$, an error term will exist in the macroscopic internal energy equation recovered from Eq. (63) [195]:

$$\partial_t \left( \rho c_v T \right) + \nabla \cdot \left( \rho c_v T \mathbf{v} \right) = \nabla \cdot \left( \lambda \nabla T + \gamma T \mathbf{F} \right), \tag{64}$$

where $\gamma = \lambda / p$. Such an issue arises from the coupling between the Chapman-Enskog analyses of Eq. (63) and the LB equation with a forcing term. To be specific, it is attributed to the evaluation of $\partial_{t0} \mathbf{v}$ in the Chapman-Enskog analysis of Eq. (63), which can be seen from Eq. (68). To eliminate the error term in Eq. (64), a correction term should be added to Eq. (63) [195]:

$$g_\alpha \left( \mathbf{x} + \mathbf{e}_\alpha \delta_t, t + \delta_t \right) - g_\alpha \left( \mathbf{x}, t \right) = -\frac{1}{\tau_g} \left( g_\alpha - g_\alpha^{eq} \right) + \delta_t \left( 1 - \frac{1}{2\tau_g} \right) \omega_\alpha c_v T \frac{\left( \mathbf{e}_\alpha \cdot \mathbf{F} \right)}{c_s^2}. \tag{65}$$

The correction term does not affect the calculation of the macroscopic internal energy as its summation over $\alpha$ is zero. In 2009, Li *et al.* [196] mentioned that a correction term is needed for Eq. (63) in the presence of a body force, but also pointed out that the correction term can be neglected. Recently, Li and Luo [195] showed that the error term arising from the external force will lead to significant numerical errors in certain cases, e.g., in the pseudopotential LB modeling of thermal flows.

### 2.4.1.2 The temperature-based DDF models

Now attention turns to the temperature-based DDF LB models, which are usually devised for solving Eq. (59). For incompressible flows with negligible density variations, the right-hand side of Eq. (59) can be rewritten as $\nabla \cdot \left( \lambda \nabla T \right) / \rho c_v \approx \nabla \cdot \left( \chi \nabla T \right)$, where $\chi$ is the thermal diffusivity. The most well-known temperature-based DDF LB model may be attributed to Shan [197], who treated the temperature as a passive scalar and introduced a temperature distribution function, which obeys

$$T_\alpha \left( \mathbf{x} + \mathbf{e}_\alpha \delta_t, t + \delta_t \right) - T_\alpha \left( \mathbf{x}, t \right) = -\frac{1}{\tau_T} \left( T_\alpha - T_\alpha^{eq} \right), \tag{66}$$

where $T_\alpha$ is the temperature distribution function, $\tau_T$ is the relaxation time for the temperature, and $T_\alpha^{eq} = T f_\alpha^{eq} / \rho$ is the equilibrium temperature distribution function, in which $f_\alpha^{eq}$ is given by Eq. (9). The thermal diffusivity is given by $\chi = c_s^2 \left( \tau_T - 0.5 \right) \delta_t$. In 2002, Guo *et al.* [198] proposed a similar temperature-based DDF model using the D2Q4 lattice and $T_\alpha^{eq}$ is given by $T_\alpha^{eq} = 0.25 T \left( 1 + 2 \mathbf{e}_\alpha \cdot \mathbf{v} / c^2 \right)$. In recent years, several temperature-based DDF LB-MRT models have also been developed [146, 217, 218], in which the LB-BGK equations of the density and temperature distribution functions were replaced by an isothermal LB-MRT equation and a thermal LB-MRT equation, respectively. Numerical results showed that these temperature-based DDF LB-MRT models have better numerical stability than their BGK counterparts [146, 217, 218].

Through Chapman-Enskog analysis, it can be found that the macroscopic temperature equation recovered from Eq. (66) with $T_\alpha^{eq} = Tf_\alpha^{eq}/\rho$ is given by [200]

$$\partial_t T + \nabla \cdot (\mathbf{v}T) = \nabla \cdot \left\{ \delta_t \left( \tau_T - 0.5 \right) \left[ \partial_{t0} \left( T\mathbf{v} \right) + \nabla \cdot \left( T\mathbf{v}\mathbf{v} \right) + c_s^2 \nabla T \right] \right\}. \tag{67}$$

Some error terms can be observed in the recovered temperature equation. The first term in the bracket, $\partial_{t0}\left( T\mathbf{v} \right)$, can be rewritten as $\partial_{t0}\left( T\mathbf{v} \right) = \mathbf{v}\partial_{t0}T + T\partial_{t0}\mathbf{v}$ in which $\partial_{t0}\mathbf{v}$ should be evaluated according to the Chapman-Enskog analysis of the LB equation for the density distribution function. For the standard LB equation with an external force, $\partial_{t0}\mathbf{v}$ is given by [195]

$$\partial_{t0}\mathbf{v} = -\mathbf{v}\cdot\nabla\mathbf{v} + \frac{1}{\rho}\left( \mathbf{F} - \nabla p \right), \tag{68}$$

where $p = \rho c_s^2$. Since $\mathbf{v}\partial_{t0}T - T\mathbf{v}\cdot\nabla\mathbf{v} + \nabla\cdot\left( T\mathbf{v}\mathbf{v} \right) = 0$, the error terms $\partial_{t0}\left( T\mathbf{v} \right) + \nabla\cdot\left( T\mathbf{v}\mathbf{v} \right)$ in Eq. (67) can be rewritten as $T\left( \mathbf{F} - \nabla p \right)/\rho$, which shows that the error terms mainly arise from $\partial_{t0}\left( T\mathbf{v} \right)$ and one error term is proportional to $\nabla p/\rho = c_s^2 \nabla\rho/\rho$, while the other error term is introduced by the external force $\mathbf{F}$. We should be aware of these error terms because in certain cases they may result in considerable numerical errors, e.g., if the hydrodynamic model is a multiphase LB model, then $\nabla\rho/\rho$ cannot be neglected. For the temperature-based thermal LB equation on the D2Q4 or D2Q5 lattice, similar error terms can be found, whether using the BGK or MRT collision operator.

A couple of improved DDF LB models [199, 200] have been recently developed to eliminate the error terms in Eq. (67). Based on the isothermal LB model proposed by He and Luo [219], Chai and Zhao [199] have devised an improved DDF LB model by introducing an additional term into the equilibrium temperature distribution function as well as a source term into the thermal LB equation. Huang and Wu [200] have developed an improved thermal LB-MRT equation by modifying the collision process of the distribution functions in the moment space, which prevents the generation of $\partial_{t0}(T\mathbf{v})$ in Eq. (67). Therefore the error terms resulting from $\partial_{t0}(T\mathbf{v})$ were removed. Meanwhile, the term $\nabla \cdot (T\mathbf{v}\mathbf{v})$ in Eq. (67) was eliminated by dropping the second-order velocity terms in the equilibria. Besides the above three types of DDF LB models, several DDF LB models have been developed based on the enthalpy and the total enthalpy [201, 202], which are similar to the models based on the internal energy and the total energy, respectively.

### 2.4.2 The thermal boundary treatments

In this subsection, the thermal boundary treatments for the DDF LB approach are briefly summarized. In LB simulations, after the streaming step, the distribution functions pointing into the fluid domain are unknown at the boundary nodes [1]. Evaluating these unknown distribution functions plays a crucial role in the LB method. For isothermal LB models, many boundary treatments have been devised [142, 220-225]. However, compared with the hydrodynamic boundary conditions, the thermal boundary conditions have not been satisfactorily addressed because the number of the unknown distribution functions at the boundary nodes is usually larger than the number of the constraints [184].

For the internal-energy-based DDF LB model, an early thermal boundary treatment was developed by He *et al*. [189]. In their work, the non-equilibrium bounce-back rule proposed by Zou and He [221] was extended to impose thermal boundary conditions. Later, D'Orazio *et al*. [226] and D'Orazio and Succi [227] devised a counter-slip thermal boundary treatment for the internal-energy-based DDF LB model through calculating the unknown internal energy distribution functions from the equilibrium internal energy distribution functions with a counter-slip internal energy. In 2002, Guo *et al*. [198] proposed a non-equilibrium extrapolation treatment for their temperature-based DDF LB model. The temperature distribution functions at the boundary nodes were decomposed into equilibrium and non-equilibrium parts. Meanwhile, the non-equilibrium part was evaluated with an extrapolation of the non-equilibrium part of the distribution functions at the neighboring nodes. Inspired by Guo *et al*.'s non-equilibrium extrapolation scheme, Tang *et al*. [204]

have developed a thermal boundary treatment for the internal-energy-based DDF LB model. In addition, Guo *et al*. [190] have extended the non-equilibrium extrapolation scheme to their total-energy-based DDF LB model.

Furthermore, Ginzburg [228] has proposed a multi-reflection scheme to mimic the Dirichlet and Neumann boundary conditions in the LB models for advection and anisotropic dispersion equations with arbitrarily shaped surfaces. In 2007, Kao and Yang [229] used the bounce-back scheme to implement the adiabatic boundary condition and employed the equilibrium distribution functions to treat boundaries with a constant temperature. Later, Kuo and Chen [230] developed a non-equilibrium mirror-reflection boundary treatment for the Dirichlet boundary condition and the adiabatic boundary condition, in which the relationship between the heat flux and the first-order moment of the non-equilibrium temperature distribution function was considered. In 2010, Liu *et al*. [231] suggested that the unknown energy distribution functions at the boundary nodes can be chosen to be functions of the local known energy distribution functions together with some corrections.

Recently, Zhang *et al*. [232] have proposed a general bounce-back scheme to implement thermal and concentration boundary conditions. In addition, Li *et al*. [233] have also developed a thermal boundary treatment based on the bounce-back scheme and the interpolation of the distribution functions. Inspired by the idea of Yin and Zhang's method [234], Chen *et al*. [235] have developed an improved bounce-back boundary treatment for thermal boundary conditions, in which the midpoint temperature value was utilized to modify the bounced-back population. It was shown that [235] the improved treatment has a simple algorithm and can easily deal with the boundaries of complex geometries. Besides the above thermal boundary treatments, several immersed-boundary-method-based thermal boundary schemes have also been proposed [236, 237]. Moreover, Huang *et al*. [238], Chen *et al*. [239], and Li *et al*. [233] have developed thermal boundary treatments for curved boundaries.

### 2.4.3 The hybrid approach

The concept of hybrid thermal LB approach was formally introduced by Lallemand and Luo in 2003 [214]. Before that time, Filippova and Hänel [212, 213] had practically implemented such an approach in modeling low Mach number combustion. Similar to the flow simulation in the DDF LB approach, in the hybrid approach the LB flow simulation is also separated from the solution of the temperature field, which is usually solved with conventional numerical methods, such as the finite-difference or finite-volume method [51, 104, 111, 212-216, 240].

Various finite-difference schemes can be found in the textbooks of computational fluid dynamics. Here we just take the second-order Runge-Kutta scheme as an example to illustrate the implementation of a finite-difference scheme in the hybrid LB approach. With a constant thermal conductivity, Eq. (59) can be written as

$$\partial_t T = -\mathbf{v} \cdot \nabla T + \frac{\lambda}{\rho c_v} \nabla^2 T \ . \tag{69}$$

Note that, if the thermal conductivity $\lambda$ is variable, the term $\nabla \cdot (\lambda \nabla T)$ in Eq. (59) should be rewritten as $\nabla \cdot (\lambda \nabla T) = \nabla \lambda \cdot \nabla T + \lambda \nabla^2 T$. For simplicity, the right-hand side of Eq. (69) is represented by $K(T)$. Using the second-order Runge-Kutta scheme, Eq. (69) can be solved as follows:

$$T(\mathbf{x}, t + \delta_t) = T(\mathbf{x}, t) + \frac{\delta_t}{2} (h_1 + h_2) , \tag{70}$$

in which $h_1$ and $h_2$ denote the calculations of the right-hand side of Eq. (69) and are given by

$$h_1 = K(T^t), \quad h_2 = K\left(T^t + \frac{\delta_t}{2} h_1\right), \tag{71}$$

where $T^t = T(\mathbf{x}, t)$. In calculating $h_2$, the density $\rho$ and the velocity $\mathbf{v}$ are still at the $t$ time level since they are given by the hydrodynamic LB model.

For the spatial discretization, a second-order difference scheme can be used. In the framework of the LB method, a widely used scheme is the second-order isotropic difference scheme [47, 241], which is constructed based on the following Taylor series expansion:

$$\phi(\mathbf{x} + \mathbf{e}_\alpha \delta_t) = \phi(\mathbf{x}) + e_{\alpha k} \delta_t \partial_k \phi(\mathbf{x}) + \frac{\delta_t^2}{2} e_{\alpha k} e_{\alpha l} \partial_k \partial_l \phi(\mathbf{x}) + \cdots, \tag{72}$$

where $\phi$ is an arbitrary quantity. According to Eq. (72), the second-order isotropic difference scheme for the spatial gradient of the quantity $\phi$ is given as follows:

$$\partial_i \phi(\mathbf{x}) \approx \frac{1}{c_s^2 \delta_t} \sum_\alpha \omega_\alpha \phi(\mathbf{x} + \mathbf{e}_\alpha \delta_t) e_{\alpha i}, \tag{73}$$

where $e_{\alpha i}$ is the $i$-component ($x$, $y$, or $z$) of $\mathbf{e}_\alpha$ and the weights $\omega_\alpha$ are the same as those in Eq. (9).

Similarly, the second-order isotropic difference scheme for the Laplacian of $\phi$ is given by

$$\nabla^2 \phi(\mathbf{x}) \approx \frac{2}{c_s^2 \delta_t^2} \sum_\alpha \omega_\alpha \left[ \phi(\mathbf{x} + \mathbf{e}_\alpha \delta_t) - \phi(\mathbf{x}) \right]. \tag{74}$$

Note that both Eq. (70) and the second-order isotropic difference schemes are only applied in the interior field, namely the whole computational field except the boundaries. For the Dirichlet thermal boundary condition, the temperature at the boundary is known, $T = T_w$. For the Neumann thermal boundary condition, the temperature at the boundary can be extrapolated from the interior flow field, where the temperature at the $t + \delta_t$ time level has been obtained according to Eq. (70).

## 3. The pseudopotential multiphase LB method

### 3.1 The basic theory

#### 3.1.1 The interaction force

The pseudopotential multiphase LB method was proposed by Shan and Chen around 1993 [31, 32]. They introduced an interparticle potential $\psi(\mathbf{x})$ to mimic the interactions among the particles on the nearest-neighboring sites. Theoretically, $\psi(\mathbf{x})$ is not a potential as it depends on space only through the intermediate of the fluid density. This is why it is now widely called "pseudopotential". For single-component systems, the interaction force acting on the particles at site $\mathbf{x}$ is given by [32, 242]

$$\mathbf{F}(\mathbf{x}, t) = -G\psi(\mathbf{x}) \sum_\alpha w_\alpha \psi(\mathbf{x} + \mathbf{e}_\alpha \delta_t) \mathbf{e}_\alpha, \tag{75}$$

where $G$ is a parameter that controls the strength of the interaction force and $w_\alpha$ are the weights. For the nearest-neighbor interactions on the D2Q9 lattice, the weights $w_\alpha = 1/3$ for $|\mathbf{e}_\alpha|^2 = 1$ and $w_\alpha = 1/12$ for $|\mathbf{e}_\alpha|^2 = 2$ [242]. In some studies, the weights $w_\alpha$ are set to $\omega_\alpha$ in $f_\alpha^{eq}$ or $Gc_s^2$ is used in Eq. (75) instead of $G$. The conversion of $G$ between these choices should be noticed.

The most distinctive feature of the pseudopotential LB method is that the phase segregation between different phases can emerge automatically as a result of the particle interactions, namely the interaction force Eq. (75). As highlighted by Succi [16], the magic of the simple interaction force lies in that it not only gives a non-monotonic equation of state supporting the phase transition but also yields non-zero surface tension. Therefore the interface between different phases can arise, deform and migrate naturally without using any techniques to track or capture the interface. Owing to its conceptual simplicity, computational efficiency and kinetic features, the pseudopotential LB method has attracted significant attention and has been applied in a variety of fields. In this section we will review some recent advances in the pseudopotential LB method as well as thermal pseudopotential LB models for simulating phase-change heat transfer.

### 3.1.2 The Shan-Chen forcing scheme

In the original pseudopotential LB model devised by Shan and Chen [31, 32], the interaction force was incorporated into the LB equation by shifting the velocity in the equilibrium density distribution function and the evolution equation is given by

$$f_\alpha\left(\mathbf{x}+\mathbf{e}_\alpha\delta_t,t+\delta_t\right)-f_\alpha\left(\mathbf{x},t\right)=-\frac{1}{\tau}\Big[f_\alpha-f_\alpha^{eq}\left(\rho,\mathbf{u}^{eq}\right)\Big]. \tag{76}$$

The shifted equilibrium velocity $\mathbf{u}^{eq}$ is defined as $\mathbf{u}^{eq}=\mathbf{u}+\tau\delta_t\mathbf{F}/\rho$, where $\mathbf{u}=\sum_\alpha f_\alpha\mathbf{e}_\alpha/\rho$. The actual fluid velocity can be defined as $\mathbf{v}=\mathbf{u}+\delta_t\mathbf{F}/\left(2\rho\right)$ by averaging the moment before and after the collision process [243]. The forcing scheme given by Eq. (76) is referred to as the Shan-Chen forcing scheme.

In 2011, Gross *et al.* [244] pointed out that a forcing term can be depicted for the Shan-Chen forcing scheme by rewriting Eq. (76) as

$$f_\alpha\left(\mathbf{x}+\mathbf{e}_\alpha\delta_t,t+\delta_t\right)-f_\alpha\left(\mathbf{x},t\right)=-\frac{1}{\tau}\Big(f_\alpha-f_\alpha^{eq}\left(\rho,\mathbf{u}\right)\Big)+\frac{1}{\tau}\Big[f_\alpha^{eq}\left(\rho,\mathbf{u}^{eq}\right)-f_\alpha^{eq}\left(\rho,\mathbf{u}\right)\Big], \tag{77}$$

which means that the forcing term of the Shan-Chen forcing scheme is given by

$$F_{\alpha,\text{SC}}=\frac{1}{\tau}\Big[f_\alpha^{eq}\left(\rho,\mathbf{u}^{eq}\right)-f_\alpha^{eq}\left(\rho,\mathbf{u}\right)\Big]. \tag{78}$$

Subsequently, Huang *et al.* [245] attained a form of $F_{\alpha,\text{SC}}$ by substituting the expression of the equilibrium distribution function into the above equation (see Eq. (16) in Ref. [245]). Later, by noting that $f_\alpha^{eq}(\rho, \mathbf{u}^{eq})$ with $\mathbf{u}^{eq} = \mathbf{u} + \tau \delta_t \mathbf{F}/\rho$ can be written as [163]

$$f_\alpha^{eq}(\rho, \mathbf{u}^{eq}) \equiv f_\alpha^{eq}(\rho, \mathbf{u}) + \rho \omega_\alpha \left[ \frac{\mathbf{e}_\alpha}{c_s^2} \cdot \frac{\tau \delta_t \mathbf{F}}{\rho} + \left( \frac{\tau \delta_t \mathbf{F}}{\rho} \frac{\tau \delta_t \mathbf{F}}{\rho} + 2\mathbf{u} \frac{\tau \delta_t \mathbf{F}}{\rho} \right) : \frac{(\mathbf{e}_\alpha \mathbf{e}_\alpha - c_s^2 \mathbf{I})}{2c_s^4} \right], \quad (79)$$

Li *et al.* [163] obtained the following forcing term of the Shan-Chen forcing scheme:

$$F_{\alpha, SC} = \omega_\alpha \delta_t \left[ \frac{\mathbf{e}_\alpha \cdot \mathbf{F}}{c_s^2} + \frac{(\mathbf{v}_{SC} \mathbf{F} + \mathbf{F} \mathbf{v}_{SC}) : (\mathbf{e}_\alpha \mathbf{e}_\alpha - c_s^2 \mathbf{I})}{2c_s^4} \right], \quad (80)$$

where $\mathbf{v}_{SC} = \mathbf{u} + \tau \delta_t \mathbf{F}/(2\rho)$. The above equation indicates that the Shan-Chen forcing term can be written in the general form of forcing schemes, i.e., Eq. (23), and the velocity in the forcing term is given by $\mathbf{v}_{SC}$. With this result, the differences between the Shan-Chen forcing scheme and Guo *et al.*'s forcing scheme, which has been demonstrated to be capable of correctly recovering the unsteady macroscopic equations at the Navier-Stokes level, can be clearly observed.

Another forcing scheme that has received much attention in the pseudopotential LB method is the EDM forcing scheme proposed by Kupershtokh *et al.* [149], which has been discussed in detail in Section 2.3.2. It has been shown that several assumptions were used in the derivation of the EDM forcing scheme. A theoretical comparison between the Shan-Chen scheme, the EDM scheme, and Guo *et al.*'s scheme has been made by Li *et al.* [163], and can be found in Table 1, where $\mathbf{u} = \sum_\alpha f_\alpha \mathbf{e}_\alpha / \rho$. From the table we can see that the Shan-Chen scheme is identical to the EDM scheme when the non-dimensional relaxation time $\tau = 1$, which was numerically reproduced by Sun *et al.* [246] (see Fig. 3 in the reference). Moreover, both the Shan-Chen and the EDM forcing schemes suffer from the discrete lattice effects and therefore cannot recover the correct macroscopic equations. In addition, it can be found that the Shan-Chen and the EDM forcing schemes share the following feature: the velocity in $f_\alpha^{eq}$ is $\mathbf{u}$, while the actual fluid velocity is defined as $\mathbf{v} = \mathbf{u} + \delta_t \mathbf{F}/(2\rho)$. In Section 2.3.2, we have mentioned that the usual Chapman-Enskog analysis should be revised for this class of forcing schemes so as to derive the actual macroscopic equations.

**Table 1**. Comparison between different forcing schemes. Adapted from Li *et al.* [163] with permission of the American Physical Society.

| scheme | velocity in $f_\alpha^{eq}$ | velocity in $F_\alpha$ | fluid velocity | discrete effects |
|---|---|---|---|---|
| Shan-Chen | $\mathbf{u}$ | $\mathbf{u} + \dfrac{\tau \delta_t \mathbf{F}}{2\rho}$ | $\mathbf{u} + \dfrac{\delta_t \mathbf{F}}{2\rho}$ | Yes |
| EDM | $\mathbf{u}$ | $\mathbf{u} + \dfrac{\delta_t \mathbf{F}}{2\rho}$ | $\mathbf{u} + \dfrac{\delta_t \mathbf{F}}{2\rho}$ | Yes |
| Guo *et al.* | $\mathbf{u} + \dfrac{\delta_t \mathbf{F}}{2\rho}$ | $\mathbf{u} + \dfrac{\delta_t \mathbf{F}}{2\rho}$ | $\mathbf{u} + \dfrac{\delta_t \mathbf{F}}{2\rho}$ | No |

Huang *et al.* [245] and Sun *et al.* [246] have numerically investigated the performances of the Shan-Chen and the EDM forcing schemes in the pseudopotential LB modeling of multiphase flows. It was found [245] that the coexistence liquid-gas densities given by the Shan-Chen forcing scheme significantly vary with the non-dimensional relaxation time $\tau$. Using the EDM forcing scheme, the effects of the relaxation time $\tau$ were greatly reduced. Nevertheless, Huang *et al.* [245] found that the density ratios obtained by the EDM forcing scheme still change with $\tau$ to some extent in the cases of large density ratios. Meanwhile, Huang *et al.* showed that the coexistence curves given by He *et al.*'s forcing scheme are essentially independent of $\tau$.

### 3.1.3 The mechanical stability condition

In this subsection, the mechanical stability condition in the pseudopotential LB method is introduced, which can be established according to the pressure tensor given by a pseudopotential LB model. When the interaction force is recovered in the momentum equation without any additional terms at the Navier-Stokes level, the pressure tensor $\mathbf{P}$ can be defined as [247]

$$\nabla \cdot \mathbf{P} = \nabla \cdot \left( \rho c_s^2 \mathbf{I} \right) - \mathbf{F} . \tag{81}$$

In 2008, Shan [242] clarified that in the pseudopotential LB method the *discrete form* of the pressure tensor must be used, which can be derived from the volume integral of Eq. (81). This point has also been demonstrated by Sbragaglia and Belardinelli [248] for multi-component systems. The corresponding discrete form of Eq. (81) is given by

$$\sum \mathbf{P} \cdot \mathbf{A} = \sum \rho c_s^2 \mathbf{I} \cdot \mathbf{A} - \sum \mathbf{F} , \tag{82}$$

where $\mathbf{A}$ is a closed area. According to Eqs. (82) and (75), the discrete form pressure tensor is [242]

$$\mathbf{P} = \rho c_s^2 \mathbf{I} + \frac{G}{2} \psi(\mathbf{x}) \sum_\alpha w_\alpha \psi(\mathbf{x} + \mathbf{e}_\alpha \delta_t) \mathbf{e}_\alpha \mathbf{e}_\alpha$$

$$= \left( \rho c_s^2 + \frac{Gc^2 \psi^2}{2} \right) \mathbf{I} + \frac{G\psi}{2} \sum_\alpha w_\alpha \left[ \psi(\mathbf{x} + \mathbf{e}_\alpha \delta_t) - \psi(\mathbf{x}) \right] \mathbf{e}_\alpha \mathbf{e}_\alpha . \tag{83}$$

Using the Taylor series expansion (see Eq. (72)), the following second-order pressure tensor can be attained for the cases of nearest-neighbor interactions [39]:

$$\mathbf{P} = \left( \rho c_s^2 + \frac{Gc^2}{2} \psi^2 + \frac{Gc^4}{12} \psi \nabla^2 \psi \right) \mathbf{I} + \frac{Gc^4}{6} \psi \nabla \nabla \psi . \tag{84}$$

The last term on the right-hand side of Eq. (84) is related to the surface tension and the non-ideal equation of state is given by

$$p_{\text{EOS}}(\rho) = \rho c_s^2 + \frac{Gc^2}{2} \psi^2 . \tag{85}$$

In some studies, $w_\alpha$ in Eq. (75) are set to $\omega_\alpha$ in $f_\alpha^{eq}$, which gives $p_{\text{EOS}}(\rho) = \rho c_s^2 + 0.5 Gc_s^2 \psi^2$.

According to Eq. (84), the normal pressure tensor of a flat interface is given by [242]

$$P_n = \rho c_s^2 + \frac{Gc^2}{2} \psi^2 + \frac{Gc^4}{4} \psi \frac{\mathrm{d}^2 \psi}{\mathrm{d}n^2} , \tag{86}$$

where $n$ denotes the normal direction of the interface. Using the following relationships [39]:

$$\frac{\mathrm{d}^2 \psi}{\mathrm{d}n^2} = \frac{1}{2} \frac{\mathrm{d}}{\mathrm{d}\psi} \left( \frac{\mathrm{d}\psi}{\mathrm{d}n} \right)^2 , \quad \frac{\mathrm{d}}{\mathrm{d}\psi} = \frac{1}{\psi'} \frac{\mathrm{d}}{\mathrm{d}\rho} , \quad \left( \frac{\mathrm{d}\psi}{\mathrm{d}n} \right)^2 = \psi'^2 \left( \frac{\mathrm{d}\rho}{\mathrm{d}n} \right)^2 , \tag{87}$$

where $\psi' = \mathrm{d}\psi / \mathrm{d}\rho$, Eq. (86) can be rewritten as

$$P_n - \rho c_s^2 - \frac{Gc^2}{2} \psi^2 = \frac{Gc^4}{8} \frac{\psi}{\psi'} \frac{\mathrm{d}}{\mathrm{d}\rho} \left[ \psi'^2 \left( \frac{\mathrm{d}\rho}{\mathrm{d}n} \right)^2 \right] . \tag{88}$$

According to Eq. (88) and the requirement that [242] at equilibrium $P_n$ should be equal to the constant static pressure in the bulk, $p_b$, the following equation can be obtained:

$$\frac{8}{Gc^4} \int_{\rho_g}^{\rho_l} \left( p_b - \rho c_s^2 - \frac{Gc^2}{2} \psi^2 \right) \frac{\psi'}{\psi} \mathrm{d}\rho = \int_{\rho_g}^{\rho_l} \mathrm{d} \left[ \psi'^2 \left( \frac{\mathrm{d}\rho}{\mathrm{d}n} \right)^2 \right] , \tag{89}$$

where $p_b = p_{\text{EOS}}(\rho_l) = p_{\text{EOS}}(\rho_g)$. Since $\mathrm{d}\rho / \mathrm{d}n$ is zero at $\rho = \rho_l$ and $\rho_g$, Eq. (89) yields [242]

$$\int_{\rho_g}^{\rho_l} \left( p_b - \rho c_s^2 - \frac{Gc^2}{2} \psi^2 \right) \frac{\psi'}{\psi} \mathrm{d}\rho = 0 . \tag{90}$$

The above equation is the so-called mechanical stability condition of the standard pseudopotential LB model with the nearest-neighbor interactions.

The mechanical stability condition plays a very important role in the pseudopotential LB method because it determines the coexistence densities of liquid and gas phases ($\rho_l$ and $\rho_g$), which can be theoretically obtained by solving Eq. (90) and the relation $p_b = p_{\text{EOS}}(\rho_l) = p_{\text{EOS}}(\rho_g)$ via numerical integration [242]. Meanwhile, in thermodynamic theory, the Maxwell equal-area construction that determines the liquid-gas coexistence densities is built in terms of $\int_{\rho_g}^{\rho_l} \left[ p_b - p_{\text{EOS}}(\rho) \right] dV = 0$, where $V \propto 1/\rho$ [12], which yields the following requirement for the pseudopotential LB models:

$$\int_{\rho_s}^{\rho_l} \left( p_b - \rho c_s^2 - \frac{Gc^2}{2}\psi^2 \right) \frac{1}{\rho^2} d\rho = 0 .$$ (91)

From Eqs. (90) and (91), we can find that the mechanical stability condition does not meet the requirement of the thermodynamic theory unless $\psi \propto \exp(-1/\rho)$, which gives $\psi'/\psi \sim 1/\rho^2$, such as $\psi(\rho) = \psi_0 \exp(-\rho_0/\rho)$ [12, 242]. With this choice, the phase separation can be achieved when an appropriate value of $G$ is adopted, e.g., for the case of $\psi_0 = 4$ and $\rho_0 = 200$ [12], $G = -40$ leads to $\rho_l \approx 514$ and $\rho_g \approx 79.5$ [12, 249]. Similarly, it can be found that, when $\psi_0 = 1$ and $\rho_0 = 1$, $G = -10/3$ gives $\rho_l \approx 2.78$ and $\rho_g \approx 0.367$ [242, 250].

Actually, the interaction force given by Eq. (75) is nothing but a finite-difference scheme, which can be seen by comparing Eq. (75) with Eq. (73). Fortunately, this formulation can lead to a non-ideal pressure tensor, Eq. (84). For a standard free energy functional in thermodynamics, the following non-ideal pressure tensor can be obtained [40]:

$$\mathbf{P}_{\text{FE}} = \left( p_{\text{EOS}} - k\rho\nabla^2\rho - \frac{k}{2}|\nabla\rho|^2 \right)\mathbf{I} + k\nabla\rho\nabla\rho .$$ (92)

The Maxwell equal-area construction can be derived from the above pressure tensor [251]. It is obvious that Eq. (84) deviates from Eq. (92) in general. However, it has been shown that, with an appropriate choice of $\psi(\rho)$, the Maxwell construction can be reproduced from Eq. (84). From this point, the pseudopotential LB method can be regarded as a "pseudo" free-energy approach [252], but note that in the pseudopotential LB method the non-ideal pressure tensor is established through the interaction force rather than changing $f_\alpha^{eq}$, which suffers from the loss of Galilean invariance.

Numerically, to guarantee thermodynamic consistency, an accurate forcing scheme should be employed even though the pseudopotential is chosen to be proportional to $\exp(-1/\rho)$. In the literature, Yu and Fan [250] have shown that the numerical coexistence densities of liquid and gas phases obtained using the pseudopotential $\psi(\rho) = \psi_0 \exp(-\rho_0/\rho)$ together with Guo *et al.*'s forcing scheme are basically independent of the non-dimensional relaxation time $\tau$ and consistent with the results given by the Maxwell construction (see Table 1 in Ref. [250]).

The problem that the Shan-Chen forcing scheme results in $\tau-$dependent coexistence curves was found to be attributed to the additional terms introduced by the scheme into the macroscopic equations, which make the mechanical stability condition dependent on $\tau$ and $\psi^2/\rho$ [163]. For the EDM forcing scheme, the mechanical stability condition still depends on $\psi^2/\rho$ due to the additional term $\nabla \cdot (\mathbf{FF}/\rho)$ in Eq. (45), although the relaxation time $\tau$ is seemingly not involved. Obviously, when an additional term is introduced into Eq. (81), the pressure tensor will be modified, and then the mechanical stability condition will be changed correspondingly.

### 3.2 Thermodynamic inconsistency

### 3.2.1 Theoretical results

The choice of $\psi(\rho) = \psi_0 \exp(-\rho_0/\rho)$ can ensure the consistency between Eqs. (90) and (91). However, by doing so the non-ideal equation of state is fixed. In 2002, He and Doolen [247] pointed out that, to reproduce a non-ideal equation of state in the thermodynamic theory, the pseudopotential should be chosen as follows (according to Eq. (85)):

$$\psi = \sqrt{\frac{2\left(p_{\text{EOS}} - \rho c_s^2\right)}{Gc^2}}. \tag{93}$$

Here $p_{\text{EOS}}$ represents a non-ideal equation of state in the thermodynamic theory, such as the van der Waals, the Carnahan-Starling, and the Peng-Robinson equations of state. Similar statements can be found in Refs. [34, 253]. In 2006, Yuan and Schaefer [33] numerically found that the achievable largest density ratio of the pseudopotential LB method can be significantly enhanced by choosing an appropriate equation of state in Eq. (93). For static cases, a liquid-gas density ratio in excess of 1000 was successfully modeled, however, with very large spurious currents at $\tau = 1$ [254]. Usually, when the relaxation time $\tau$ decreases, the spurious currents will be further enlarged. Moreover, Eq. (93) will lead to *thermodynamic inconsistency*: the liquid-gas coexistence densities given by the mechanical stability condition Eq. (90) are inconsistent with the results of the Maxwell equal-area construction.

In 2012, Li *et al.* [163] pointed out that, when using Eq. (93), the thermodynamic consistency can be approximately restored by adjusting the mechanical stability condition

$$\int_{\rho_g}^{\rho_l} \left( p_b - \rho c_s^2 - \frac{Gc^2}{2} \psi^2 \right) \frac{\psi'}{\psi^{1+\varepsilon}} \mathrm{d}\rho = 0 \,, \tag{94}$$

where $\varepsilon$ is a produced parameter that controls the mechanical stability condition. For the standard pseudopotential LB model with the nearest-neighbor interactions, we can obtain the numerical results of the case $\varepsilon = 0$ (namely Eq. (90)) by employing an accurate forcing scheme in the standard LB approach, such as Guo *et al.*'s scheme or the MRT forcing scheme in Section 2.3.3. It can be found that the results of $\varepsilon = 0$ significantly deviate from those given by the Maxwell construction when Eq. (93) is applied. The Shan-Chen and the EDM forcing schemes usually yield a positive $\varepsilon$ ( $\varepsilon > 0$ ); however, the specific value of $\varepsilon$ depends on $\tau$ and/or $\psi^2/\rho$ . Analytically, the liquid-gas coexistence densities can be obtained by solving Eq. (94). For example, when $\varepsilon = 2$ , Eq. (94) gives

$$\int_{\rho_g}^{\rho_l} \left( p_b - \rho c_s^2 - \frac{Gc^2}{2} \psi^2 \right) \frac{\psi'}{\psi^3} \mathrm{d}\rho = 0 \,. \tag{95}$$

Since $\psi' \mathrm{d}\rho = \mathrm{d}\psi$ , Eq. (95) can be transformed to [163]

$$\left( p_b - \rho c_s^2 \right) \left( -\frac{1}{2\psi^2} \right) \Big|_{\rho_g}^{\rho_l} - \frac{Gc^2}{2} \ln \psi \Big|_{\rho_g}^{\rho_l} + \int_{\rho_g}^{\rho_l} c_s^2 \left( -\frac{1}{2\psi^2} \right) \mathrm{d}\rho = 0 \,. \tag{96}$$

With Eq. (96) and the relationship $p_b = p_{\mathrm{EOS}}(\rho_l) = p_{\mathrm{EOS}}(\rho_g)$ , the analytical solution ( $p_b$ , $\rho_l$ , and $\rho_g$ ) of the mechanical stability condition at $\varepsilon = 2$ can be obtained via numerical integration. In a similar way, the solution of the Maxwell construction $\int_{\rho_g}^{\rho_l} \left[ p_b - p_{\text{EOS}}(\rho) \right] d\rho / \rho^2 = 0$ can also be attained.

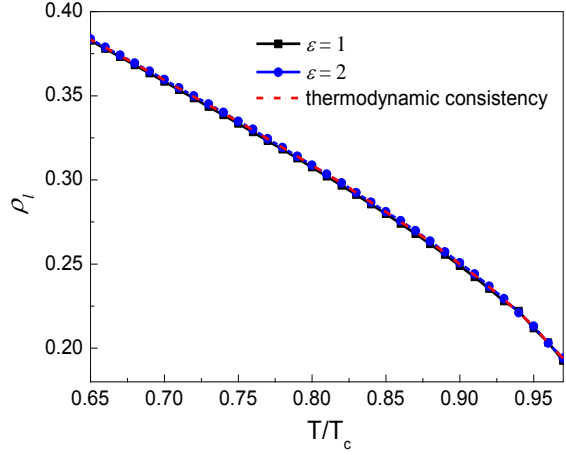

(a) Density of the liquid phase

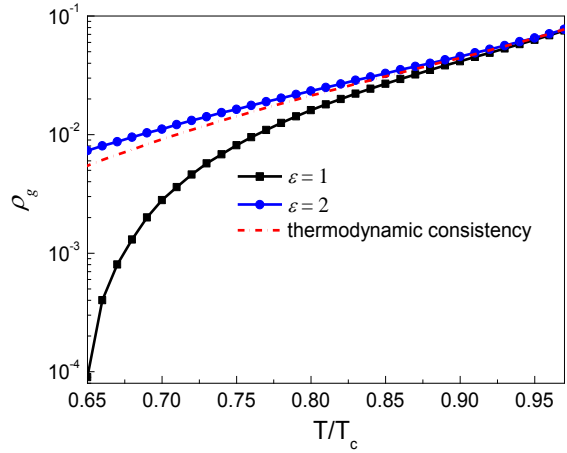

(b) Density of the gas phase

**Fig. 2**. Analytical solutions of the mechanical stability condition at $\varepsilon = 1$ and $\varepsilon = 2$ for the Carnahan-Starling equation of state. Reprinted from Li *et al.* [163] with permission of the American Physical Society.

By taking the Carnahan-Starling equation of state as an example, Li *et al.* [163] have compared the analytical solutions of the mechanical stability condition at $\varepsilon = 1$ and $2$ with the solution given by the thermodynamic consistency requirement (i.e., the Maxwell construction). The results can be found in Fig. 2. It can be seen that there are nearly no differences in the liquid-phase density between the analytical solutions of $\varepsilon = 1$ and $\varepsilon = 2$, which are both in good agreement with the results given by the Maxwell construction. Meanwhile, it can be observed that the gas-phase density given by the Maxwell construction is larger than the gas-phase density of $\varepsilon = 1$ but is smaller than that of $\varepsilon = 2$, which indicates that there exists a value of $\varepsilon$ $(1 < \varepsilon < 2)$ that can make the mechanical stability solution approximately identical to the solution given by the Maxwell construction [163]. In other words, thermodynamic consistency can be approximately restored by producing a suitable value of $\varepsilon$ in the mechanical stability condition. According to Fig. 2, the value of $\varepsilon$ that can provide thermodynamically consistent results falls into the interval $\varepsilon \in [1, 2]$ and is close to 2.

### 3.2.2 Numerical treatments

Based on theoretical analysis, Li *et al.* [163] proposed a modified forcing scheme as follows:

$$F_\alpha^{new} = \omega_\alpha \delta_t \left( 1 - \frac{1}{2\tau} \right) \left[ \frac{\mathbf{F} \cdot \mathbf{e}_\alpha}{c_s^2} + \frac{\left( \mathbf{v}_{new} \mathbf{F} + \mathbf{F} \mathbf{v}_{new} \right) : \left( \mathbf{e}_\alpha \mathbf{e}_\alpha - c_s^2 \mathbf{I} \right)}{2 c_s^4} \right], \tag{97}$$

which is identical to Guo *et al.*'s forcing scheme except that the velocity in the forcing term is modified as $\mathbf{v}_{new} = \mathbf{v} + \sigma \mathbf{F} / \left[ (\tau - 0.5) \delta_t \psi^2 \right]$, where $\sigma$ is a constant. With this change, the second-order moment of the forcing term is given by

$$\sum_\alpha \mathbf{e}_\alpha \mathbf{e}_\alpha F_\alpha^{new} = \sum_\alpha \mathbf{e}_\alpha \mathbf{e}_\alpha F_{\alpha, Guo} + \frac{2\sigma \mathbf{F} \mathbf{F}}{\tau \psi^2}. \tag{98}$$

The first term on the right-hand side of Eq. (98) is defined by Eq. (27). From the Taylor series expansion of Eq. (75), it can be found that $\mathbf{F} \approx -G c^2 \psi \nabla \psi$, which yields $\mathbf{F} \mathbf{F} / \psi^2 \approx G^2 c^4 \nabla \psi \nabla \psi$. According to Chapman-Enskog analysis, the new second-order pressure tensor is given by [163]

$$\mathbf{P} \approx \mathbf{P}_{original} + 2 G^2 c^4 \sigma \nabla \psi \nabla \psi, \tag{99}$$

where $\mathbf{P}_{original}$ denotes the original pressure tensor given by Eq. (83). With the modification of the pressure tensor, the mechanical stability condition is changed and the produced *parameter* $\varepsilon$ is given by $\varepsilon = -16 G \sigma$ [163]. Note that, when the pseudopotential $\psi$ is calculated via Eq. (93), the only requirement for $G$ is to ensure that the whole term inside the square root in Eq. (93) is positive [33, 34] and the value of $|G|$ does not affect the numerical results. In many cases, $G$ can be set to $G = -1$. For the EDM forcing scheme, an additional term that is proportional to $\left( \psi^2 / \rho \right) \nabla \psi \nabla \psi$ will be introduced into the pressure tensor owing to the term $\nabla \cdot \left( \mathbf{F} \mathbf{F} / \rho \right)$ in Eq. (45). As a result, the parameter $\varepsilon$ yielded by the EDM forcing scheme will be dependent on $\psi^2/\rho$.

In Ref. [149], Kupershtokh *et al*. have proposed a mixed interaction force to adjust the coexistence densities of liquid and gas phases, which is given by

$$\mathbf{F}(\mathbf{x},t) = -G\psi(\mathbf{x})\sum_\alpha w_\alpha \left[(1-2A)\psi(\mathbf{x}+\mathbf{e}_\alpha\delta_t)\mathbf{e}_\alpha\right] - G\sum_\alpha w_\alpha \left[A\psi^2(\mathbf{x}+\mathbf{e}_\alpha\delta_t)\mathbf{e}_\alpha\right], \qquad (100)$$

where the parameter $A$ is utilized to tune the mixed interaction force. Such a treatment has been applied in several studies [96, 149, 255], but no physical explanation was given about the reason why it is able to adjust liquid-gas coexistence densities. Actually, using the Taylor series expansion and evaluating the discrete form pressure tensor, one can find that Eq. (100) will introduce a term proportional to $\nabla\psi\nabla\psi$ into the pressure tensor. In this regard, the treatments given by Eqs. (97) and (100) share the same feature, i.e., thermodynamic inconsistency is approximately eliminated by modifying the mechanical stability condition. Similarly, they also share the same weakness, namely the surface tension will be affected because a term proportional to $\nabla\psi\nabla\psi$ has been added to the anisotropic part of the pressure tensor. Recently, Hu *et al*. have conducted theoretical and numerical analyses of the mixed interaction force given by Eq. (100) and the details can be found in Ref. [256].

To overcome the above weakness, Li *et al*. [39] in 2013 proposed an improved forcing scheme based on the LB-MRT equation, which utilizes the following forcing term:

$$\mathbf{S} = \begin{bmatrix} 0 \\ 6\left(v_x F_x + v_y F_y\right) + \dfrac{12\sigma|\mathbf{F}|^2}{\psi^2\delta_t\left(\tau_e - 0.5\right)} \\ -6\left(v_x F_x + v_y F_y\right) - \dfrac{12\sigma|\mathbf{F}|^2}{\psi^2\delta_t\left(\tau_\varsigma - 0.5\right)} \\ F_x \\ -F_x \\ F_y \\ -F_y \\ 2\left(v_x F_x - v_y F_y\right) \\ \left(v_x F_y + v_y F_x\right) \end{bmatrix}, \qquad (101)$$

where $|\mathbf{F}|^2 = \left(F_x^2 + F_y^2\right)$. Obviously, when $\sigma = 0$, Eq. (101) reduces to Eq. (51). According to the Chapman-Enskog analysis, the discrete form pressure tensor resulting from Eq. (101) is given by [39]

$$\mathbf{P} \approx \mathbf{P}_{\text{original}} + 2G^2 c^4 \sigma |\nabla \psi|^2 \mathbf{I}. \qquad (102)$$

The $\varepsilon$ produced in the mechanical stability condition is still given by $\varepsilon = -16G\sigma$, but here it can be seen that the mechanical stability condition is tuned by changing the *isotropic part* of the pressure tensor rather than the *anisotropic part*, therefore avoiding affecting the surface tension term. Moreover, it can be found that $\mathbf{F}$ in the additional terms of $\mathbf{v}_{\text{new}}$ and Eq. (101) actually serves as an approximation for $\psi \nabla \psi$. At large density ratios, the higher-order terms in $\mathbf{F}$ will also exert an influence on the liquid-gas coexistence densities. For the BGK scheme, the additional term in $\mathbf{v}_{\text{new}}$ is proportional to $\mathbf{F}/(\tau - 0.5)$. As a result, the coexistence densities at large density ratios will vary with viscosity to some extent when $\tau$ changes with viscosity. Using the MRT scheme, the influence of viscosity at large density ratios can be significantly reduced because only $\tau_v$ changes with viscosity and the other relaxation times including $\tau_\varepsilon$ and $\tau_\varsigma$ in Eq. (101) remain unchanged.

For the Carnahan-Starling equation of state, Li *et al*. [39] numerically found that thermodynamic consistency can be approximately achieved when $\varepsilon \sim 1.7$, which agrees with the theoretical analysis that $1 < \varepsilon < 2$ and $\varepsilon$ is close to 2. The numerical coexistence curves obtained by the improved MRT forcing scheme for the Carnahan-Starling equation of state can be found in Fig. 3, which shows that the numerical coexistence curves are in good agreement with the results given by the Maxwell construction and are essentially independent of the viscosity (the relaxation time $\tau_v$). Using the improved MRT forcing scheme, Li *et al*. [39] have simulated droplet splashing on a thin liquid film at the density ratio $\rho_l/\rho_g \sim 750$ ($T/T_c = 0.5$) and $\text{Re} = 1000$. Several treatments have been introduced to reduce the spurious currents and increase the achievable highest Reynolds number. Some numerical results can be found in Fig. 4. The approach has been applied to simulate droplet impact on a solid surface at a large density ratio by Zhang *et al*. [70], who have extended the improved MRT forcing scheme to three dimensions using the D3Q19 lattice [71]. The three-dimensional scheme has been applied to study coalescence-induced droplet jumping on superhydrophobic textured surfaces and liquid condensate adhesion on slit and plain fins [257]. Moreover, Xu *et al*. [258] have proposed an alternative three-dimensional version using the D3Q15 lattice.

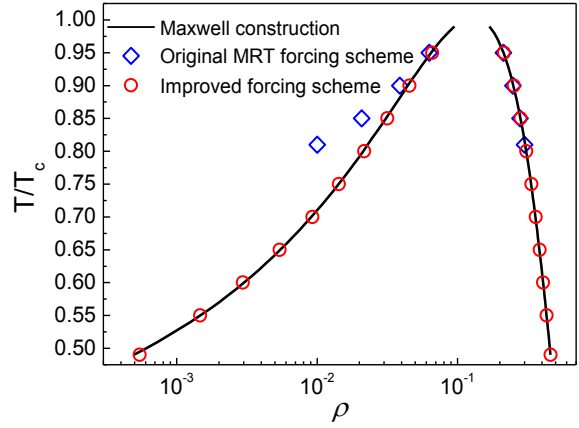

(a) $\tau_v = 0.6$

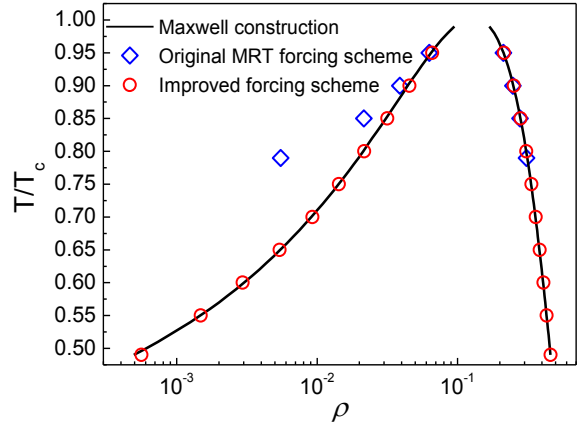

(b) $\tau_v = 0.8$

**Fig. 3**. Comparison of the numerical coexistence curves between the original and the improved MRT forcing schemes. Reprinted from Li *et al*. [39] with permission of the American Physical Society.

Finally, it should be stressed that adjusting the mechanical stability condition is an approximate approach to achieving thermodynamic consistency for Eq. (93) because a fitting procedure is required. Recently, Khajepor *et al*. [259] proposed a multi-pseudopotential interaction force to eliminate thermodynamic inconsistency, which consists of several exponential pseudopotential interactions, $G_i \exp(-\lambda_i/\rho)$, where the subscript $i = 1, 2, \cdots, n$ ($n$ is the total number of the pseudopotential interactions). A fitting procedure is utilized to adjust $G_i$ and $\lambda_i$ so as to match the results given by the prescribed equation of state such as the Carnahan-Starling equation of state (see Appendix B in Ref. [259]). Because the exponential pseudopotential interaction does not contain the temperature, the parameters $G_i$ and $\lambda_i$ should change with the reduced temperature $T/T_c$.

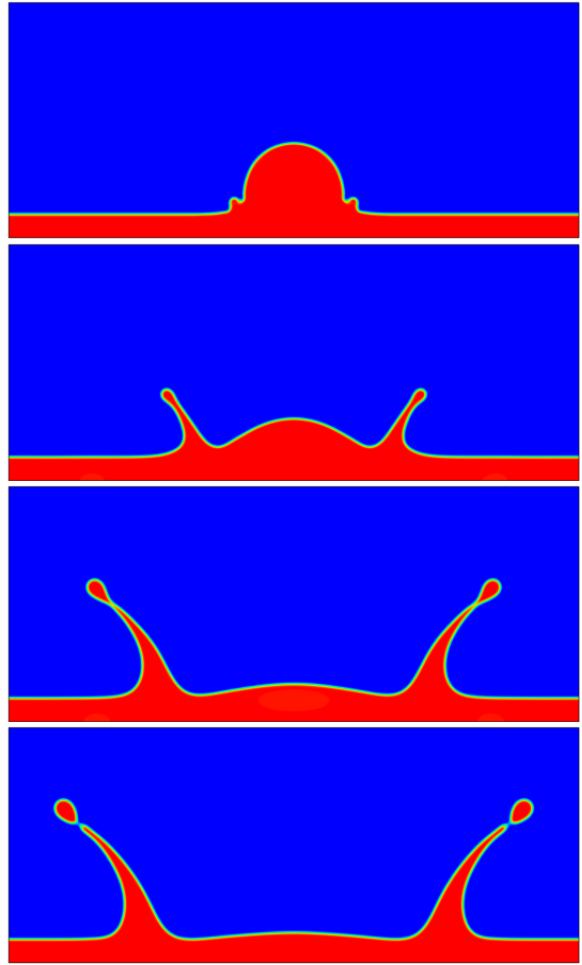

**Fig. 4**. Numerical results of droplet splashing at $\rho_l/\rho_g \sim 750$ and $\mathrm{Re} = 1000$. From top to bottom: $t^* = 0.25$, 0.75, 1.5, and 1.9. Reprinted from Li *et al*. [39] with permission of the American Physical Society.

### 3.3 Spurious currents and the interface thickness

Despite its undeniable success, the pseudopotential LB method has also drawn much criticism because of the large spurious currents produced near curved interfaces [36]. The spurious currents, which are also called spurious velocities, have been observed in almost all the simulations of multiphase flows involving curved interfaces. When the spurious currents are as large as the characteristic velocities of the problem under study, it is difficult to distinguish the physical velocity from the spurious velocity and then the flow physics may be misinterpreted [260]. Therefore it is very important to suppress or reduce the spurious currents when simulating multiphase flows. In 2006, Shan [261] pointed out that the spurious currents produced by the pseudopotential LB models mainly arise from insufficient isotropy of the discrete gradient operator and can be reduced by using high-order isotropic gradient operators in calculating the interaction force.

Later, Sbragaglia *et al.* [36] extended the idea of Shan to include more neighbors. They showed that the spurious currents can be further reduced when higher-order isotropic operators are employed to calculate the interaction force. However, it was also pointed out that [36, 260], when additional neighbors are included, the implementation of boundary conditions (particularly for wall boundaries) will become complex. Moreover, Sbragaglia and Shan [262] found that, when using the high-order isotropic operators, the pseudopotential should be modified from $\psi(\rho) = \psi_0 \exp(-\rho_0/\rho)$ to

$$\psi(\rho) = \left( \frac{\rho}{\varepsilon + \rho} \right)^{1/\varepsilon}, \tag{103}$$

so as to satisfy the thermodynamic consistency, where $\varepsilon$ is determined by the mechanical stability condition of the higher-order interactions.

In 2009, Yu and Fan [38] developed an MRT version of the pseudopotential LB model and found that the MRT collision model can reduce the spurious currents as compared with the BGK collision model, which can be seen in Fig. 5. For the studied case, it was shown that the maximum magnitudes of the spurious currents given by the BGK and MRT models are 0.028 and 0.0053, respectively [38].

Besides the above approaches, a refinement of curved interfaces is also capable of reducing the spurious currents, which is based on the fact that a better resolved interface provides a better resolution for the density gradient and thus can give a reduction of spurious currents. Sbragaglia *et al.* [36] found that, for the pseudopotential $\psi(\rho) = \sqrt{\rho_0} \left[ 1 - \exp(-\rho/\rho_0) \right]$, the interface thickness can be adjusted by tuning $\rho_0$. They stressed that increasing the interface thickness is more effective than the use of high-order isotropic gradient operators in terms of reducing the spurious currents, as the former treatment is found to give a numerical reduction of the maximal spurious current up to a factor of 10 by only doubling the interface thickness, while a 10th-order (or even higher) isotropic gradient operator is required to reach similar level of accuracy [36].

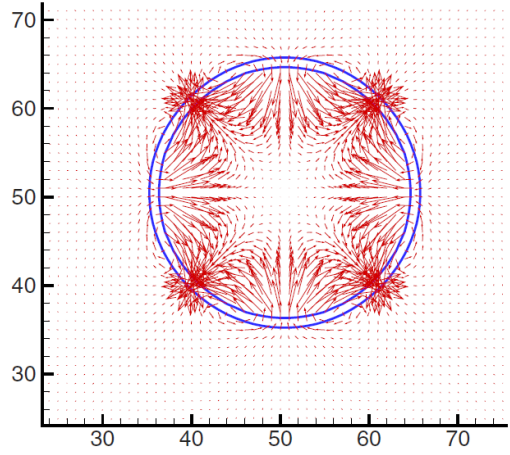

(a) BGK model

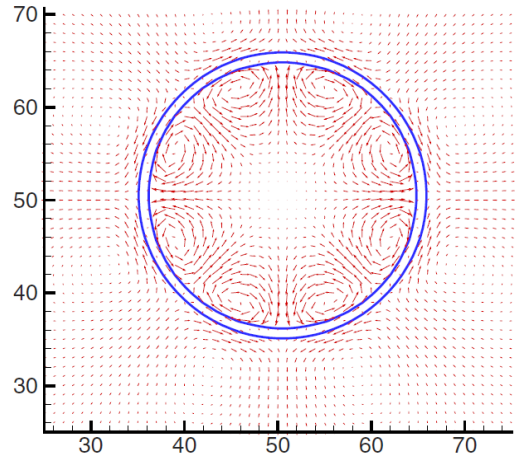

(b) MRT model

**Fig. 5**. Comparison of the spurious currents between the BGK and MRT collision models. Reprinted from Yu and Fan [38] with permission of the American Physical Society.

For the pseudopotential $\psi = \sqrt{2\left(p_{\text{EOS}} - \rho c_s^2\right)/Gc^2}$, the interface thickness was found to be affected by $p_{\text{EOS}}$. In the LB community, Wagner and Pooley [263] first conducted research on the influence of the equation of state on interface thickness. They introduced a pre-factor $p_0$ into the pressure tensor, which gives $p_{\text{EOS}}^{\text{new}} = p_0 p_{\text{EOS}}$. It was found that lowering $p_0$ can increase the interface thickness and the achievable largest density ratio can be enhanced by widening the interface thickness, which can be seen in Fig. 6. A similar treatment has also been proposed by Hu *et al*. [264] for the pseudopotential LB method.

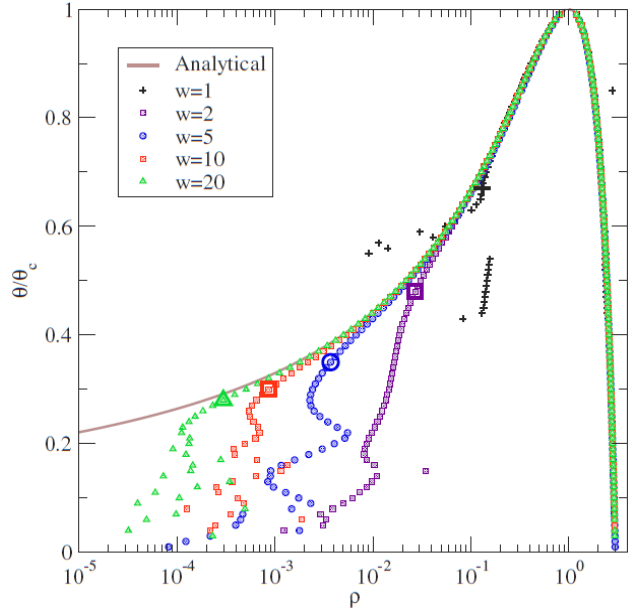

**Fig. 6**. The numerical van der Waals phase diagram with different interface thickness (w, in lattice unit). Here $\theta/\theta_c$ is the reduced temperature $T/T_c$. Reprinted from Wagner and Pooley [263] with permission of the American Physical Society.

Moreover, Huang *et al*. [245] found that the parameters $a$ and $b$ in the equations of state [33] affect the interface thickness. Similarly, Li *et al*. [39] found that the interface thickness is approximately proportional to $1/\sqrt{a}$ when the other parameters are fixed. Meanwhile, it was shown that the interface thickness decreases when the reduced temperature $T/T_c$ decreases (see Fig. 3 in Ref. [39]). With the usual choice of the parameters $a$ and $b$, such as $a=1$ and $b=4$ for the Carnahan-Starling equation of state [33], the interface will become very sharp in the cases of small $T/T_c$ (corresponding to large density ratios). Using a sharp interface, the spurious currents will be magnified dramatically. This is one of the reasons why the spurious currents shown in Yuan and Schaefer's work [33] are very large at high density ratios. Montessori *et al*. [265] recently also found that the spurious currents can be significantly reduced when decreasing the parameter $a$ (see Fig. 5 in the reference). Here it should also be mentioned that, at a given density ratio, the interface thickness produced by different equations of state is often different, unless an adjustment of the interface thickness has been made.

Using $p_{EOS}^{new} = p_0 p_{EOS}$ or tuning the parameter $a$ in the equation of state will not change the liquid-gas coexistence densities given by the Maxwell construction [39, 264]. However, the coexistence densities produced by the pseudopotential LB models will be affected, because the mechanical stability condition will be changed due to the term $\psi'/\psi$ in Eq. (90). Accordingly, the constant $\sigma$ in Eq. (101), which is used to tune the mechanical stability condition, should be slightly changed with the parameter $a$ [39].

Furthermore, it should be noted that the sound speed (related to $\partial p/\partial\rho$) in both the liquid and gas phases will decrease when $p_0$ is lowered [263]. A similar trend can be found when decreasing the parameter $a$ in the equation of state (see Eq. (33) in Ref. [39] and notice that $p_c \sim a/b^2$), which will lead to a relatively strong dependence of the coexistence densities on the droplet/bubble radius according to Laplace's law [266]. For circular droplets, Laplace's law is given by $p_l - p_g = \vartheta/r$, where $r$ is the radius of the droplet; $p_l$ and $p_g$ are the pressures of the liquid and gas phases, respectively, which can be described as [266]:

$$p_l = p_l^e + \frac{\rho_l^e}{\rho_l^e - \rho_g^e}\left(\frac{\vartheta}{r}\right), \quad p_g = p_g^e + \frac{\rho_g^e}{\rho_l^e - \rho_g^e}\left(\frac{\vartheta}{r}\right), \tag{104}$$

where the superscript "e" denotes the equilibrium properties of flat interfaces ($r \to \infty$). The pressure difference $\Delta p_g = \left(p_g - p_g^e\right)$ can be defined as $\Delta p_g = \left(\rho_g - \rho_g^e\right)\left(\sqrt{\left(\partial p/\partial\rho\right)_g}\right)^2$ [267], in which $\sqrt{\left(\partial p/\partial\rho\right)_g}$ is the gas-phase sound speed. With this relationship, Li and Luo [267] obtained

$$\rho_g - \rho_g^e = \frac{1}{\left(\partial p/\partial\rho\right)_g}\frac{\rho_g^e}{\rho_l^e - \rho_g^e}\left(\frac{\vartheta}{r}\right). \tag{105}$$

According to Eq. (105), it is obvious that, when $\left(\partial p/\partial\rho\right)_g$ decreases, the variation of the gas-phase density $\rho_g$ with the droplet radius $r$ will be enlarged. Similar results can be found for $\rho_l$.

In other words, for the classical equations of state in the thermodynamic theory such as the van der Waals and the Carnahan-Starling equations of state, the variations of the liquid-gas coexistence densities with the droplet/bubble radius (or the curvature of the liquid-gas interface) will be magnified when the interface is widened by lowering $\left(\partial p/\partial\rho\right)$. To reduce such an influence, for isothermal systems an alternative choice may be adopting a piecewise equation of state proposed by Colosqui *et al.* [268], through which $(\partial p/\partial \rho)$ can be controlled separately in every single phase region and the mixed region ($\partial p/\partial \rho < 0$). Li and Luo [267] showed that, using a piecewise equation of state, the interface can be widened by adjusting $(\partial p/\partial \rho)$ in the mixed region. Meanwhile, they found that [267] the sound speeds in the liquid and gas regions, $\sqrt{(\partial p/\partial \rho)_l}$ and $\sqrt{(\partial p/\partial \rho)_g}$, should be of the same order of magnitude as the lattice sound speed $c_s$ so that the dependence of the coexistence densities on the droplet size can be significantly reduced.

### 3.4 The surface tension treatment

Another drawback of the original pseudopotential LB model is that the surface tension given by the model cannot be tuned independently of the density ratio [36]. This problem can be found from the definition of the surface tension. For a flat interface, the surface tension $\vartheta$ can be evaluated as

$$\vartheta = \int_{-\infty}^{+\infty} \left( p_n - p_T \right) \mathrm{d}n , \tag{106}$$

where $p_n$ is the normal pressure tensor and $p_T$ is the transversal pressure tensor. Taking the $x$ direction as the normal direction of the flat interface, we have $p_n = P_{xx}$ and $p_T = P_{yy}$ [242]. According to the discrete form of the pressure tensor, i.e., Eq. (84), Eq. (106) gives

$$\vartheta = \frac{Gc^4}{6} \int_{-\infty}^{+\infty} \psi \frac{\mathrm{d}^2 \psi}{\mathrm{d}x^2} \mathrm{d}x \equiv \frac{Gc^4}{6} \int_{-\infty}^{+\infty} \psi \mathrm{d}\left( \frac{\mathrm{d}\psi}{\mathrm{d}x} \right) . \tag{107}$$

Using integration by parts and noting that $\mathrm{d}\psi/\mathrm{d}x = 0$ at $x = \pm\infty$, one can obtain

$$\vartheta = -\frac{Gc^4}{6} \int_{-\infty}^{+\infty} \left( \frac{\mathrm{d}\psi}{\mathrm{d}x} \right) \mathrm{d}\psi = -\frac{Gc^4}{6} \int_{\rho_g}^{\rho_l} \psi'^2 \left( \frac{\mathrm{d}\psi}{\mathrm{d}\rho} \right) \mathrm{d}\rho , \tag{108}$$

where $\psi' = \mathrm{d}\psi/\mathrm{d}\rho$. It seems that the surface tension can be adjusted by $G$ in Eq. (108). However, in the original pseudopotential LB model, $G$ controls the interaction strength. Different values of $G$ correspond to different liquid-gas coexistence densities. For the square-root pseudopotential given by Eq. (93), $G$ in front of the integral in Eq. (108) and $1/G$ in $\psi'^2$ will cancel each other out. As a result, $G$ has no influence on the surface tension. This conclusion is also applicable to circular and spherical interfaces, although Eq. (108) is the analytical expression for flat interfaces. For circular and spherical interfaces, the surface tension can be numerically evaluated via Laplace's law. Here it should also be noted that the definition of the surface tension in the pseudopotential LB method is different from that in thermodynamic theory, in which the surface tension is defined as $\vartheta \propto \int_{-\infty}^{+\infty} \left( \mathrm{d}\rho/\mathrm{d}x \right)^2 \mathrm{d}x$ [34, 247] for flat interfaces.

Several attempts have been made to enable a tunable surface tension. A well-known approach is the multi-range interaction force proposed by Sbragaglia $et\ al.$ [36]

$$\mathbf{F}(\mathbf{x},t) = -\psi(\mathbf{x}) \sum_{\alpha} w_{\alpha} \left[ G_1 \psi(\mathbf{x}+\mathbf{e}_{\alpha}\delta_t) + G_2 \psi(\mathbf{x}+2\mathbf{e}_{\alpha}\delta_t) \right] \mathbf{e}_{\alpha} \ . \tag{109}$$

When $G_1 = G$ and $G_2 = 0$, the multi-range interaction force reduces to the standard interaction force given by Eq. (75). Later, based on Sbragaglia $et\ al.$'s work, Falcucci $et\ al.$ [37] and Chibbaro $et\ al.$ [269] proposed a two-belt multi-range interaction force. In addition, Falcucci $et\ al.$ [270] have shown that the cooperation between the short- and mid-range interactions permits the achievement of phase-separation at liquid-gas density ratios in excess of 500. Using the Taylor expansion, Sbragaglia $et\ al.$ have obtained the following $continuum\ form$ pressure tensor of the multi-range interaction force [36]:

$$\mathbf{P} = \left[ p(\rho) + A_2 \left( \frac{c^4}{12} |\nabla \psi|^2 + \frac{c^4}{6} \psi \nabla^2 \psi \right) \right] \mathbf{I} - \frac{A_2 c^4}{6} \nabla \psi \nabla \psi \ , \tag{110}$$

where $A_2 = G_1 + 8G_2$ and $p(\rho) = \rho c_s^2 + A_1 c^2 \psi^2 / 2$ is the equation of state with $A_1 = G_1 + 2G_2$. With a fixed $A_1$, the surface tension can be adjusted by tuning $A_2$.

**Table 2**. Density ratios given by the multi-range interaction force for different surface tensions. Reprinted from Huang $et\ al.$ [245] with permission of the American Physical Society.

| $T/T_c$ | $\sqrt{A_2/A_1}$ | $\vartheta(\times 10^{-3})$ | $\rho_l/\rho_g$ |
|---------|------------------|------------------------------|-----------------|
| 0.85 | 1.0 | 2.683 | 0.2768 / 0.02250 |
| 0.85 | 1.458 | 4.199 | 0.2785 / 0.02587 |
| 0.85 | 1.904 | 5.461 | 0.2795 / 0.02775 |
| 0.775 | 1.0 | 5.661 | 0.3175 / 0.003745 |
| 0.775 | 1.458 | 7.939 | 0.3191 / 0.008159 |
| 0.775 | 1.904 | 10.57 | 0.3201 / 0.010520 |

In 2011, Huang *et al.* [245] found that the density ratio of the system changes when the multi-range interaction force is employed to adjust the surface tension. The results can be found in Table 2, which shows that the multi-range interaction force can provide a tunable surface tension at a given $T/T_c$; nevertheless, the density ratio changes considerably with the surface tension. In fact, the multi-range interaction force is capable of separating the equation of state from the surface tension. However, in the pseudopotential LB method, the liquid-gas coexistence densities are not only related to the equation of state but also dependent on the mechanical stability condition.

In 2013, following the statement of Shan [242], i.e., in the pseudopotential LB method the *discrete form* pressure tensor should be used, Li and Luo [249] derived the discrete form pressure tensor of the multi-range interaction force and found that the mechanical stability condition given by the multi-range interaction force depends on the parameters $G_1$ and $G_2$, which may be the reason why the density ratio of the system changes when the multi-range interaction force is used to adjust the surface tension.

An alternative approach was therefore proposed by Li and Luo [249] based on the LB-MRT equation. The basic idea is adding a source term into the LB-MRT equation and then Eq. (49) becomes

$$\mathbf{m}^* = \bar{\mathbf{m}} - \mathbf{\Lambda}\left(\bar{\mathbf{m}} - \mathbf{m}^{eq}\right) + \delta_t\left(\mathbf{I} - \frac{\mathbf{\Lambda}}{2}\right)\mathbf{S} + \delta_t\mathbf{C}, \qquad (111)$$

where the source term $\mathbf{C}$ is given by [249]

$$\mathbf{C} = \begin{bmatrix} 0 \\ 1.5\tau_e^{-1}\left(Q_{xx} + Q_{yy}\right) \\ -1.5\tau_\varsigma^{-1}\left(Q_{xx} + Q_{yy}\right) \\ 0 \\ 0 \\ 0 \\ 0 \\ -\tau_v^{-1}\left(Q_{xx} - Q_{yy}\right) \\ -\tau_v^{-1}Q_{xy} \end{bmatrix}. \qquad (112)$$

The discrete lattice effect [153] has been included to the source term $\mathbf{C}$, which actually takes the form $\mathbf{C} = \left(\mathbf{I} - 0.5\mathbf{\Lambda}\right)\mathbf{C}'$. The three variables $Q_{xx}$, $Q_{yy}$, and $Q_{xy}$ in Eq. (112) are calculated via [249]

$$\boldsymbol{Q} = \kappa\frac{G}{2}\psi\left(\mathbf{x}\right)\sum_{\alpha=1}^{8} w_\alpha\left[\psi\left(\mathbf{x} + \mathbf{e}_\alpha\delta_t\right) - \psi\left(\mathbf{x}\right)\right]\mathbf{e}_\alpha\mathbf{e}_\alpha, \qquad (113)$$

where the constant $\kappa$ is used to tune the surface tension. It can be seen that $\boldsymbol{Q}$ is based on the standard discrete form pressure tensor defined by Eq. (83) and is consistent with the calculation of the pseudopotential interaction force, Eq. (75). Through Chapman-Enskog analysis, it can be found that Eqs. (111) and (112) lead to the following second-order discrete form pressure tensor [249]:

$$\mathbf{P}_{\text{new}} = \left[ p_{\text{EOS}}(\rho) + (1+2\kappa)\frac{Gc^4}{12}\psi\nabla^2\psi \right]\mathbf{I} + (1-\kappa)\frac{Gc^4}{6}\psi\nabla\nabla\psi , \qquad (114)$$

where $p_{\text{EOS}}(\rho) = \rho c_s^2 + 0.5Gc^2\psi^2$. The surface tension can be tuned via $\kappa$ since the coefficient in front of the term $\psi\nabla\nabla\psi$ has been modified from 1 to $(1-\kappa)$. Meanwhile, the coefficient in front of the term $\psi\nabla^2\psi\mathbf{I}$ has also been changed, which ensures that the normal pressure tensor given by Eq. (114) for flat interfaces is the same as Eq. (86). In other words, the coefficient in front of the term $\psi\,\mathrm{d}^2\psi/\mathrm{d}n^2$ in Eq. (86) remains unchanged. Thus the mechanical stability condition, which determines the coexistence densities, will not be affected when the surface tension is adjusted [249].

Numerical results showed that the surface tension can be adjusted over a wide range and the density ratio can be kept essentially unchanged (see Tables II and III in Ref. [249]). The surface tension was found to be approximately proportional to the coefficient $(1-\kappa)$ in the cases of $(1-\kappa) > 0.1$. The linear relationship may not be valid when $(1-\kappa) < 0.1$ because Eq. (114) is the second-order pressure tensor. Some higher-order terms, which are truncation error terms for the surface tension and not shown in Eq. (114), may noticeably contribute to the surface tension when the coefficient in front of the term $\psi\nabla\nabla\psi$ is very small. Recently, following the above approach, Xu *et al.* [258] have devised a three-dimensional pseudopotential LB-MRT model with flexible surface tension. In addition, the approach has been employed by Hu *et al.* [271] to adjust the surface tension given by the mixed interaction force, Eq. (100).

It is worth noting that the theoretical frameworks of all the existing multiphase LB methods are based on second-order Chapman-Enskog analysis. Wagner [161] argued that second-order Chapman-Enskog analysis may be insufficient for a multiphase LB model because the higher-order derivatives are seemingly not derivable with a second-order expansion. By taking the free-energy multiphase LB method as an example, Wagner [161] has conducted a high-order Taylor expansion analysis to identify the high-order terms recovered in the macroscopic equations (see Eq. (49) in the reference). Nevertheless, Wagner also stressed that there is still much debate about whether a Taylor expansion analysis is equivalent to a Chapman-Enskog analysis in high orders. Moreover, Wagner [161] has proposed a forcing scheme based on the Taylor expansion analysis, which has been introduced in Section 2.3.2 and shown to be identical to Guo *et al.*'s forcing scheme.

Recently, based on Wagner's work, Lycett-Brown and Luo [272] have performed a high-order Taylor expansion analysis for the standard pseudopotential LB model. An assumption $\mathbf{F} = c_s^2 \nabla \rho$ was employed (see Eq. (A8) in Ref. [272]). The surface tension was tuned in a way that is similar to Eq. (113). In addition, a forcing scheme similar to Eq. (97) was utilized to adjust the mechanical stability condition. Without the adjustment of the surface tension, the forcing term in Ref. [272] satisfies the following relationship (see Eqs. (34) and (37) in the reference):

$$\sum_\alpha \mathbf{e}_\alpha \mathbf{e}_\alpha F_{\alpha, \mathrm{LyB}} = \mathbf{uF} + \mathbf{Fu} + \left(1 - \frac{1}{4\tau}\right)\frac{\mathbf{FF}}{\rho} - \frac{\varepsilon_0}{8Gc_s^2}\frac{\mathbf{FF}}{\tau\psi^2}. \tag{115}$$

The first three terms on the right-hand side of Eq. (115) are the results of Wagner's forcing term ($\delta_t = 1$), which can be seen from Eq. (33). The last terms in Eqs. (98) and (115) are similar since $\sigma$ in Eq. (98) is given by $\sigma = -\varepsilon/16G$ and $\varepsilon_0$ in Eq. (115), which was numerically determined by Eq. (35) in Ref. [272], corresponds to $\varepsilon = \varepsilon_0$. The appearance of $c_s^2$ in Eq. (115) is due to its existence in the interaction force in Ref. [272].

### 3.5 The contact angle treatment

Wetting phenomena are widespread in natural and industrial processes. The wettability of a solid surface by a liquid is often quantified via the contact angle [273-275]. There have been many studies of wetting phenomena using the pseudopotential LB model since its emergence. The first attempt was made by Martys and Chen in 1996 [276]. In their study, a fluid-solid interaction force was introduced to describe the interaction between the fluid and the solid wall

$$\mathbf{F}_{\mathrm{fs}} = -G_w \rho(\mathbf{x})\sum_\alpha \omega_\alpha s(\mathbf{x} + \mathbf{e}_\alpha \delta_t)\mathbf{e}_\alpha, \tag{116}$$

where $G_w$ is an adsorption parameter for adjusting the value of the contact angle and $s(\mathbf{x}+\mathbf{e}_\alpha\delta_t)$ is a "switch" function, which is equal to 1 or 0 for a solid or a fluid phase, respectively. It means that an adhesion force is acted on the lattice nodes that are located in the fluid with neighboring solid nodes.

Later, Raiskinmäki *et al*. [277, 278] proposed another type of fluid-solid interaction, which was reformulated by Sukop and Thorne as follows [12]:

$$\mathbf{F}_{fs} = -G_w\psi(\mathbf{x})\sum_\alpha \omega_\alpha s(\mathbf{x}+\mathbf{e}_\alpha\delta_t)\mathbf{e}_\alpha \ . \tag{117}$$

The difference between Eq. (116) and Eq. (117) lies in that the pre-sum factor in Eq. (116) is the density $\rho(\mathbf{x})$, while in Eq. (117) the pre-sum factor is the pseudopotential $\psi(\mathbf{x})$. For simplicity, these two types of fluid-solid interaction are referred to as the $\rho-$based interaction and the $\psi-$based interaction, respectively. In the literature, Kang *et al*. [279, 280] have extended Martys and Chen's $\rho-$based fluid-solid interaction to the D2Q9 lattice model.

In 2006, Benzi *et al*. [281, 282] introduced a parameter $\psi(\rho_w)$ to fix the pseudopotential at the solid wall via a virtual wall density $\rho_w$. By introducing a suitable value for $\rho_w$, the contact angle can be adjusted between $0^\circ$ and $180^\circ$. Some numerical results obtained with such a treatment can be found in Fig. 7. It can be seen that the contact angle approaches $0^\circ$ and $180^\circ$ when the virtual wall density $\rho_w$ is close to the liquid-phase density $\rho_l$ and the gas-phase density $\rho_g$, respectively.

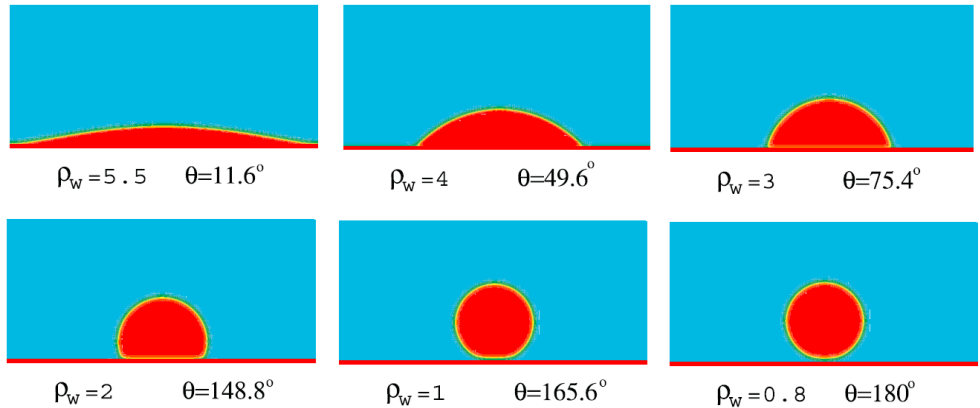

**Fig. 7**. Simulation of contact angles with a virtual wall density $\rho_w$. The corresponding liquid and gas densities are $\rho_l = 6.06$ and $\rho_g = 0.5$, respectively. Reprinted from Huang *et al*. [283] with permission of John Wiley & Sons, Inc.

In 2013, Colosqui *et al.* [284] proposed a new fluid-solid interaction, which is composed of a repulsive core and an attractive tail

$$\mathbf{F}_{fs} = \rho(\mathbf{x}) \sum_{\alpha} \omega_{\alpha} \psi_{FS} \left(\mathbf{x} + \mathbf{e}_{\alpha} \delta_t\right) \mathbf{e}_{\alpha} + \Delta \mathbf{F}_S (\mathbf{x}), \qquad (118)$$

where $\Delta \mathbf{F}_S (\mathbf{x})$ introduces a momentum exchange between the fluid and solid molecules due to the short-range interactions in the region adjacent to the solid surface and $\psi_{FS}(\mathbf{x}) = G_R \bar{\psi}_R (\mathbf{x}) + G_A \bar{\psi}_A (\mathbf{x})$, in which $\bar{\psi}_R$ and $\bar{\psi}_A$ are the repulsive and attractive potentials, respectively, and $G_R$ and $G_A$ are the attraction parameters. With the new fluid-solid interaction force, Colosqui *et al.* [284] successfully simulated both static contact angles and contact angle hysteresis; nevertheless, they pointed out that their approach is stable and robust at low and moderate density ratios ($\rho_l / \rho_g \sim 10$).

Recently, Li *et al.* [285] have conducted a comparative study between the $\rho$-based interaction and the $\psi$-based interaction at a high density ratio ($\rho_l / \rho_g = 500$). For liquid contact angles, it was found that the $\psi$-based interaction works well for $G_w < 0$. However, when $G_w > 0$, the $\rho$-based interaction performs much better than the $\psi$-based interaction in terms of the achievable contact angle and the deviations (from equilibrium) of the maximum and minimum densities. A modified $\psi$-based interaction was therefore formulated in Ref. [285] for liquid contact angles with $G_w > 0$

$$\mathbf{F}_{fs} = -G_w \psi(\mathbf{x}) \sum_{\alpha} \omega_{\alpha} S \left(\mathbf{x} + \mathbf{e}_{\alpha} \delta_t\right) \mathbf{e}_{\alpha}, \qquad (119)$$

where $S(\mathbf{x} + \mathbf{e}_{\alpha} \delta_t) = \phi(\mathbf{x}) s(\mathbf{x} + \mathbf{e}_{\alpha} \delta_t)$, in which $\phi(\mathbf{x})$ can be chosen as $\phi(\mathbf{x}) = \psi(\mathbf{x})$. It was shown that the modified $\psi$-based interaction is superior over the $\rho$-based interaction in light of the deviations (from equilibrium) of the maximum and minimum densities.

When the fluid-solid interaction force is exerted on the lattice nodes adjacent to the solid wall, it can be found that the mechanical stability condition at these lattice nodes will be affected. As a result, the liquid-gas coexistence densities at these lattice nodes will be changed. This is the reason why the maximum and minimum densities deviate from the equilibrium liquid and gas densities, respectively [285]. Moreover, it should be noted that the status of the fluid-fluid interaction force (namely Eq. (75)) at the solid wall will also affect the contact angle according to Young's law. For a given contact angle, the values of $G_w$ are usually different between the cases with and without the fluid-fluid interaction force at the solid wall.

Although substantial progress has been made in simulating contact angles with the pseudopotential LB method, many more efforts are still required to address several critical issues, such as the large spurious currents around the three-phase contact line in the cases with low relaxation times and the problem that the value of the contact angle cannot be simply prescribed before simulations. Recently, Hu *et al.* [286] presented an alternative treatment to implement contact angles in the pseudopotential LB modeling of wetting based on the geometric formulation approach, which was proposed by Ding and Spelt [287] within the framework of the phase-field method and has been demonstrated to be capable of exactly reproducing a given contact angle. The geometric formulation approach will be reviewed in Section 4.4.

### 3.6 Thermal models for liquid-vapor phase change

In this subsection, the thermal pseudopotential LB models for simulating liquid-vapor phase change are reviewed. Historically, the first thermal pseudopotential LB model may be attributed to Zhang and Chen [94]. They devised a new force expression and successfully modeled the nucleate boiling. Later, an important contribution was made by Házi and Márkus [95, 96], who established the target governing equation of the temperature for the thermal pseudopotential LB method.

Házi and Márkus [95] started from the local balance law for entropy (the viscous heat dissipation was neglected) [288]

$$\rho T \frac{Ds}{Dt} = \nabla \cdot \left( \lambda \nabla T \right),$$ (120)

where $s$ is the entropy, $\lambda$ is the thermal conductivity, and $D(\bullet)/Dt = \partial_t(\bullet) + \mathbf{v} \cdot \nabla(\bullet)$ is the material derivative. Using the thermodynamic relationships of non-ideal gases [95], Márkus and Házi [96] obtained the following target temperature equation:

$$\partial_t T + \mathbf{v} \cdot \nabla T = \frac{1}{\rho c_v} \nabla \cdot \left( \lambda \nabla T \right) - \frac{T}{\rho c_v} \left( \frac{\partial p_{\text{EOS}}}{\partial T} \right)_\rho \nabla \cdot \mathbf{v}.$$ (121)

In the thermal pseudopotential LB method, the liquid-vapor phase change is driven by the equation of state $p_{EOS}$. Hence no artificial phase-change terms need to be added to the temperature equation. A similar treatment can also be found in Refs. [289, 290]. With such a treatment, the rate of the liquid-vapor phase change is a computational output rather than a prerequisite input.

Since $\mathbf{v} \cdot \nabla T$ and $\nabla \cdot (\lambda \nabla T)/(\rho c_v)$ in Eq. (121) cannot be directly recovered with a thermal LB equation, Márkus and Házi [96] rewrote Eq. (121) as follows:

$$\partial_t T + \nabla \cdot (\mathbf{v}T) = \nabla \cdot (k\nabla T) + \frac{1}{\rho c_v}\nabla \cdot (\lambda \nabla T) - \nabla \cdot (k\nabla T) + \left[T - \frac{T}{\rho c_v}\left(\frac{\partial p_{EOS}}{\partial T}\right)_\rho\right]\nabla \cdot \mathbf{v}. \quad (122)$$

To mimic Eq. (122), the following temperature-based thermal LB equation was employed [96]:

$$T_\alpha\left(\mathbf{x} + \mathbf{e}_\alpha \delta_t, t + \delta_t\right) - T_\alpha\left(\mathbf{x}, t\right) = -\frac{1}{\tau_T}\left(T_\alpha - T_\alpha^{eq}\right) + S_\alpha, \quad (123)$$

where the equilibrium temperature distribution function $T_\alpha^{eq}$ is taken as $T_\alpha^{eq} = \omega_\alpha T\left[1 + (\mathbf{e}_\alpha \cdot \mathbf{v})/c_s^2\right]$ and the source term $S_\alpha$ is utilized to recover all the terms on the right-hand side of Eq. (122) except the first term $\nabla \cdot (k\nabla T)$.

Using the model, Márkus and Házi [96] have reproduced nucleate boiling. However, the error terms in the recovered temperature equation should be noticed. The macroscopic temperature equation recovered from the standard temperature-based thermal LB equation, namely Eq. (123) without the source term, has been discussed in Section 2.4.1. Details can be found in the paragraph below Eq. (67). Using the equilibrium distribution $T_\alpha^{eq} = \omega_\alpha T\left[1 + (\mathbf{e}_\alpha \cdot \mathbf{v})/c_s^2\right]$, the error term $\nabla \cdot (T\mathbf{v}\mathbf{v})$ in Eq. (67) can be eliminated. However, the main error term $\partial_{t0}(T\mathbf{v})$, which yields an error proportional to $\left(\mathbf{F} - c_s^2\nabla\rho\right)/\rho$, still exists. Furthermore, it is worth mentioning that the discrete lattice effects [153] should also be considered when incorporating a source term into a thermal LB equation.

In 2012, Biferale *et al.* [98] proposed a three-dimensional thermal pseudopotential LB model. The target macroscopic temperature equation is similar to Eq. (121). The main difference is that the exponential-form pseudopotential $\psi(\rho) = \exp(-1/\rho)$ is adopted in Biferale *et al.*'s work and the equation of state is given by $p_{EOS} = \rho RT + 0.5Gc^2\psi^2(\rho)$. Hence $(\partial p_{EOS}/\partial T)_\rho = \rho R$. Using the model, Biferale *et al.* have performed three-dimensional numerical simulations of boiling phenomena at the Rayleigh number $Ra \sim 10^7$. Some numerical results can be found in Fig. 8. In Ref. [98], the ideal part of the equation of state, $p_{id} = \rho RT$, was incorporated into the model by modifying the equilibrium density distribution function. Similar treatments can be found in the free-energy and the color-gradient multiphase LB models. It is now widely accepted that, when the equation of state or the pressure is changed by modifying the equilibrium density distribution function, the Galilean invariance is lost [42, 43]. Some correction terms, which usually involve the derivatives of the density, should be added to either the equilibrium distribution function or the LB equation to restore the Galilean invariance [42, 43, 137].

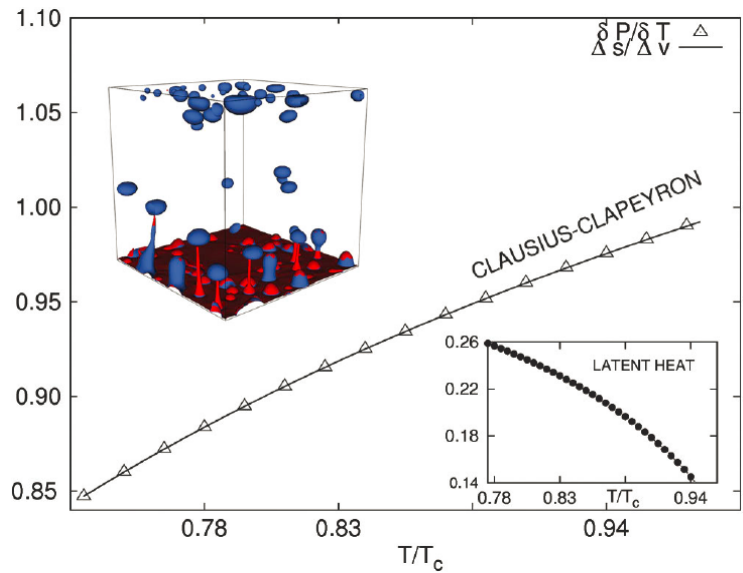

**Fig. 8**. Three-dimensional thermal pseudopotential LB modeling of nucleate boiling. Bubbles are in blue. Regions with high temperature are in red. Reprinted from Biferale *et al.* [98] with permission of the American Physical Society.

Gong and Cheng [99, 100] have also proposed a thermal pseudopotential LB model and simulated bubble growth and departure. The model has been applied to simulate film condensation by Liu and Cheng [101]. In 2013, Kamali *et al.* [102] constructed a thermal pseudopotential LB model based on the energy conservation equation of mixtures and successfully simulated the evaporation of a liquid film upon heating. Similarly, in Kamali *et al.*'s work the energy equation was mimicked by a thermal

LB equation with a source term. Here it is stressed again that, when a thermal LB equation is employed together with a multiphase LB model, special attention should be paid to the correct recovery of the target temperature or energy equation. Some error terms, which are very small in single-phase incompressible flows, are non-negligible in multiphase flows. Li and Luo [195] have shown that, in the pseudopotential LB modeling of thermal flows, the error term in the recovered energy equation will lead to significant numerical errors. Recently, Li *et al.* [104] developed a hybrid thermal pseudopotential LB model, in which the pseudopotential LB method was used to simulate the density and velocity fields and a finite-difference scheme was employed to solve Eq. (121). Its applications will be discussed in Section 5.3.

### 3.7 Multi-component formulations

Besides its use in modeling single-component multiphase flows, the pseudopotential LB method has also been widely applied to simulate multi-component multiphase flows. Here we briefly summarize the multi-component formulations of the pseudopotential LB method. For multi-component systems, the fluid-fluid interaction force is given by [12, 31, 243]

$$\mathbf{F}_k(\mathbf{x}, t) = -\psi_k(\mathbf{x}) \sum_{\bar{k}} G_{k\bar{k}} \sum_{\alpha} w_\alpha \psi_{\bar{k}}(\mathbf{x} + \mathbf{e}_\alpha \delta_t) \mathbf{e}_\alpha , \qquad (124)$$

where $G_{k\bar{k}}$ is a parameter reflecting the interactive strength and satisfies $G_{k\bar{k}} = G_{\bar{k}k}$, and the subscripts $k$ and $\bar{k}$ denote the components, in which $k$ is a free index and $\bar{k}$ is a dummy index, e.g., for a two-component system with components A and B, the interaction force $\mathbf{F}_A$ is given by

$$\mathbf{F}_A(\mathbf{x}, t) = -\psi_A(\mathbf{x}) \left[ G_{AA} \sum_{\alpha} w_\alpha \psi_A(\mathbf{x} + \mathbf{e}_\alpha \delta_t) \mathbf{e}_\alpha + G_{AB} \sum_{\alpha} w_\alpha \psi_B(\mathbf{x} + \mathbf{e}_\alpha \delta_t) \mathbf{e}_\alpha \right]. \qquad (125)$$

The force $\mathbf{F}_B$ can be formulated similarly. Note that, when $\bar{k} = k$, $G_{k\bar{k}}$ represents the interaction within each component. For ideal gases, $G_{kk}$ can be set to zero.

In the original pseudopotential LB model, the LB equation for each component is given by [243]

$$f_\alpha^k(\mathbf{x} + \mathbf{e}_\alpha \delta_t, t + \delta_t) - f_\alpha^k(\mathbf{x}, t) = -\frac{1}{\tau_k} \left[ f_\alpha^k - f_\alpha^{eq,k}(\rho, \mathbf{u}_k^{eq}) \right]. \qquad (126)$$

Obviously, the above equation is a direct extension of Eq. (76) for multi-component systems. The shifted equilibrium velocity of each component is defined as $\rho_k \mathbf{u}_k^{eq} = \rho_k \mathbf{u}' + \tau_k \mathbf{F}_k$, in which $\rho_k = \sum_\alpha f_\alpha^k$ and $\mathbf{u}'$ is given as follows [243]:

$$\mathbf{u}' = \sum_k \frac{\rho_k \mathbf{u}_k}{\tau_k} \Big/ \sum_k \frac{\rho_k}{\tau_k} , \qquad (127)$$

where $\rho_k \mathbf{u}_k = \sum_\alpha f_\alpha^k \mathbf{e}_\alpha$. Previously it has been mentioned that the Shan-Chen forcing scheme, i.e., Eq. (76), will introduce some error terms into the macroscopic equations and lead to $\tau$-dependent numerical results for single-component multiphase flows. In 2012, Porter $et\ al.$ [291] found that Eqs. (126) and (127) also yield $\tau$-dependent properties (see Fig. 4 in the reference). To solve this problem, they extended He $et\ al.$'s forcing scheme to multi-component systems [291]

$$f_\alpha^k \left( \mathbf{x} + \mathbf{e}_\alpha \delta_t, t + \delta_t \right) - f_\alpha^k \left( \mathbf{x}, t \right) = -\frac{f_\alpha^k - f_\alpha^{eq,k} \left( \rho_k, \mathbf{u}^{eq} \right)}{\tau_k} + \delta_t \left( 1 - \frac{1}{2\tau_k} \right) \frac{\left( \mathbf{e}_\alpha - \mathbf{u}^{eq} \right) \cdot \mathbf{F}_k}{\rho_k c_s^2} f_\alpha^{eq,k} , \quad (128)$$

where the equilibrium velocity $\mathbf{u}^{eq}$ in $f_\alpha^{eq,k}$ and the forcing term is given by $\mathbf{u}^{eq} = \mathbf{u}'$ with $\rho_k \mathbf{u}_k$ in Eq. (127) being replaced by $\rho_k \mathbf{u}_k = \sum_\alpha f_\alpha^k \mathbf{e}_\alpha + 0.5 \delta_t \mathbf{F}_k$. The corresponding mixture velocity is defined via $\rho \mathbf{u} = \sum_k \rho_k \mathbf{u}_k$, where $\rho = \sum_k \rho_k$. Here it is seen that $\mathbf{u}^{eq}$ in Eq. (128) is not equal to the mixture velocity $\mathbf{u}$. Meanwhile, we also notice that, in the studies of Chai and Zhao [292] and Sbragaglia and Belardinelli [248], where the MRT version of Guo $et\ al.$'s forcing scheme was employed, the velocity used in the equilibrium distribution function and the forcing term is defined as $\mathbf{u} = \sum_k \left( \sum_\alpha f_\alpha^k \mathbf{e}_\alpha + 0.5 \delta_t \mathbf{F}_k \right) \Big/ \rho$, which is just the mixture velocity. The latter treatment is seemingly more reasonable, but further studies are still needed to clarify this issue.

In Section 3.1.3 we have mentioned that Sbragaglia and Belardinelli [248] demonstrated that the discrete form of the pressure tensor should also be used in the multi-component pseudopotential LB models. For single-component models, it has been shown that the discrete form pressure tensor leads to a mechanical stability condition like Eq. (90), from which the coexistence densities can be analytically obtained via numerical integration. Meanwhile, the form of the pseudopotential can be determined according to thermodynamic consistency or the requirement of reproducing a non-ideal equation of state in the thermodynamic theory. When the thermodynamic consistency is satisfied or approximately restored, the coexistence densities obtained from the Maxwell construction can serve as the initial densities in numerical simulations. However, for the multi-component pseudopotential LB models, there have been no such principles for determining the form of the pseudopotential $\psi_k$ and the initial densities. Some empirical choices of $\psi_k$ were therefore employed in the literature, such as $\psi_k = \rho_k$ [248, 293], $\psi_k = 1 - \exp(-\rho_k)$ [248, 292], and the square-root form adopted in several studies [294, 295]. In spite of the fact that much progress has been achieved in the multi-component pseudopotential LB method, significant efforts are still required as its theory is far from being complete.

## 4. The phase-field multiphase LB method

### 4.1 The phase-field theory

#### 4.1.1 Governing equations

The phase-field theory is a descendant of van der Waals [296] and Cahn-Hilliard's [297, 298] classical field theoretical approaches, in which the local state of matter is represented continuously by a single variable known as the order parameter $\phi(\mathbf{x}, t)$ that monitors the transition between different states or phases. The set of values of the order parameter over the whole occupied space is the so-called *phase-field*.

For a fluid, its thermodynamic behavior can be expressed by a free energy that is functional of the order parameter $\phi(\mathbf{x}, t)$ as follows [299]:

$$\mathcal{F}(\phi) = \int_V \left( \mathcal{E}(\phi) + \frac{1}{2} k |\nabla \phi|^2 \right) dV , \tag{129}$$

where $V$ is the region of space occupied by the system. The term $0.5k |\nabla \phi|^2$ denotes the interfacial energy density, in which $k$ is a positive constant, and $\mathcal{E}(\phi)$ is the bulk energy density which has two minima corresponding to the two phases of the fluid. The chemical potential $\mu_\phi$ is defined as the variation of the free energy with respect to the order parameter [65, 300]

$$\mu_\phi = \frac{\delta \mathcal{F}(\phi)}{\delta \phi} = \mathcal{E}'(\phi) - k \nabla^2 \phi, \tag{130}$$

where $\mathcal{E}'(\phi) = \mathrm{d}\mathcal{E}(\phi)/\mathrm{d}\phi$. van der Waals [296] hypothesized that the equilibrium interface profiles can be obtained by minimizing $\mathcal{F}(\phi)$, i.e., the equilibrium profiles satisfy $\mu_\phi = \mathcal{E}'(\phi) - k \nabla^2 \phi = 0$ [300], which can be viewed as the governing equation for the order parameter at equilibrium. Later, Cahn and Hilliard [297, 298] generalized the time-dependent governing equation for $\phi$ by approximating the interfacial diffusion flux $\mathbf{J}$ as being proportional to $\nabla \mu_\phi$

$$\frac{\partial \phi}{\partial t} = \nabla \cdot \mathbf{J} = \nabla \cdot \left( M \nabla \mu_\phi \right), \tag{131}$$

where $M$ is the mobility coefficient. Then the convective Cahn-Hilliard equation is given by [300]

$$\frac{\partial \phi}{\partial t} + \mathbf{v} \cdot \nabla \phi = \nabla \cdot \left( M \nabla \mu_\phi \right). \tag{132}$$

The fluid dynamics is described by the Navier-Stokes equations with a surface tension term

$$\nabla \cdot \mathbf{v} = 0, \tag{133}$$

$$\rho \left( \frac{\partial \mathbf{v}}{\partial t} + \mathbf{v} \cdot \nabla \mathbf{v} \right) = -\nabla p + \nabla \cdot \mathbf{\Pi} + \mathbf{F}_s, \tag{134}$$

where $\mathbf{\Pi}$ is the viscous stress tensor and $\mathbf{F}_s$ is the surface tension force, which can be given by $\mathbf{F}_s = -\phi \nabla \mu_\phi$ [65] or $\mathbf{F}_s = \mu_\phi \nabla \phi$ [300]. Note that these two-type surface tension forces are theoretically equivalent since $\mu_\phi \nabla \phi \equiv \nabla \left( \mu_\phi \phi \right) - \phi \nabla \mu_\phi$, in which $\nabla \left( \mu_\phi \phi \right)$ can be absorbed into the pressure gradient term in Eq. (134). Nevertheless, the numerical performances of these two forms are usually different as the variations of $\phi$ and $\mu_\phi$ across the interface are different. There are also some other forms of the surface tension force, which can be found in Ref. [301]. Khan remarked that [302] the existence of so many forms of the surface tension force is because further efforts are needed to simulate the problems with large density and viscosity ratios.

The coupled Cahn-Hilliard-Navier-Stokes equations, Eqs. (132), (133), and (134), have been demonstrated to be valid for both density-matched binary fluids [65, 303] and two-phase fluid flows with a density contrast [304]. For density-matched binary fluids, the density $\rho$ in Eq. (134) is the average density, which is nearly a constant, while for two-phase fluid flows the density $\rho$ is functional of the order parameter $\phi$.

### 4.1.2 Interface properties

The equilibrium interface profiles can be obtained by solving $\mu_\phi = \mathcal{E}'(\phi) - k\nabla^2\phi = 0$. In the phase-field method, the bulk energy density is usually chosen to have a double-well form [300]

$$\mathcal{E}(\phi) = \beta\left(\phi - \phi^*\right)^2\left(\phi + \phi^*\right)^2, \tag{135}$$

where $\phi^*$ is a constant that defines the two phases of the fluid at $\phi = \pm\phi^*$ and $\beta$ is a parameter that can be used together with $k$ in the chemical potential to control the surface tension and the interface thickness. According to Eqs. (130) and (135), the chemical potential is given by

$$\mu_\phi = 4\beta\left(\phi^3 - \phi^{*2}\phi\right) - k\nabla^2\phi. \tag{136}$$

The one-dimensional (e.g., along the $x$-direction) solution of $\mu_\phi = 0$ is [300]

$$\phi(x) = \phi^*\tanh\left(\frac{x}{\sqrt{2}w}\right), \tag{137}$$

where $w = \left(1/2\phi^*\right)\sqrt{k/\beta}$. In some studies [48, 305], the interface thickness is defined from $x = -\sqrt{2}w$ to $x = \sqrt{2}w$, making the total interface thickness equal to $W = 2\sqrt{2}w = \left(1/\phi^*\right)\sqrt{2k/\beta}$. Obviously, such a choice defines the interface thickness from $\phi = -\left(\tanh 1\right)\phi^* \approx -0.76\phi^*$ to $\phi \approx 0.76\phi^*$. In Refs. [65, 300, 306], the interface thickness is calculated from $\phi = -0.9\phi^*$ to $\phi = 0.9\phi^*$, with the corresponding formulation $W = 2\sqrt{2}w\tanh^{-1}\left(0.9\right) \approx 4.164w$. According to Fig. 9, the latter treatment is more reasonable.

At equilibrium the surface tension is equal to the integral given by Eq. (129) along the interface [300]. For a flat interface along the $x$-direction, the surface tension can be calculated by

$$\vartheta = \int_{-\infty}^{+\infty}\left[\mathcal{E}(\phi) + \frac{k}{2}\left(\frac{\mathrm{d}\phi}{\mathrm{d}x}\right)^2\right]\mathrm{d}x. \tag{138}$$

Meanwhile, the requirement $\mu_\phi = \mathcal{E}'(\phi) - k\nabla^2\phi = 0$ gives

$$\frac{\mathrm{d}\mathcal{E}(\phi)}{\mathrm{d}\phi} = k\frac{\mathrm{d}^2\phi}{\mathrm{d}x^2} \equiv \frac{k}{2}\frac{\mathrm{d}}{\mathrm{d}\phi}\left(\frac{\mathrm{d}\phi}{\mathrm{d}x}\right)^2. \tag{139}$$

Since $\mathcal{E}\left(\pm\phi^*\right)=0$ and $\mathrm{d}\phi/\mathrm{d}x=0$ at $\phi=\pm\phi^*$, integrating Eq. (139) will lead to the following relationship:

$$\mathcal{E}\left(\phi\right)=\frac{k}{2}\left(\frac{\mathrm{d}\phi}{\mathrm{d}x}\right)^2 . \tag{140}$$

Substituting Eq. (140) as well as Eq. (137) into Eq. (138) yields

$$\vartheta=\int_{-\infty}^{+\infty}k\left(\frac{\mathrm{d}\phi}{\mathrm{d}x}\right)^2\mathrm{d}x=k\phi^{*2}\frac{4}{3}\frac{1}{\sqrt{2}w}=\frac{4\phi^{*3}}{3}\sqrt{2k\beta} . \tag{141}$$

The above formulations of the interface thickness and the surface tension are based on Eq. (135). For a general choice of $\mathcal{E}\left(\phi\right)$, such as $\mathcal{E}\left(\phi\right)=\beta\left(\phi-\phi_l\right)^2\left(\phi-\phi_g\right)^2$, the constant $\phi^*$ in Eq. (141) can be replaced with $\left(\phi_l-\phi_g\right)/2$ [307]. It can be seen that both the interface thickness and the surface tension are related to $k$ and $\beta$. In numerical simulations, when the interface thickness and the surface tension are given, the parameters $k$ and $\beta$ can be determined.

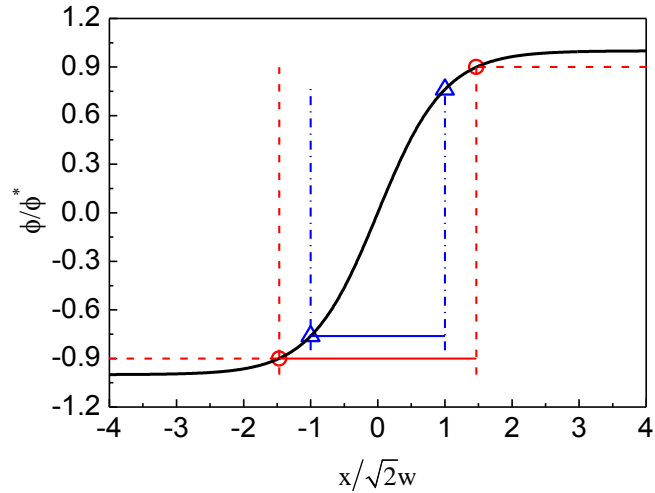

**Fig. 9**. The interface profile given by Eq. (137). The solid red line denotes the interface thickness defined from $\phi=-0.9\phi^*$ to $\phi=0.9\phi^*$, while the solid blue line represents the interface thickness calculated from $\phi\approx-0.76\phi^*$ to $\phi\approx0.76\phi^*$.

### 4.2 The isothermal phase-field LB models

#### 4.2.1 Early models

In 1999, He *et al.* [45] proposed an incompressible LB model for simulating multiphase flows, in which they used an index function to track the interface between different phases. Two distribution functions were employed in the model, the index distribution function $f_\alpha$ and the pressure distribution function $g_\alpha$, which satisfy the following evolution equations [45, 308]:

$$f_\alpha\left(\mathbf{x}+\mathbf{e}_\alpha\delta_t, t+\delta_t\right)-f_\alpha(\mathbf{x}, t)=-\frac{1}{\tau}\left(f_\alpha-f_\alpha^{eq}\right)-\delta_t\left(1-\frac{1}{2\tau}\right)\frac{\Gamma_\alpha(\mathbf{v})}{c_s^2}(\mathbf{e}_\alpha-\mathbf{v})\cdot\nabla\varphi(\phi), \quad (142)$$

$$g_\alpha\left(\mathbf{x}+\mathbf{e}_\alpha\delta_t, t+\delta_t\right)-g_\alpha(\mathbf{x}, t)=-\frac{1}{\tau}\left(g_\alpha-g_\alpha^{eq}\right)+\delta_t\left(1-\frac{1}{2\tau}\right)(\mathbf{e}_\alpha-\mathbf{v})\cdot\mathbf{\Phi}, \quad (143)$$

where $\varphi(\phi)=p_{\text{EOS}}(\phi)-\phi c_s^2$ and $\mathbf{\Phi}=\Gamma_\alpha(\mathbf{v})\left(\mathbf{F}_s+\mathbf{F}_g\right)-\left[\Gamma_\alpha(\mathbf{v})-\Gamma_\alpha(0)\right]\nabla\varphi(\rho)$, in which $\mathbf{F}_s$ is the surface tension force, $\mathbf{F}_g$ is the gravity force, and $\varphi(\rho)$ and $\Gamma_\alpha(\mathbf{v})$ are given by, respectively

$$\varphi(\rho)=p-\rho c_s^2, \quad (144)$$

$$\Gamma_\alpha(\mathbf{v})=\omega_\alpha\left(1+\frac{(\mathbf{e}_\alpha\cdot\mathbf{v})}{c_s^2}+\frac{(\mathbf{e}_\alpha\cdot\mathbf{v})^2}{2c_s^4}-\frac{|\mathbf{v}|^2}{2c_s^2}\right), \quad (145)$$

where $p$ is the hydrodynamic pressure and the weights $\omega_\alpha$ are the same as those in Eq. (9). The variable $\phi$ in $\varphi(\phi)$ is an order parameter and is calculated via $\phi=\sum_\alpha f_\alpha$. The recovered macroscopic hydrodynamic equations are

$$\frac{1}{\rho c_s^2}\left(\frac{\partial p}{\partial t}+\mathbf{v}\cdot\nabla p\right)+\nabla\cdot\mathbf{v}=0, \quad (146)$$

$$\frac{\partial(\rho\mathbf{v})}{\partial t}+\nabla\cdot(\rho\mathbf{v}\mathbf{v})=-\nabla p+\nabla\cdot\mathbf{\Pi}+\mathbf{F}_s+\mathbf{F}_g. \quad (147)$$

For incompressible flows, $\partial_t p$ is very small and $\mathbf{v}\cdot\nabla p$ is the order of $O(\text{Ma}^3)$, where Ma is the Mach number, thus approximately satisfying the divergence-free condition ($\nabla\cdot\mathbf{v}=0$). Meanwhile, the recovered governing equation for $\phi$ is [309]

$$\frac{\partial\phi}{\partial t}+\nabla\cdot(\phi\mathbf{v})=\nabla\cdot\left[\lambda_\phi\left(\nabla p_{\text{EOS}}(\phi)-\frac{\phi}{\rho}\nabla p\right)\right], \quad (148)$$

where $\lambda_\phi$ is related to the relaxation time. In Ref. [309], Eq. (148) was interpreted as a level-set-like equation. Lee and Lin [310] pointed out such an interpretation is not appropriate. It is more appropriate to regard Eq. (148) as a Cahn-Hilliard-like equation. From this point, He *et al.*'s model is the first phase-field LB model for incompressible multiphase flows. Nevertheless, this model was found to be limited to moderate density ratios.

Later, an incompressible multiphase LB model was proposed by Inamuro *et al.* [46] for multiphase flows at large density ratios. Inamuro *et al.*'s model also utilized two different particle distribution functions: one for the order parameter and the other for solving the velocity field without the pressure gradient. The model is then supplemented by a relationship between the velocity correction and the pressure gradient, which is determined by solving a Poisson equation at every collision-streaming step. Inamuro *et al.* [46] showed that their model recovers the following macroscopic equations:

$$\frac{\partial \phi}{\partial t} + \mathbf{v} \cdot \nabla \phi = \theta_M \nabla \cdot (\nabla \cdot \mathbf{P}),$$ (149)

$$\nabla \cdot \mathbf{v} = 0,$$ (150)

$$\frac{\partial \mathbf{v}}{\partial t} + \mathbf{v} \cdot \nabla \mathbf{v} = -\frac{1}{\rho} \nabla p + \frac{1}{\rho} \nabla \cdot \mathbf{\Pi} + \nabla \cdot \left[ \frac{k_g}{\rho} \left( \nabla \rho \nabla \rho - |\nabla \rho|^2 \mathbf{I} \right) \right],$$ (151)

where $\theta_M$ is the mobility coefficient, $\mathbf{P}$ is the thermodynamic pressure tensor, and $k_g$ is a constant determining the surface tension. Equation (149) is the recovered phase-field advection-diffusion equation for the order parameter $\phi$. Using the model, Inamuro *et al.* successfully simulated rising bubbles in a square duct at the density ratio $\rho_l / \rho_g = 1000$.

In 2005, Lee and Lin [47] devised a three-stage stabilized LB model for simulating incompressible multiphase flows at large density ratios. Similarly, two particle distribution functions were used in their model: the density distribution function $f_\alpha$ and the pressure distribution function $g_\alpha$, which obey the following discrete Boltzmann-BGK equations, respectively

$$\frac{\partial f_\alpha}{\partial t} + \mathbf{e}_\alpha \cdot \nabla f_\alpha = -\frac{1}{\tau} \left( f_\alpha - f_\alpha^{eq} \right) + \frac{\Gamma_\alpha(\mathbf{v})}{c_s^2} (\mathbf{e}_\alpha - \mathbf{v}) \cdot \left[ \nabla(\rho c_s^2) - \rho \nabla \left( \mu_\phi - k \nabla^2 \rho \right) \right],$$ (152)

$$\frac{\partial g_\alpha}{\partial t} + \mathbf{e}_\alpha \cdot \nabla g_\alpha = -\frac{1}{\tau} \left( g_\alpha - g_\alpha^{eq} \right) + \frac{1}{c_s^2} (\mathbf{e}_\alpha - \mathbf{v}) \cdot \mathbf{\Phi},$$ (153)

where $\mu_\phi$ is the chemical potential and $\mathbf{\Phi} = \Gamma_\alpha(\mathbf{v}) \mathbf{F}_s + \left[ \Gamma_\alpha(\mathbf{v}) - \Gamma_\alpha(0) \right] \nabla(\rho c_s^2)$. It was shown that the following macroscopic equations can be recovered from Eqs. (152) and (153) [47]:

$$\frac{\partial \rho}{\partial t} + \nabla \cdot (\rho \mathbf{v}) = \nabla \cdot \left[ \left( \mu / c_s^2 \right) \nabla \mu_\phi \right],$$ (154)

$$\frac{1}{\rho c_s^2} \frac{\partial p}{\partial t} + \nabla \cdot \mathbf{v} = 0,$$ (155)

$$\frac{\partial(\rho\mathbf{v})}{\partial t} + \nabla\cdot(\rho\mathbf{v}\mathbf{v}) = -\nabla p + \nabla\cdot\mathbf{\Pi} + \mathbf{F}_s. \tag{156}$$

Lee and Lin pointed out that $\mu/c_s^2$ in Eq. (154) acts as the mobility in the Cahn-Hilliard-like equation [47]; thus in this model the density $\rho$ serves as the order parameter. To stabilize the LB scheme at large density ratios, the second-order discretizations of Eqs. (152) and (153) were solved in three steps and the gradient terms were evaluated using different forms before and after the streaming step [47]. A second-order mixed difference scheme, which is a combination of a second-order central difference scheme and a second-order biased difference scheme, was introduced to evaluate some spatial derivatives, e.g., for the directional derivatives (of a variable $\varphi$), the second-order central and biased difference schemes are given by, respectively [47]

$$\delta_t \mathbf{e}_\alpha \cdot \nabla^{\mathrm{CD}} \varphi \big|_{(\mathbf{x},t)} = \frac{1}{2}\big[\varphi(\mathbf{x}+\mathbf{e}_\alpha\delta_t) - \varphi(\mathbf{x}-\mathbf{e}_\alpha\delta_t)\big], \tag{157}$$

$$\delta_t \mathbf{e}_\alpha \cdot \nabla^{\mathrm{BD}} \varphi \big|_{(\mathbf{x},t)} = \frac{1}{2}\big[-\varphi(\mathbf{x}+2\mathbf{e}_\alpha\delta_t) + 4\varphi(\mathbf{x}+\mathbf{e}_\alpha\delta_t) - 3\varphi(\mathbf{x})\big]. \tag{158}$$

The mixed difference scheme can be given by $\delta_t \mathbf{e}_\alpha \cdot \nabla^{\mathrm{MD}} \varphi = \big(\delta_t \mathbf{e}_\alpha \cdot \nabla^{\mathrm{CD}} \varphi + \delta_t \mathbf{e}_\alpha \cdot \nabla^{\mathrm{BD}} \varphi\big)/2$ [49].

Numerical results showed that Lee and Lin's model is capable of simulating multiphase flows at the density ratio $\rho_l/\rho_g = 1000$. Nevertheless, Chiappini *et al*. [311] analytically demonstrated that the different discretization of the streaming operator along the directions of molecular versus fluid motion is in principle non-conservative. Lou *et al*. [312] also found that the second-order mixed difference scheme, which plays a critical role in Lee and Lin's model for enhancing the numerical stability at large density ratios, is non-Galilean invariant and does not conserve the total mass of the system. Meanwhile, it was shown that [312] the second-order isotropic difference scheme (see the Appendix in Ref. [47] or Eqs. (73) and (74) in the present paper) can ensure that the total mass of the system is conserved, although the isotropic difference scheme is seemingly inefficient at large density ratios.

### 4.2.2 Improved models

In 2006, Zheng *et al*. [48] pointed out that the above three models cannot correctly recover the Cahn-Hilliard equation, which can be seen by comparing Eqs. (148), (149), and (154) with Eq. (132).

A new model was therefore developed in Ref. [48]. To recover the Cahn-Hilliard equation, Zheng *et al*.

[48] proposed a modified LB equation for the order parameter distribution function

$$f_\alpha\left(\mathbf{x}+\mathbf{e}_\alpha\delta_t,t+\delta_t\right)-f_\alpha\left(\mathbf{x},t\right)=-\frac{1}{\tau_\phi}\left(f_\alpha-f_\alpha^{eq}\right)+\left(1-q\right)\left[f_\alpha\left(\mathbf{x}+\mathbf{e}_\alpha\delta_t,t\right)-f_\alpha\left(\mathbf{x},t\right)\right],\quad(159)$$

where $q=1/\left(\tau_\phi+0.5\right)$. Furthermore, the particle distribution function for the average density

$n=\left(n_A+n_B\right)/2$, in which $n_A$ and $n_B$ are the densities of fluid A and fluid B respectively, satisfies

the following evolution equation:

$$g_\alpha\left(\mathbf{x}+\mathbf{e}_\alpha\delta_t,t+\delta_t\right)-g_\alpha\left(\mathbf{x},t\right)=-\frac{1}{\tau_n}\left(g_\alpha-g_\alpha^{eq}\right)+\delta_t F_\alpha\,,\qquad(160)$$

where the forcing term $F_\alpha$ is given by

$$F_\alpha=\left(1-\frac{1}{2\tau_n}\right)\omega_\alpha\left[\frac{\mathbf{e}_\alpha-\mathbf{v}}{c_s^2}+\frac{\left(\mathbf{e}_\alpha\cdot\mathbf{v}\right)\mathbf{e}_\alpha}{c_s^4}\right]\cdot\left(\mathbf{F}_s+\mathbf{F}_g\right).\qquad(161)$$

Zheng *et al*. showed [48] that their model recovers the following macroscopic equations:

$$\frac{\partial\phi}{\partial t}+\nabla\cdot\left(\phi\mathbf{v}\right)=\nabla\cdot\left(\theta_M\nabla\mu_\phi\right),\qquad(162)$$

$$\frac{\partial n}{\partial t}+\nabla\cdot\left(n\mathbf{v}\right)=0\,,\qquad(163)$$

$$\frac{\partial\left(n\mathbf{v}\right)}{\partial t}+\nabla\cdot\left(n\mathbf{v}\mathbf{v}\right)=-\nabla\left(p+\phi\mu_\phi\right)+\nabla\cdot\mathbf{\Pi}+\mathbf{F}_s+\mathbf{F}_g\,,\qquad(164)$$

where $\theta_M$ is the mobility coefficient.

Previously it has been mentioned that the phase-field theory is valid for both density-matched

binary fluids and two-phase fluids with a density contrast. For density-matched binary fluids, the

density $\rho$ in Eq. (134) is the average density. The first LB model for binary fluids was devised by

Swift *et al*. [41]. A couple of LB models for binary fluids can also be found in Xu *et al*.'s studies

[313-315]. Actually, Zheng *et al*.'s model is also an LB model for density-matched binary fluids since

the density in Eqs. (163) and (164) is the average density $n=\left(n_A+n_B\right)/2$. This point was demonstrated by Fakhari and Rahimian [305], who showed that Zheng *et al*.'s model gives the same results for the cases with an equal average density, e.g., for the case $n_A=1$ and $n_B=1000$ with the case $n_A=500$ and $n_B=501$, which can be seen in Fig. 10.

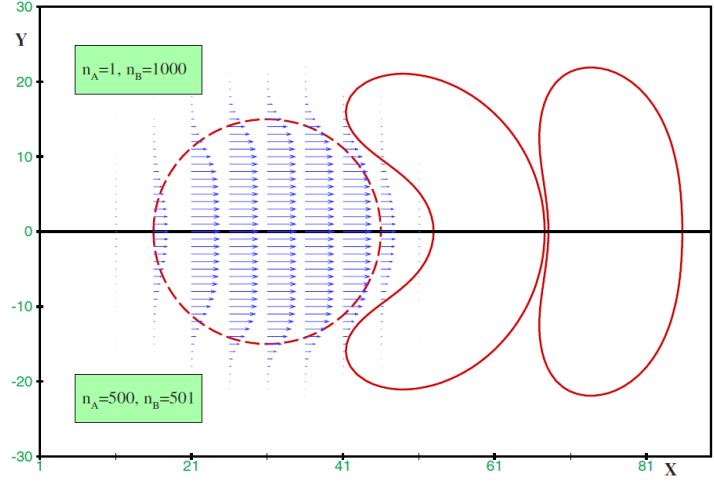

**Fig. 10**. Flow pattern around a moving droplet for cases with the same average density. Top: $n_A = 1$ and $n_B = 1000$; bottom: $n_A = 500$ and $n_B = 501$. Reprinted from Fakhari and Rahimian [305] with permission of the American Physical Society.

To simulate two-phase fluid flows with a density contrast, Fakhari and Rahimian [305] presented a modified phase-field LB model based on Zheng *et al*.'s work. A similar LB equation was used to mimic the Cahn-Hilliard equation. Meanwhile, following the line of He *et al*.'s work [45], Fakhari and Rahimian employed the pressure distribution function $g_\alpha$ to solve the incompressible Navier-Stokes equation, which obeys the following evolution equation [305]:

$$g_\alpha\left(\mathbf{x}+\mathbf{e}_\alpha\delta_t, t+\delta_t\right) - g_\alpha\left(\mathbf{x},t\right) = -\frac{1}{\tau_g}\left(g_\alpha - g_\alpha^{eq}\right) - \delta_t\left(1 - \frac{1}{2\tau_g}\right)\left(\mathbf{e}_\alpha - \mathbf{v}\right)\cdot\boldsymbol{\Phi}. \tag{165}$$

Here $\boldsymbol{\Phi} = \Gamma_\alpha\left(\mathbf{v}\right)\left(\mathbf{F}_s + \mathbf{F}_g\right) + \left[\Gamma_\alpha\left(\mathbf{v}\right) - \Gamma_\alpha\left(0\right)\right]\nabla\left(\rho c_s^2\right)$, in which $\Gamma_\alpha\left(\mathbf{v}\right)$ is still given by Eq. (145). The density $\rho$ is defined as $\rho = \rho_g + \left(\phi + \phi^*\right)\left(\rho_l - \rho_g\right)/\left(2\phi^*\right)$. Fakhari and Rahimian [305] showed that their model works well at the density ratio $\rho_l/\rho_g = 1000$ for static cases. For dynamic multiphase flows, the achievable highest liquid-gas density ratio decreases to around 10.

In 2010, Lee and Liu [49] proposed a modified phase-field LB model based on Lee and Lin's model [47]. In the modified model, the pressure distribution function $g_\alpha$ still obeys the discrete Boltzmann equation given by Eq. (153), but the density distribution function $f_\alpha$ is replaced by an order parameter distribution function $h_\alpha$, which satisfies the following equation so as to recover the Cahn-Hilliard equation [49]:

$$\frac{\partial h_\alpha}{\partial t} + \mathbf{e}_\alpha \cdot \nabla h_\alpha = -\frac{1}{\tau}\left(h_\alpha - h_\alpha^{eq}\right) + M\nabla^2 \mu_\phi \Gamma_\alpha\left(\mathbf{v}\right)$$
$$+ \left(\mathbf{e}_\alpha - \mathbf{v}\right) \cdot \left[\nabla\phi - \frac{\phi}{\rho c_s^2}\left(\nabla p + \phi \nabla \mu_\phi\right)\right]\Gamma_\alpha\left(\mathbf{v}\right). \qquad (166)$$

To enhance the numerical stability at large density ratios, the second-order mixed difference scheme was employed to evaluate some spatial derivatives; thus the weakness of the mixed difference scheme [312] was retained.

Wang *et al.* [67] have developed a multiphase LB flux solver for simulating incompressible multiphase flows at large density ratios and high Reynolds numbers. The LB flux solver was based on the following hydrodynamic equations:

$$\left(\frac{\partial p}{\partial t} + \mathbf{v} \cdot \nabla p\right) + \rho c_s^2 \nabla \cdot \mathbf{v} = 0, \qquad (167)$$

$$\frac{\partial\left(\rho \mathbf{v}\right)}{\partial t} + \nabla \cdot \left(\rho \mathbf{v} \mathbf{v}\right) = -\nabla p + \nabla \cdot \mathbf{\Pi} + \mathbf{F}_s. \qquad (168)$$

The above equations were then transformed to the governing equation of the density distribution function, which was solved by the finite-volume method and the flow variables were updated using a third-order Total-Variation-Diminishing (TVD) Runge-Kutta temporal scheme. The convective Cahn-Hilliard equation was solved with the finite-difference method and the convective term was discretized by a fifth-order Weighted Essentially Non-Oscillatory (WENO) scheme [67]. Owing to the use of these contact discontinuity capturing schemes, which are widely adopted in the simulations of compressible flows with shock waves [179, 316-318], the multiphase LB flux solver works well at the density ratio of 1000 [67]. Recently, Wang *et al.* [319] have improved their multiphase LB flux solver and developed a three-dimensional version using the D3Q15 lattice. Some numerical results can be found in Fig. 11.

Similarly, in a recent study carried out by Shao and Shu [68], the convective Cahn-Hilliard equation was solved with a third-order TVD Runge-Kutta scheme for the temporal discretization and an upwind WENO scheme for discretizing the convection term. The incompressible Navier-Stokes equations were still mimicked by an LB model [68]. Using such a hybrid phase-field LB model, Shao and Shu [68] successfully simulated droplet splashing at $\rho_l/\rho_g = 1000$ and $\text{Re} = 500$. Their work implies that the numerical stability of a phase-field LB model at large density ratios may mainly depend on the solver for the Cahn-Hilliard equation.

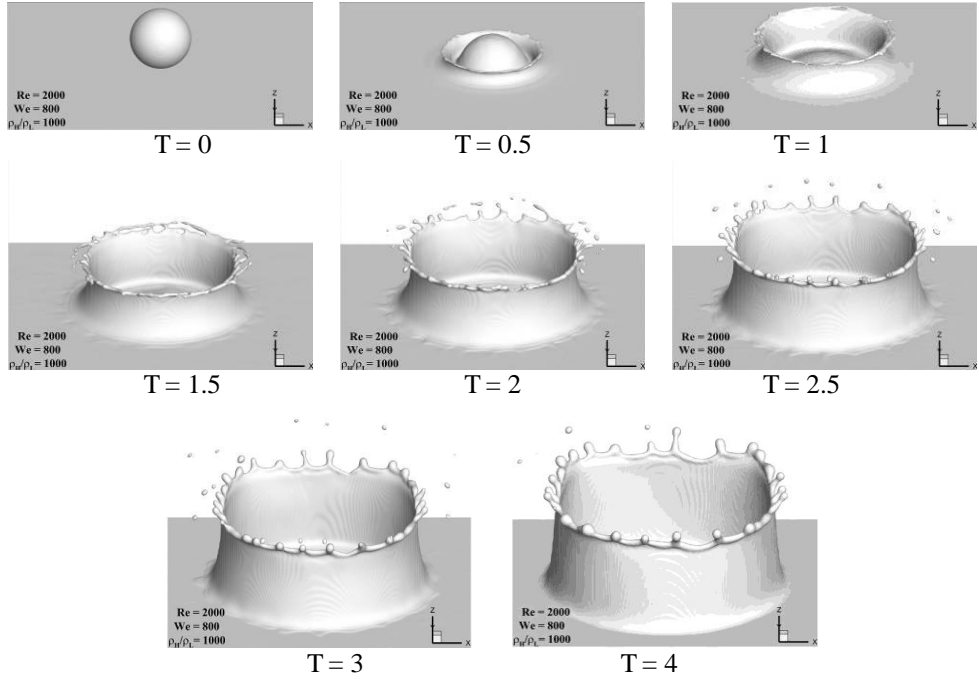

**Fig. 11**. Simulation of three-dimensional droplet splashing at $\rho_l / \rho_g = 1000$ and $\mathrm{Re} = 2000$ with a multiphase LB flux solver. Here "T" denotes non-dimensional time. Reprinted from Wang *et al.* [319] with permission of Elsevier.

### 4.3 Hydrodynamic inconsistency

In 2012, Li *et al.* [50] found that, in most of the previous incompressible multiphase LB models, the recovered momentum equation is inconsistent with the target momentum equation of incompressible multiphase flows, i.e., Eq. (134). Most of the above-mentioned models use the pressure distribution function $g_\alpha$ to simulate the Navier-Stokes equations and the equilibrium pressure distribution function $g_\alpha^{eq}$ usually satisfies

$$\frac{1}{c_s^2} \sum_\alpha g_\alpha^{eq} \mathbf{e}_\alpha = \rho \mathbf{v}\,, \quad \frac{1}{c_s^2} \sum_\alpha g_\alpha^{eq} \mathbf{e}_\alpha \mathbf{e}_\alpha = \rho \mathbf{v}\mathbf{v} + p\mathbf{I}\,. \tag{169}$$

Correspondingly, the recovered momentum equation takes the following form:

$$\frac{\partial(\rho \mathbf{v})}{\partial t} + \nabla \cdot (\rho \mathbf{v}\mathbf{v}) = \mathrm{R.\,H.\,S.}\,, \tag{170}$$

where "R. H. S." denotes the right-hand side of the recovered momentum equation, which is often the same as the right-hand side of Eq. (134). Nevertheless, the left-hand side of Eq. (170) deviates from that of Eq. (134) because [50]

$$\frac{\partial(\rho\mathbf{v})}{\partial t} + \boldsymbol{\nabla}\cdot(\rho\mathbf{v}\mathbf{v}) \equiv \rho\left(\frac{\partial\mathbf{v}}{\partial t} + \mathbf{v}\cdot\boldsymbol{\nabla}\mathbf{v}\right) + \mathbf{v}\left[\frac{\partial\rho}{\partial t} + \boldsymbol{\nabla}\cdot(\rho\mathbf{v})\right]. \tag{171}$$

When $\partial_t\rho + \boldsymbol{\nabla}\cdot(\rho\mathbf{v}) = 0$, the left-hand sides of Eqs. (170) and (134) will be identical. However, in the phase-field theory for incompressible multiphase flows, the density $\rho$ is an affine function of the order parameter $\phi$. As a result, the usual mass conservation $\partial_t\rho + \boldsymbol{\nabla}\cdot(\rho\boldsymbol{u}) = 0$ is no longer a consequence of the incompressibility condition ($\boldsymbol{\nabla}\cdot\boldsymbol{u} = 0$) since [50]

$$\partial_t\rho + \boldsymbol{\nabla}\cdot(\rho\boldsymbol{u}) = \frac{\mathrm{d}\rho}{\mathrm{d}\phi}\left(\partial_t\phi + \boldsymbol{u}\cdot\boldsymbol{\nabla}\phi\right) = \frac{\mathrm{d}\rho}{\mathrm{d}\phi}\boldsymbol{\nabla}\cdot\left(M\boldsymbol{\nabla}\mu_\phi\right). \tag{172}$$

It can be found that, when the phase-field theory with a Cahn-Hilliard-like interface capturing equation is applied, the incompressibility condition and the usual mass conversation cannot be simultaneously satisfied at the interface (the mixing layer) [65, 300, 304].

Hence the recovered momentum equation (170) should be rewritten as

$$\rho\left(\frac{\partial\mathbf{v}}{\partial t} + \mathbf{v}\cdot\boldsymbol{\nabla}\mathbf{v}\right) = -\mathbf{v}\left[\frac{\partial\rho}{\partial t} + \boldsymbol{\nabla}\cdot(\rho\mathbf{v})\right] + \text{R. H. S.} \tag{173}$$

Obviously, an additional term $\mathbf{F}_a = -\mathbf{v}\left[\partial_t\rho + \boldsymbol{\nabla}\cdot(\rho\mathbf{v})\right]$ is included in the recovered momentum equation as compared with Eq. (134). According to Eq. (172), $\mathbf{F}_a$ is zero in every single-phase region but nonzero at the interface. Li *et al.* [50] pointed out that $\mathbf{F}_a$ can be interpreted as an additional *interfacial* force and can be eliminated by adding $-\mathbf{F}_a$ into the forcing term of the LB model. It was numerically found that [50], with the increase of the velocity or the Reynolds number, the additional interfacial force gradually has an important influence on the evolution of the interface and affects the numerical accuracy, which can be seen in Fig. 12.

In some recent studies of the phase-field multiphase LB method [51, 320, 321], the elimination of the additional interfacial force was taken into account. In addition, Liu *et al.* [51] found that a similar error term exists in the recovered Cahn-Hilliard equation of Lee and Liu's model if the additional interfacial force is not eliminated (see Eq. (30)-Eq. (32) in Ref. [51]). Moreover, Liang *et al.* [320] have proposed an improved LB-MRT model to solve the Cahn-Hilliard equation. A source term was added to the LB-MRT equation to remove the error term in the recovered Cahn-Hilliard equation. Liang *et al.*

showed that their LB-MRT model is better than some previous models in terms of numerical accuracy and stability. Recently, Zheng *et al*. [322] have also constructed an improved LB model for the Cahn-Hilliard equation.

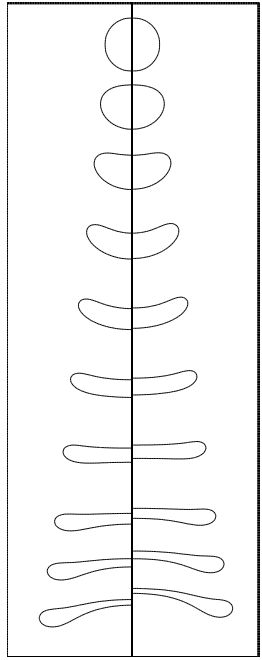

**Fig. 12**. Evolution of a falling droplet under gravity. Left: with the additional interfacial force. Right: the additional interfacial force is eliminated. Reprinted from Li *et al*. [50] with permission of the American Physical Society.

Finally, it should be noted that, although mass conservation $\partial_t \rho + \nabla \cdot (\rho \mathbf{v}) = 0$ is not satisfied at the interface when the incompressibility condition is imposed, the total mass of the whole system can be conserved as long as there is no interfacial diffusion flux across the boundaries ($M\nabla\mu_\phi$ denotes the interfacial diffusion flux), which can be found by integrating Eq. (172) over the occupied space and then using the divergence theorem [304]. Hence, in phase-field models, the no-flux boundary condition should be applied to the chemical potential $\mu_\phi$, i.e., $\mathbf{n} \cdot \nabla\mu_\phi = 0$, where $\mathbf{n}$ is the normal unit vector.

### 4.4 The contact angle treatment

In this section, the approaches used in the phase-field method for simulating contact angles are briefly reviewed.

### 4.4.1 The surface energy density

In the phase-field method, the first study about the implementation of contact angles might be attributed to Cahn [323], who assumed that the interaction between the solid surface and the fluid is sufficiently short-range so that it contributes a surface term to the total free energy of the system. With such an assumption, the free energy defined by Eq. (129) can be rewritten as [324, 325]

$$\mathcal{F}(\phi) = \int_V \left( \mathcal{E}(\phi) + \frac{1}{2}k\left|\nabla\phi\right|^2 \right) dV + \int_S \mathcal{H}(\phi_s)\,dS \ , \tag{174}$$

where $S$ is the solid surface bounding the volume $V$, $\phi_s$ is the value of $\phi$ on the solid surface, and the surface energy density $\mathcal{H}(\phi_s)$ represents the influence of the surface on the free energy.

To implement wetting boundaries in the phase-field method, the key issue is to find the relationship between the boundary condition of the order parameter $\phi$ and the value of the contact angle $\theta_W$. According to Ref. [324], the natural boundary condition of $\phi$ is given by

$$k\mathbf{n}\cdot\nabla\phi = \frac{d\mathcal{H}(\phi_s)}{d\phi_s}, \tag{175}$$

where $k$ is the constant in the free energy $\mathcal{F}(\phi)$ and $\mathbf{n}$ is the unit vector normal to the boundary. Many forms of $\mathcal{H}(\phi_s)$ have been proposed in the literature. The simplest $\mathcal{H}(\phi_s)$ is linearly proportional to $\phi_s$ [324, 326]: $\mathcal{H}(\phi_s) = -\lambda_c\phi_s$, where $\lambda_c$ is a constant. Then Eq. (175) becomes

$$k\mathbf{n}\cdot\nabla\phi = -\lambda_c \ . \tag{176}$$

From Eq. (176) it can be seen that the boundary condition of $\phi$ has been linked to $\lambda_c$. Meanwhile, the contact angle can be determined by the Young's law [324]

$$\cos\theta_W = \frac{\vartheta_{sg} - \vartheta_{sl}}{\vartheta}, \tag{177}$$

where $\vartheta$, $\vartheta_{sg}$, and $\vartheta_{sl}$ are the liquid-gas, solid-gas, and solid-liquid surface tensions, respectively. For the linear surface energy density $\mathcal{H}(\phi_s) = -\lambda_c\phi_s$, the surface tensions $\vartheta_{sg}$ and $\vartheta_{sl}$ can be related to $\lambda_c$ as follows (Details can be found in Refs. [324, 325]):

$$\vartheta_{sg} = -\lambda_c \frac{(\phi_l + \phi_g)}{2} + \frac{\vartheta}{2} - \frac{\vartheta}{2}(1-\Omega)^{3/2}, \tag{178}$$

$$\vartheta_{sl} = -\lambda_c \frac{(\phi_l + \phi_g)}{2} + \frac{\vartheta}{2} - \frac{\vartheta}{2}(1+\Omega)^{3/2}, \tag{179}$$

where $\Omega = 4\lambda_c \left/ \left(\phi_l - \phi_g\right)^2 \sqrt{2k\beta}\right.$ is the wetting potential.

Substituting Eqs. (178) and (179) into Eq. (177), one can find [324, 325, 327]

$$\cos\theta_w = \frac{(1+\Omega)^{3/2} - (1-\Omega)^{3/2}}{2}. \tag{180}$$

For a given contact angle, the value of $\lambda_c$ can be obtained from Eq. (180); thus the boundary condition of $\phi$ can be determined via Eq. (176) and the following formulations can be used [328]

$$\left.\frac{\partial\phi}{\partial z}\right|_{z=0} = -\frac{\lambda_c}{k}, \tag{181}$$

$$\left.\frac{\partial^2\phi}{\partial z^2}\right|_{z=0} \approx \frac{1}{2}\left(-3\left.\frac{\partial\phi}{\partial z}\right|_{z=0} + 4\left.\frac{\partial\phi}{\partial z}\right|_{z=1} - \left.\frac{\partial\phi}{\partial z}\right|_{z=2}\right), \tag{182}$$

where $z$ is the perpendicular direction to the wall. Briant $et\ al.$ [328] suggested that $\partial\phi/\partial z|_{z=1}$ can be calculated with a central finite-difference scheme. In addition, they found that the best choice for $\partial\phi/\partial z|_{z=2}$ is a left-handed finite-difference scheme, i.e., $\partial\phi/\partial z|_{z=2} \approx 0.5\left(3\phi|_{z=2} - 4\phi|_{z=1} + \phi|_{z=0}\right)$.

Moreover, Pooley $et\ al.$ [329] have investigated the performances of the BGK and MRT collisions models when the above approach is employed to implement the contact angles. It was found that, for the BGK collision model, strong spurious currents appear when the relaxation time $\tau \neq 1$. Pooley $et\ al.$ demonstrated that the spurious currents mainly arise from two effects: the long-range contribution to the equilibrium distribution function near the contact line and the bounce-back boundary treatment. They suggested using the MRT collision model for cases when the simulated two components/phases have different viscosities. Recently, Taghilou and Rahimian [330] successfully applied the above approach to simulate the penetration of a liquid droplet in a porous media with wetting boundaries. Some numerical results can be found in Fig. 13.

Besides the linear form of the surface energy density, some other forms can be found in Refs. [331-333]. For example, a quadratic form surface energy density and a cubic form surface energy density were proposed in Ref. [331], with the relationship between the boundary condition of $\phi$ and the contact angle being derived in a similar way. Numerical investigations on the performances of these different forms have been conducted by Liu and Lee [331] and Huang $et\ al.$ [333].

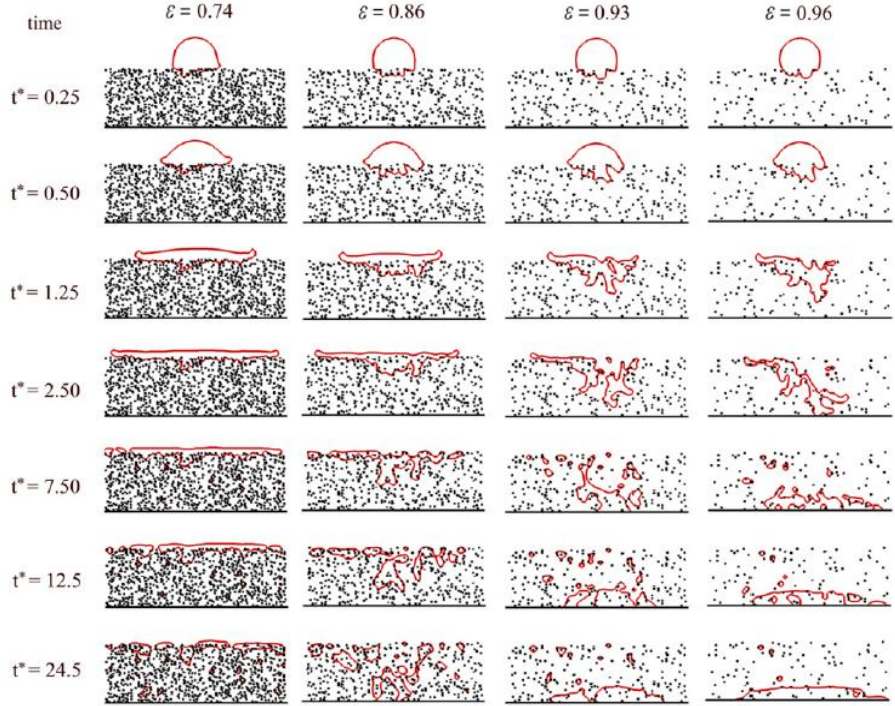

**Fig. 13**. Simulation of impact of droplet on the permeable surface at different porosities. The contact angle was implemented via Eqs. (181) and (182). Reprinted from Taghilou and Rahimian [330] with permission of Elsevier.

### 4.4.2 The geometric formulation

In 2007, Ding and Spelt [287] found that the approach based on the surface energy density cannot give a slope of the interface that is completely consistent with the prescribed value of the contact angle. The detailed explanations can be found in Section II.B in Ref. [287]. To solve this problem, Ding and Spelt proposed a geometric formulation to implement wetting boundaries in the phase-field method.

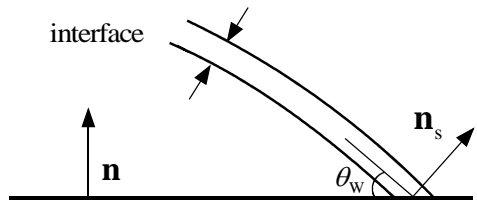

**Fig. 14**. The geometric information of a contact angle.

According to the geometric information in Fig. 14, the following equation was obtained [287]:

$$\tan\left(\frac{\pi}{2} - \theta_{\mathrm{w}}\right) = \frac{\mathbf{n}_s \cdot \mathbf{n}}{\left|\mathbf{n}_s - (\mathbf{n}_s \cdot \mathbf{n})\mathbf{n}\right|} = \frac{-\nabla\phi \cdot \mathbf{n}}{\left|\nabla\phi - (\nabla\phi \cdot \mathbf{n})\mathbf{n}\right|},$$

(183)

where $\mathbf{n}$ is the unit vector normal to the wall and $\mathbf{n}_s = -\nabla\phi/|\nabla\phi|$ is the normal unit vector.

Equation (183) is the so-called *geometric formulation* for computing the contact angle $\theta_w$. Its discretized form (using a central finite-difference scheme) is given by [287]

$$\phi_{i,0} = \phi_{i,2} + \tan\left(\frac{\pi}{2} - \theta_w\right)\left|\phi_{i+1,1} - \phi_{i-1,1}\right|.$$  (184)

Here the index $i$ represents the coordinate along the solid wall. Ding and Spelt [287] pointed out that the geometric formulation can provide a good approximation for the contact angle $\theta_w$ as long as enough grid points (usually 4–8) are used to resolve the interface and the diffuse interface is at equilibrium or near equilibrium at the solid substrate. It was found [287] that the geometric formulation gives more accurate results in comparison with the surface-energy-density-based approach. Nevertheless, it should also be noted that the geometric formulation requires more neighboring information; therefore its application to rough surfaces may need more careful treatment.

### 4.5 Thermal models for liquid-vapor phase change

In this section, the thermal phase-field LB models for simulating liquid-vapor phase change are reviewed. The existing thermal phase-field LB models can be classified into three types. The first-type thermal phase-field LB model was proposed by Dong *et al.* [106] based on Zheng *et al.*'s binary-fluid model, in which the phase change was defined by adding a source term to the Cahn-Hilliard equation:

$$\frac{\partial\phi}{\partial t} + \nabla\cdot(\phi\mathbf{v}) = \nabla\cdot\left(M\nabla\mu_\phi\right) - \frac{\rho_l^2}{\rho_g}\frac{\text{Ja}}{\text{Pe}}\nabla^2 T,$$  (185)

where $\text{Ja} = c_{p,l}\left(T_{sat} - T_\infty\right)/h_{fg}$ is the Jacob number and $\text{Pe} = UL/\chi_l$ is the Peclet number, in which $h_{lg}$ is the latent heat of vaporization, $U$ is the characteristic velocity, $L$ is the characteristic length, and $\chi_l$ is the thermal diffusivity of liquid phase. The temperature equation was simulated by a temperature distribution function $T_\alpha$, which obeys [106]

$$T_\alpha\left(\mathbf{x} + \mathbf{e}_\alpha\delta_t, t + \delta_t\right) - T_\alpha\left(\mathbf{x}, t\right) = -\frac{1}{\tau_T}\left(T_\alpha - T_\alpha^{eq}\right) + \omega_\alpha\frac{\rho_g}{\rho_l\left(\rho_l - \rho_g\right)}\frac{\phi'}{\text{Ja}},$$  (186)

where $\phi' = \Delta\phi/\Delta t$. The last term on the right-hand side of Eq. (186) is the latent heat term defined by Dong *et al.* [106]. The order parameter $\phi$ ranges from $-\phi^*$ to $\phi^*$, but the complete Eqs. (185) and

(186) are only applied in the region $\phi < 0$; when $\phi \geq 0$, the source terms are ignored. In other words, the phase change terms are considered in the vapor region only [106]. The model has been employed to simulate bubble growth on and departure from a superheated wall [106-108] and has been extended to three dimensions by Sun and Li [109]. Some numerical results can be found in Fig. 15.

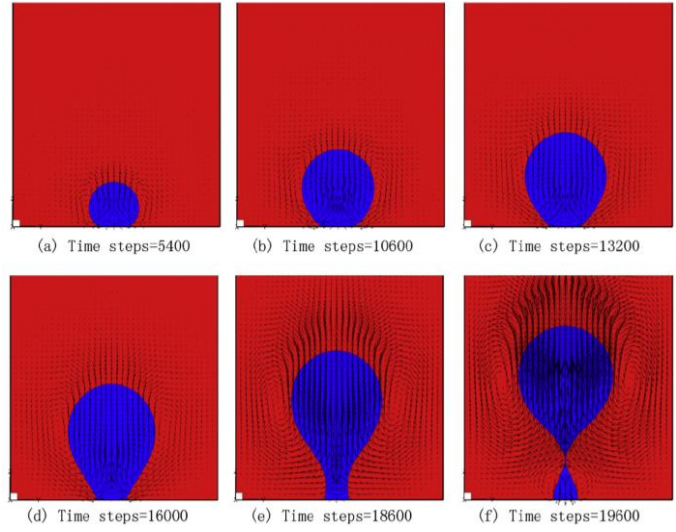

**Fig. 15**. Bubble growth on and departure from a superheated wall. Reprinted from Sun and Li [109] with permission of Elsevier.

The second-type thermal phase-field LB model was developed by Safari *et al.* [110, 111] based on Lee and Liu's isothermal phase-field LB model [49]. The liquid-vapor phase change was incorporated by redefining the divergence of the velocity field. From the continuity equations of local densities, the following velocity divergence was obtained [110]

$$\nabla \cdot \mathbf{v} = m' \left( \frac{1}{\rho_g} - \frac{1}{\rho_l} \right),$$ (187)

where $m'$ is the mass flux due to the liquid-vapor phase change. The above formulation can also be found in Refs. [334, 335]. When the density is defined as $\rho = \rho_g + \left( \phi - \phi_g \right) \left( \rho_l - \rho_g \right) / \left( \phi_l - \phi_g \right)$, the Cahn-Hilliard equation should be rewritten as

$$\frac{\partial \phi}{\partial t} + \nabla \cdot \left( \phi \mathbf{v} \right) = \nabla \cdot \left( M \nabla \mu_\phi \right) + \left( \frac{\phi_g}{\rho_g} - \frac{\phi_l}{\rho_l} \right) m'.$$ (188)

If we choose $\phi_l = 1$ and $\phi_g = -1$, the last term on the right-hand side of Eq. (188) will yield

$-\left(\rho_l + \rho_g\right)m'/\rho_l\rho_g$ [335]. In Safari *et al*.'s model, the mass flux $m'$ is defined as

$$m' = \frac{\lambda\nabla T \cdot \nabla\phi}{h_{lg}},\tag{189}$$

where $\lambda$ is the thermal conductivity and $h_{lg}$ is the latent heat of vaporization. In Ref. [110], the model was validated by simulating a one-dimensional Stefan problem and two-dimensional droplet evaporation. Recently, this model has been applied by Begmohammadi *et al*. to simulate bubble growth on and departure from a superheated wall [112]. Using Lee and Liu's mixed difference scheme, Begmohammadi *et al*. successfully simulated bubble growth at a large density ratio ($\rho_l/\rho_g = 1000$). With the mass flux defined by Eq. (189), an initial small bubble is often placed on the wall for simulating bubble growth because there will be no liquid-vapor phase change when the order parameter $\phi$ is uniform in the whole computational domain.

The third type of thermal phase-field LB model was devised by Tanaka *et al*. [113] on the basis of Inamuro *et al*.'s isothermal phase-field model [46]. Similar to the thermal pseudopotential LB models, Tanaka *et al*.'s model also employs the temperature $T$ in the equation of state $p_0 = \phi T/(1-b\phi) - a\phi^2$ [113] to drive the phase change from the liquid state to vapor state. Hence in Tanaka *et al*.'s model the rate of the liquid-vapor phase change is also a computational output, while in the above two types of thermal phase-field LB models, the phase change rate is a prerequisite input. Owing to this feature, the bubble growth and departure can be simulated by Tanaka *et al*.'s model without placing an initial bubble on the superheated wall [113, 114].

Finally, it is noticed that in some of the above thermal phase-field LB models the target macroscopic temperature equation takes the following form:

$$\partial_t T + \mathbf{v}\cdot\nabla T = \nabla\cdot\left(\chi\nabla T\right) + S_R,\tag{190}$$

where $\chi$ is the thermal diffusivity and $S_R$ is a source term. Different models use different source terms: for instance, $S_R$ is given by $S_R = -m'h_{lg}/\left(\rho c_P\right)$ in Ref. [111]. Although in the phase-field theory there has not been a unified definition for the source term $S_R$, it is believed the first term on the right-hand side of Eq. (190) should at least be

$$\frac{1}{\rho c_v}\nabla\cdot\left(\lambda\nabla T\right) \quad\text{or}\quad \frac{1}{\rho c_p}\nabla\cdot\left(\lambda\nabla T\right), \tag{191}$$

which can be observed in the studies of thermal phase-field algorithms with conventional numerical methods: see Eq. (10d) in Ref. [334] and Eq. (3) in Ref. [335]. Since the density usually varies significantly across the liquid-vapor interface, it is inappropriate to replace Eq. (191) with $\nabla\cdot\left(\chi\nabla T\right)$, which can also be seen from Eq. (121).

## 5. Some applications of multiphase and thermal LB methods

In the above sections, we have introduced the fundamentals of the LB method and reviewed the advances in the theories of two popular multiphase LB methods. In this section we shall review some applications of these two multiphase LB methods as well as the application of thermal LB approaches in energy storage with phase change materials.

### 5.1 Fuel cells and batteries

A fuel cell, which consists of an anode, a cathode, a membrane, and an electrolyte, is an energy conversion device that converts the chemical energy stored in fuels and oxidants into electricity through electrochemical reactions [336, 337]. Fuel cells can be classified into different types according to the electrolyte in use, among which the proton exchange membrane (PEM) fuel cell, also known as the polymer electrolyte membrane fuel cell, is widely regarded as one of the most promising types of fuel cells [336, 338-340] due to its advantages such as low operating temperature, high power density, and zero/low emission. There are many phenomena involved in a PEM fuel cell, e.g., heat transfer, species and charge transport, multiphase flows, and electrochemical reactions. In particular, the water-gas two-phase flow in the gas diffusion layer (GDL) has been identified as an important phenomenon because it affects the amount of water in the catalyst layer and membrane, and plays a crucial role in the water management of PEM fuel cells.

In the literature, the first attempt of applying a multiphase LB model to simulate the water-gas transport in the GDL of a PEM fuel cell was attributed to Niu *et al*. [73] and Sinha *et al*. [74] in 2007.

By employing a three-dimensional MRT version of Zheng *et al.*'s binary-fluid phase-field LB model, Niu *et al.* [73] investigated the effects of the pressure drop, the wettability, and the viscosity ratio on the liquid-gas transport process in the GDL of a PEM fuel cell. They observed that an increase in the viscosity ratio results in a decreased relative permeability of the liquid phase. Meanwhile, Sinha *et al.* [74] utilized a multi-component pseudopotential LB model to simulate the liquid water distribution in an initially air-saturated carbon paper GDL. The influence of the capillary pressure was investigated and some results can be found in Fig. 16, which show that, at a low capillary pressure, the invading front overcomes the barrier pressure only at some preferential locations depending upon the pore size. With the increase of the capillary pressure, it can be seen that the liquid water gradually penetrates into the domain occupied by air. Moreover, the difference of the liquid water distribution between a hydrophobic GDL and a mixed wettability GDL has also been studied [74].

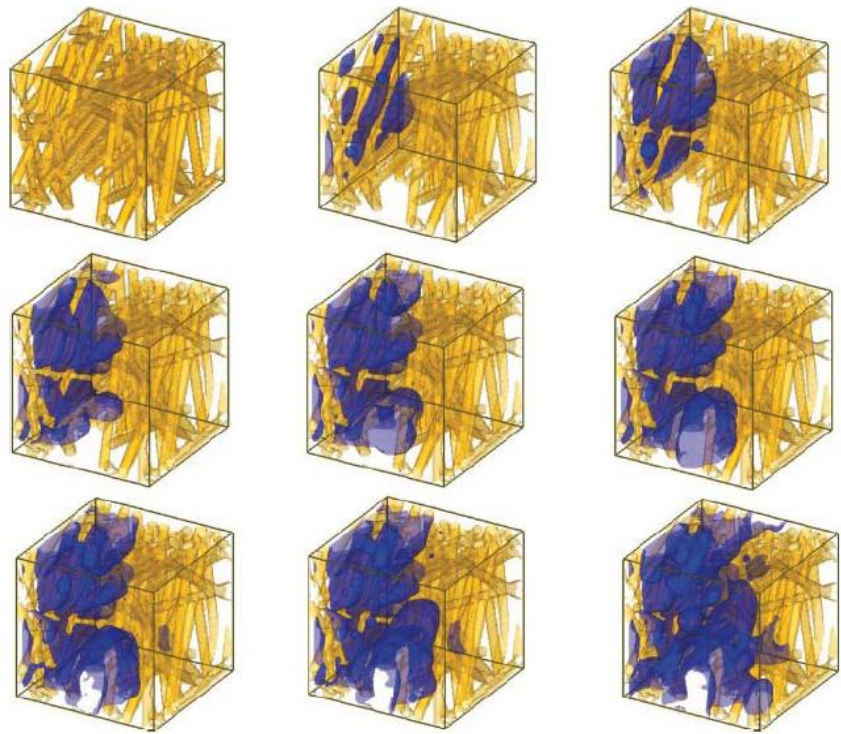

**Fig. 16**. Numerical simulation of the liquid water distribution in a reconstructed non-woven GDL microstructure with increasing capillary pressure. The pseudopotential LB method was used. Reprinted from Sinha *et al.* [74] with permission of the Royal Society of Chemistry.

Later, Koido *et al.* [75] employed a combination of the standard single-phase LB method and the pseudopotential LB method to predict the relative permeabilities of liquid and gas phases versus the saturation of the GDL of a PEM fuel cell. The shape of the curve of the experimentally measured gas-phase relative permeability was correctly captured by the numerical prediction. The effect of wettability on the liquid water distribution in the GDL was also investigated. It was found that the liquid water tends to be present in large pores as spheres in a hydrophobic GDL and in small pores as thin films in a hydrophilic GDL [75]. In the meantime, by adopting the pseudopotential LB method, Park and Li [76] carried out a numerical study on the two-phase flow through a complicated porous medium, which represents a carbon paper GDL.

In 2009, using Inamuro *et al.*'s phase-field LB model, Tabe *et al.* [77] studied the dynamic behavior of condensed water and gas flow in a PEM fuel cell with a simplified GDL. Their results showed that the wettability of the flow channel has a strong effect on the two-phase flow in PEM fuel cells. Tabe *et al.* suggested that controlling the wettability of the porous separator can be used to produce an efficient distribution of the water and gas flow. At the same time, Mukherjee *et al.* [78] employed the pseudopotential LB method coupled with the stochastic microstructure reconstruction method to investigate the influences of the pore structure and the surface wettability on the liquid water transport, and the flooding dynamics in the catalyst layer and the GDL of a PEM fuel cell. In addition, Hao and Cheng [79] used a binary-fluid phase-field LB model to simulate the formation and removal of water droplets in the micro-gas flow channel of a PEM fuel cell. It was found that increasing the gas flow velocity and enhancing the GDL hydrophobicity can promote the formation of small droplets and thus reduce the time of water droplets adhering to the GDL surface near the emergence pore. Moreover, Hao and Cheng [80, 81] have also studied the effect of wettability on the water transport dynamics in a carbon paper GDL. It was shown that, for high hydrophobicity, the water transport in the GDL falls into the regime of capillary fingering, whereas for neutral wettability, the water transport exhibits stable displacement characteristics even for a capillary force dominated flow.

Zhou and Wu [82] have applied the pseudopotential LB method to study the liquid water transport mechanism in the GDLs of PEM fuel cells. It was found that the distribution of the fibers and the spatial mixed-wettability play an important role in the liquid water transport. Chen *et al*. [83] investigated the pore-scale flow and mass transport in the GDL of an interdigitated PEM fuel cell using the pseudopotential LB method. The GDL was reconstructed with the stochastic microstructure reconstruction method. They showed that the pore-scale behavior of the liquid water in the GDL can be classified as slow creeping in the regions with slow air flow and quick moving in the regions with fast air flow. Ben Salah *et al*. [84] utilized Inamuro *et al*.'s phase-field LB model to study the droplet behavior in the gas flow channel of a PEM fuel cell. They found that hydrophilic channel walls result in better gas transport characteristics than hydrophobic walls. Using the pseudopotential LB method, Han *et al*. [85] have also studied the liquid droplet dynamics in the gas flow channel of a PEM fuel cell. The development and interaction of two liquid droplets were simulated. It was found that increasing the gas stream velocity and the liquid pore distance can prevent the interaction of liquid droplets and enhance the removal of droplets. In addition, Han and Meng [86] have simulated the liquid water transport in two different types of serpentine gas channels of PEM fuel cells. They found that a smooth U-shaped turning region is beneficial to the liquid water removal in a serpentine gas channel.

Moreover, Han and Meng [87] have applied the pseudopotential LB method to investigate the interfacial phenomena of liquid water transport in the porous diffusion layers of PEM fuel cells. They showed that the liquid water moves preferentially through the regions with large perforated pores. In addition, Gao *et al*. [88] have simulated the water and gas flow in porous GDLs reconstructed from micro-tomography and found that, with an increase in hydrophobicity, the liquid water transport in GDLs changes from piston flow to channel flow. Recently, using the pseudopotential LB method, Molaeimanesh and Akbari [89] studied the two-phase flow through the GDL of a PEM fuel cell with the non-homogeneous and anisotropic transport properties being considered. Furthermore, they have studied the pore-scale flows in three reconstructed GDLs with different anisotropic parameters using a three-dimensional pseudopotential LB model [90]. Some results can be found in Fig. 17. In case **a** the carbon fibers are mostly oriented normal to the catalyst layer, while in case **c** the carbon fibers are mostly oriented parallel to the catalyst layer. The results showed that, when the carbon fibers are more likely oriented normal to the catalyst layer, the distributions of the oxygen and water vapor become more disturbed. Moreover, Kim *et al*. [91] have investigated the effects of micro-porous layers (MPLs) in PEM fuel cells using the pseudopotential LB method. It was found that the liquid water inside the GDL is reduced as the thickness of the MPLs increases.

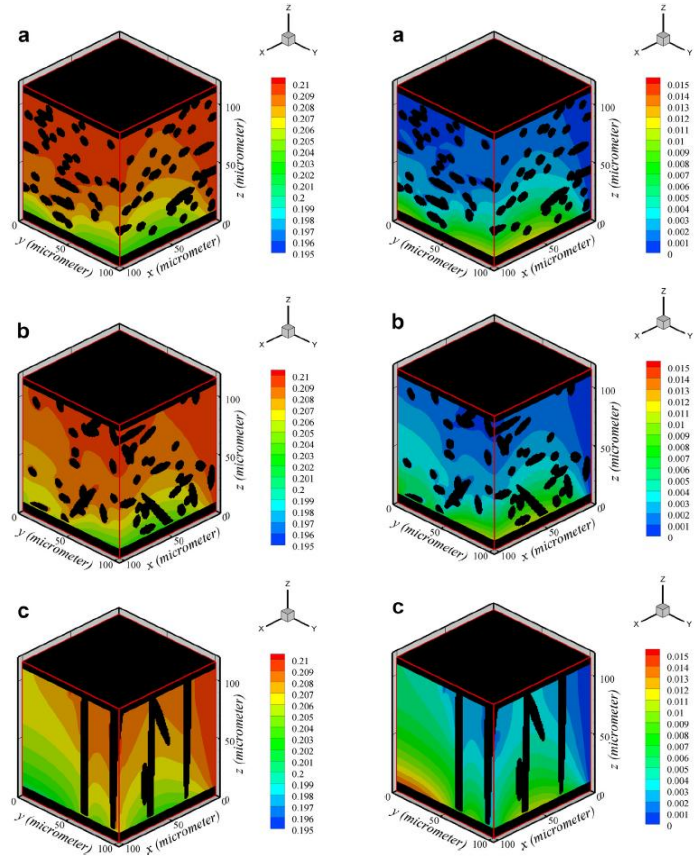

**Fig. 17**. Oxygen mole fraction distribution (left) and water vapor mole fraction distribution (right) in three different reconstructed GDLs. Reprinted from Molaeimanesh and Akbari [90] with permission of Elsevier.

Recently, the LB method has also been applied to batteries [92, 93, 341]. The first study was conducted by Qiu *et al*. [341]. In their study, the flow of electrolyte through the pore space in a vanadium redox flow battery [342] was modeled by the standard single-phase LB method with the species and charge transport being solved using the finite-volume method. Recently, Lee *et al*. [92] successfully applied the pseudopotential multiphase LB method to simulate the dynamic behavior of a liquid electrolyte in the porous electrode of a lithium-ion battery, which is a member of rechargeable batteries family. In lithium-ion batteries, the lithium ions move from the negative electrode to the positive electrode during the discharge process and move back when charging. Lee *et al*. found that the wettability in a porous electrode was strongly affected by the two-phase liquid electrolyte and gas flow, and enhanced wettability can be achieved through controlling the material properties. Furthermore, Lee *et al*. [93] have studied the influence of the electrode compression ratio on the wettability of a lithium-ion battery. Some results can be found in Fig. 18: it can be seen that a higher compression ratio leads to a reduced liquid electrolyte distribution.

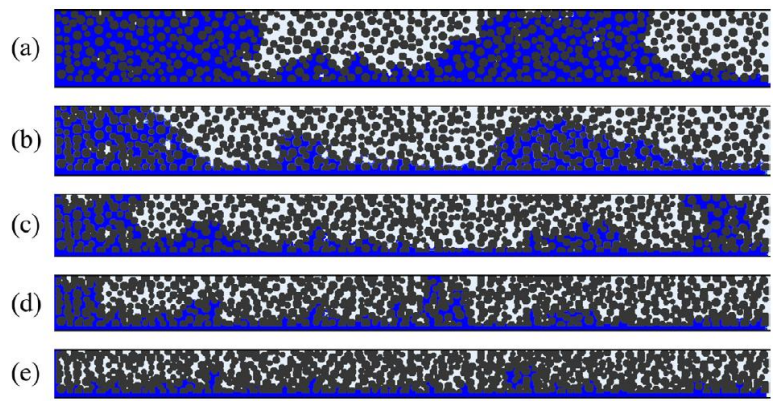

**Fig. 18**. The liquid electrolyte in the cathode with the compression ratios being (a) 0%, (b) 10%, (c) 20%, (d) 30%, and (e) 40%. Reprinted from Lee *et al*. [93] with permission of Elsevier.

### 5.2 Droplet collisions

Understanding droplet collisions is of crucial importance in many industrial processes such as inkjet printing, spray cooling, and spray combustion [343]. Here droplet collisions include droplet-droplet collisions and droplet-wall collisions. The first study of droplet collisions in the LB community may be attributed to Schelkle and Frohn [344], who applied the pseudopotential LB method to simulate binary collisions between equal-size droplets. Owing to the limitations of the original pseudopotential LB model, the liquid-gas density ratio, the Reynolds number, and the Weber number in Schelkle and Frohn's study are relatively low.

In 2004, Inamuro *et al*. [46, 345] studied binary droplet collisions at a high Reynolds number ($\mathrm{Re} = 2000$) with their phase-field LB model. The liquid-gas density ratio is 50. Three types of collisions were reproduced, i.e., coalescence collision, reflexive separation collision, and stretching separation collision. The results of a stretching separation collision can be found in Fig. 19. The impact parameter is defined as $B = X/(R_1 + R_2)$, where $R_1$ and $R_2$ are the initial radii of the two droplets, respectively, and $X$ is the distance between the centers of the droplets, perpendicular to their direction of motion. It can be seen that at $t^* = 1.38$ only a portion of the droplets contacts directly due to the high impact parameter. The remaining portions of the droplets still move in the direction of their initial velocities and the collision finally leads to two major droplets with a small satellite droplet [345].

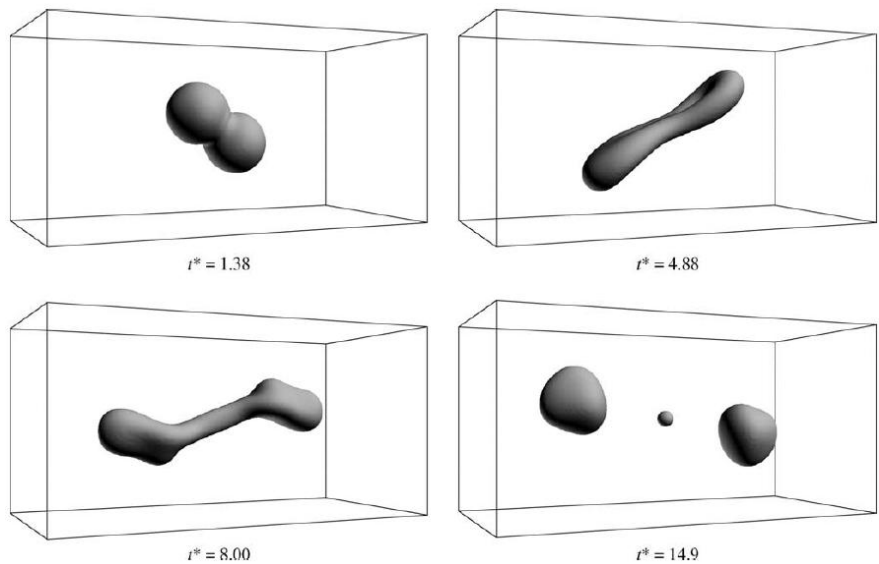

**Fig. 19**. A stretching separation collision between off-center binary droplets. The Weber number is 79.5, the Reynolds number is 2000, and the impact parameter is 0.5. Reprinted from Inamuro *et al*. [345] with permission of Elsevier.

Later, Premnath and Abraham [346] studied binary droplet collisions using both axisymmetric and three-dimensional MRT versions of He *et al*.'s multiphase LB model. For head-on collisions, it was shown that the droplet coalescence occurs at low Weber numbers and the coalesced droplet entraps a micro-bubble. When the Weber number increases, the colliding droplets separate rather than coalescing permanently. Furthermore, the effects of the droplet size ratio were also studied in Premnath and Abraham's work [346]. Similarly, Sakakibara and Inamuro [347] have investigated binary collisions of two unequal-size droplets with a three-dimensional version of Inamuro *et al*.'s phase-field LB model.

They showed that the available theoretical prediction of the boundary between the coalescence collision and the stretching separation collision fails in the case of a low diameter ratio ( $0.25$ ). Furthermore, Mukherjee and Abraham [348] have simulated collisions between a droplet and a flat solid wall with an axisymmetric version of He *et al.*'s model. They reported two outcomes of droplet-wall collisions: deposition and rebound, and found that the transition between deposition and rebound is influenced by the Weber number, the Ohnesorge number, and the advancing and receding static contact angles.

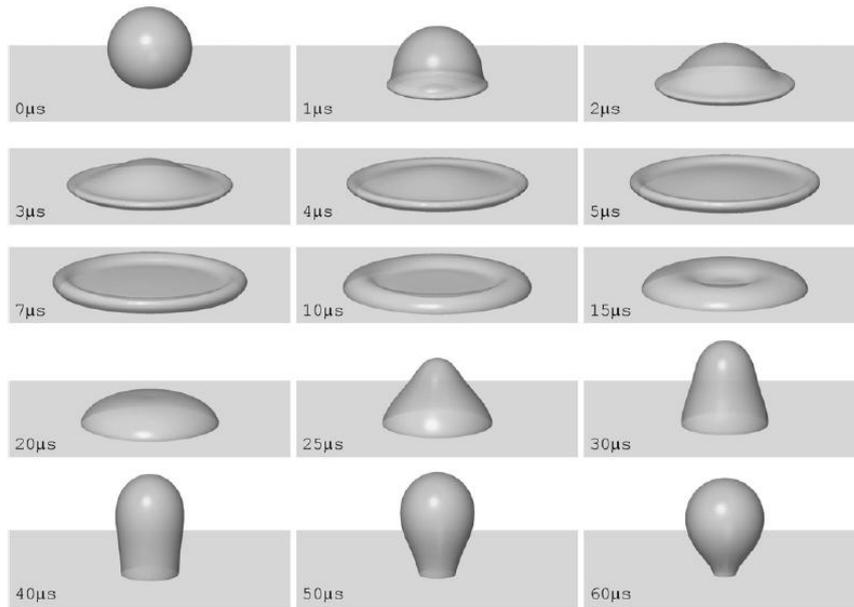

**Fig. 20**. Simulation of micro-scale droplet-wall collisions. The initial droplet diameter is approximately $50 \, \mu m$ and the static contact angle is $\theta = 107^{\circ}$ . Reprinted from Lee and Liu [49] with permission of Elsevier.

In 2010, Gupta and Kumar [349] employed a three-dimensional pseudopotential LB model to simulate droplet collisions on a dry solid wall. The effects of the impact Reynolds number, the Weber number, and the Ohnesorge number were analyzed with the density ratio around $10$ . It was found that the maximum spreading factor obeys the power law $0.5 \, \mathrm{Re}^{0.25}$ . In the meantime, Lee and Liu [49] investigated micro-scale droplet-wall collisions at a large density ratio ( $\rho_l / \rho_g \approx 842$ ) using their phase-field LB model. Some results can be found in Fig. 20, from which the spreading and retraction processes can be clearly observed. In addition, it can be seen that the droplet rebounds in the final stage but fails to lift off from the solid surface. Lee and Liu [49] found that the influence of the Weber number emerges in the later stage of spreading. Lycett-Brown *et al*. [66] have studied binary droplet collisions using a multi-speed pseudopotential LB model. The simulations were conducted at a large density ratio but small Reynolds and Weber numbers.

Furthermore, Huang *et al*. [350] have investigated the phenomenon of bubble entrapment during the collision of a droplet on a solid wall with a phase-field LB model. Four types of entrapment were observed under different Weber numbers, Reynolds numbers, and surface wettability. They found that large Ohnesorge numbers ( $Oh = \sqrt{We}/Re$ ) can prevent the phenomenon of bubble entrapment from occurring. In addition, using Inamuro *et al*.'s phase-field LB model, Tanaka *et al*. [351] have simulated the collision of a falling droplet with a stationary droplet on a solid surface. The influence of the Weber number on the dynamic behavior of the droplets and the mixing process of the collision were studied. Shen *et al*. [352] have investigated droplet collisions on the surface of a circular pipe with the pseudopotential LB method. They found that the droplets drip down on hydrophilic surfaces and splash when the impact velocity increases, while on hydrophobic surfaces the droplets splash even at a small impact velocity.

Zhang *et al*. [70] have also simulated droplet collisions on a circular pipe. A high density ratio ( $\rho_l/\rho_g = 580$ ) was achieved using Li *et al*.'s improved MRT forcing scheme for the pseudopotential LB models. According to the dynamic behavior of the droplets, three temporal states were observed: the initial droplet deformation state, the inertia dominated state, and the viscosity dominated state. The effects of the Reynolds number, the Weber number, and the Galileo number on the dynamic behavior were analyzed. Later, Zhang *et al*. [71, 353] studied three-dimensional droplet collisions on flat and spherical surfaces using a three-dimensional pseudopotential LB model. Recently, Lycett-Brown *et al*. [354] investigated three-dimensional binary droplet collisions with a cascaded pseudopotential LB model. Head-on and off-center collisions were mimicked under various Weber and Reynolds numbers. Moreover, Lycett-Brown *et al*. [354] have studied the effects of the droplet size ratio and found that the numerical errors (compared with the theoretical analysis) decrease as the droplet size ratio increases.

Sun *et al*. [355] have investigated head-on collisions of equal-size droplets and unequal-size droplets using an axisymmetric phase-field LB model. They found that, in the axial deformation cases, "coalescence" occurs at a low Weber number and "reflexive separation with satellite droplets" appears at a high Weber number. Moreover, by incorporating Li *et al*.'s axisymmetric LB-MRT scheme [143] into Lee *et al*.'s phase-field LB model, Sun *et al*. [72] successfully simulated head-on collisions of unequal-size droplets at a large density ratio ($\rho_l / \rho_g = 830$) and a low viscosity. It was shown that, in the cases of low Ohnesorge numbers, the smaller droplet tends to penetrate deeply into the larger one. With the increase of the Ohnesorge number, the smaller droplet spreads on the surface of the larger one. Recently, Zhou *et al*. [356] studied collisions between multiple droplets and a solid wall with a phase-field LB model. Some results of two droplets can be found in Fig. 21. It can be seen that in the early stage the two droplets behave independently of each other. Later, during the spreading process the two droplets contact each other. Subsequently, droplet coalescence occurs and a large droplet is formed. The collisions between an array of droplets and a solid wall has also been studied in Zhou *et al*.'s work.

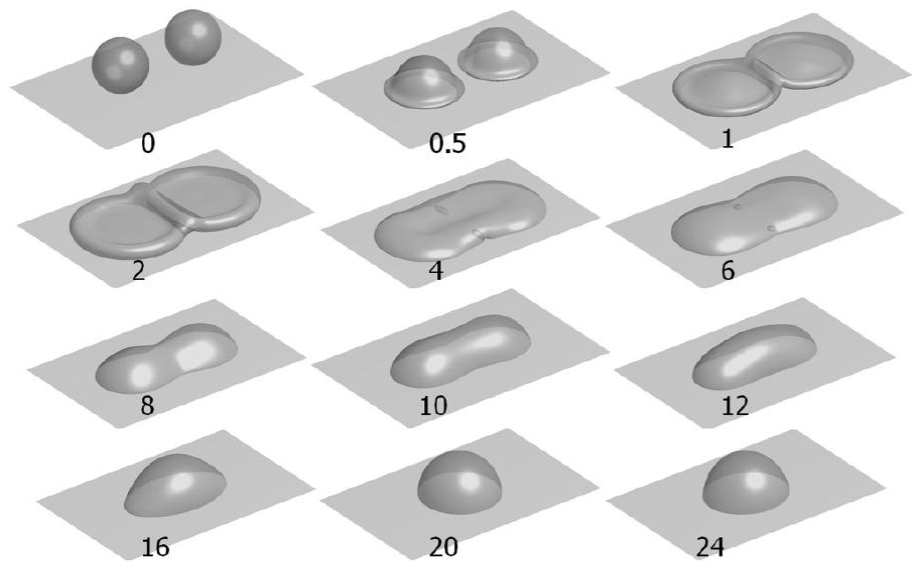

**Fig. 21**. Simulation of the collision between two droplets and a solid wall. The Weber number is $We = 100$ and the Ohnesorge number is $Oh = 0.04$. Reprinted from Zhou *et al*. [356] with permission of the American Physical Society.

*5.3 Boiling and evaporation*

Boiling heat transfer is used in a wide field of applications: from cooking activities in everyday life to various energy conversion and heat exchange systems as well as cooling of high-energy-density power/electronic devices [357, 358], such as boiling water reactors in nuclear power plants. While boiling phenomena can be encountered in daily life, the boiling processes are extremely complex and elusive because many physical components are involved and interrelated, such as the nucleation, growth, departure, and coalescence of vapor bubbles, the transport of latent heat, and the instability of liquid-vapor interfaces [290].

Boiling of stationary or non-flowing fluid is known as pool boiling with three boiling stages: nucleate boiling, transition boiling, and film boiling [359]. Nucleate boiling, which is a very efficient mode of heat transfer, occurs when the temperature of the heating surface is higher than the saturated fluid temperature by a certain amount but the heat flux is below the critical heat flux. Film boiling, which is of great interest to certain applications such as quenching of steel and spray cooling of very hot surfaces, appears when the surface temperature is increased above the so-called Leidenfrost temperature. In film boiling, the heating surface is completely covered with a continuous vapor film. Transition boiling is an unstable boiling mode between nucleate boiling and film boiling [359].

In spite of the fact that boiling heat transfer has been intensively studied in the past years, many aspects of boiling are still not well understood. For example, the physical mechanism causing the critical heat flux has not been satisfactorily addressed, because the correlations developed by experiments rely heavily on empirical parameters that are only valid in a narrow parameter range and numerical simulations based on traditional numerical methods mostly involve many assumptions and empirical correlations [360]. In Sections 3.6 and 4.5, we have introduced the thermal models for simulating liquid-vapor phase change within the frameworks of the pseudopotential and the phase-field multiphase LB methods. In this section, some applications of these models are reviewed. In particular, we put our focus on boiling and evaporation. The simulations of bubble growth and departure using point heating will not be discussed here.

The first numerical simulation of boiling using the LB method was reported by Zhang and Chen [94]. By considering a standard Rayleigh-Bénard setup, in which the upper and lower solid walls obey the no-slip boundary condition and the horizontal boundary condition is periodic, they successfully reproduced a nucleate boiling process, which can be seen in Fig. 22. To enhance bubble formation, small temperature fluctuations were added to the equation of state in the first grid point layer near the bottom solid wall. From the figure two nucleation sites can be clearly observed. In addition, the bubble rising and coalescence phenomena can also be seen in Fig. 22. Zhang and Chen [94] have analyzed the corresponding temperature distribution and found that the temperature is relatively low in the vapor-phase domains near the liquid-vapor interfaces.

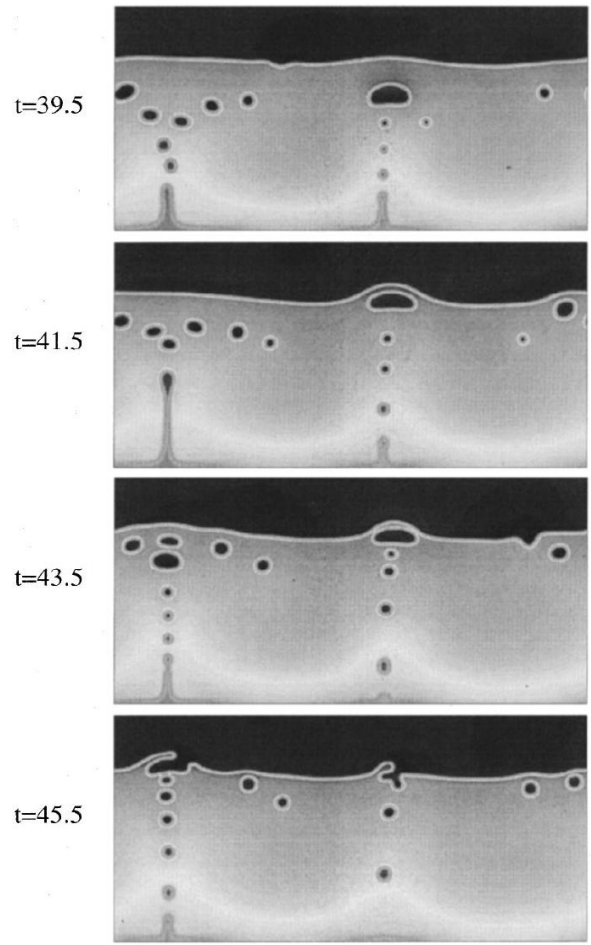

**Fig. 22**. Snapshots of a nucleate boiling process. Light gray and dark gray represent the liquid and vapor phases, respectively. Reprinted from Zhang and Chen [94] with permission of the American Physical Society.

In 2011, Márkus and Házi [96] simulated a heterogeneous boiling process using their thermal pseudopotential LB model. A uniform heat flux was applied at the bottom of a plate and a cavity was placed at the center of the plate with the heat conduction in the heated plate being taken into account. The bubble formation from the cavity was well simulated. Later, Márkus and Házi [97] investigated the influences of the heated plate configuration. Three different configurations were considered and a film boiling process was reproduced in the configuration when the heated plate is represented by a constant heat flux boundary. The results can be found in Fig. 23, which shows the variation of the surface temperature against the heat flux. From the figure a thin vapor film can be observed in the case with a high heat flux. However, an important feature of boiling heat transfer was not captured in Márkus and Házi's study. It is well-known that, in a surface-heat-flux-controlled boiling system, the boiling process will instantly enter into the film boiling regime (see Fig. 25(b), from point "C" to point "E" along the straight line) when the surface heat flux reaches the critical heat flux, which will result in a sharp and large increase in the surface temperature. Nevertheless, there is no such feature in Fig. 23.

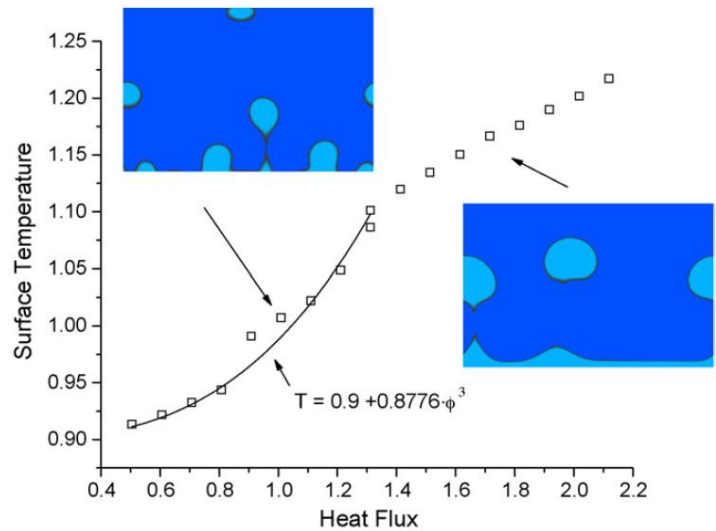

**Fig. 23**. Variation of the surface temperature against the heat flux. The heated plate is represented by a constant heat flux boundary. Reprinted from Márkus and Házi [97] with permission of Elsevier.

In 2012, by conducting a three-dimensional simulation of nucleate boiling, Biferale *et al*. [98] demonstrated that the pseudopotential LB model with a non-ideal thermodynamic pressure tensor can lead to a consistent definition of latent heat with the Clausius-Clapeyron relation being satisfied. Using

Tanaka *et al*.'s thermal phase-field LB model, Sattari *et al*. [114] have also simulated a boiling process (see Fig. 11 in the reference). They found that the boiling phenomenon begins with generating bubbles at the center of a surface and then starts to spread along the surface. Moreover, Gong and Cheng [103] have investigated the effects of mixed surface wettability on nucleate boiling heat transfer using their model proposed in Ref. [99]. The corresponding target temperature equation is similar to Eq. (190), namely the first term on the right-hand side of Eq. (121) was replaced with $\nabla \cdot \left( \chi \nabla T \right)$. The mixed wettability was obtained by placing hydrophobic spots on smooth hydrophilic surfaces. The influences of the size of the hydrophobic spots and the pitch distance between the hydrophobic spots were analyzed. Gong and Cheng [103] showed that mixed wettability surfaces can promote bubble nucleation, enhance boiling heat transfer, and reduce the nucleation time as compared with uniform hydrophilic surfaces.

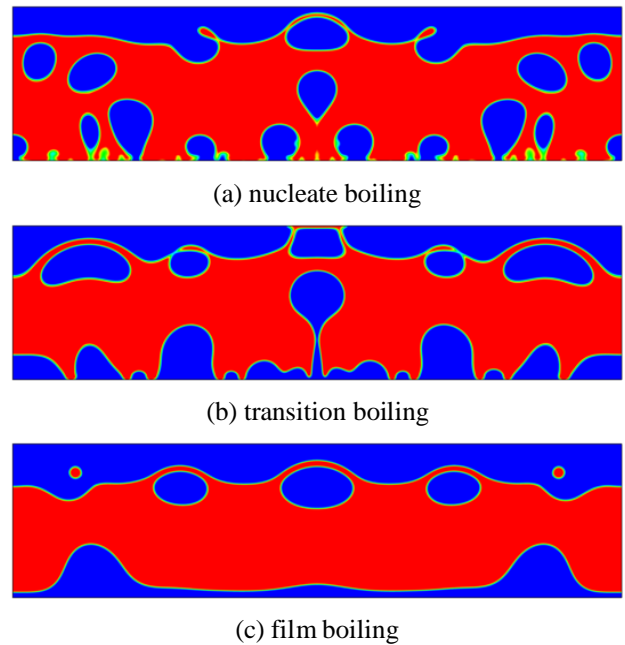

(a) nucleate boiling

(b) transition boiling

(c) film boiling

**Fig. 24**. The hybrid thermal pseudopotential LB modeling of boiling heat transfer. The three boiling stages: (a) nucleate boiling, (b) transition boiling, and (c) film boiling. Reprinted from Li *et al*. [104] with permission of Elsevier.

Recently, Li *et al*. [104] reproduced the three boiling stages of pool boiling using a hybrid thermal pseudopotential LB model, as shown in Fig. 24. In transition boiling, it can be seen that a great portion of the heating surface is covered by vapor patches, which essentially insulate the bulk liquid from the heating surface, while in film boiling the whole surface is covered with a continuous vapor film. Some other features of pool boiling heat transfer were also captured, such as the severe fluctuation of transient heat flux in the transition boiling and the feature that the maximum heat transfer coefficient lies at a lower wall superheat than that of the critical heat flux. A boiling curve obtained in Li *et al*.'s study can be seen in Fig. 25(a), which reproduces the features of a typical boiling curve of pool boiling in Fig. 25(b). Moreover, they have investigated the effects of the heating surface wettability on the boiling curve and found [104] that an increase in contact angle promotes the onset of boiling but reduces the critical heat flux, and makes the boiling process enter into the film boiling regime at a lower wall superheat. Similar trends have also been found by Gong and Cheng [105].

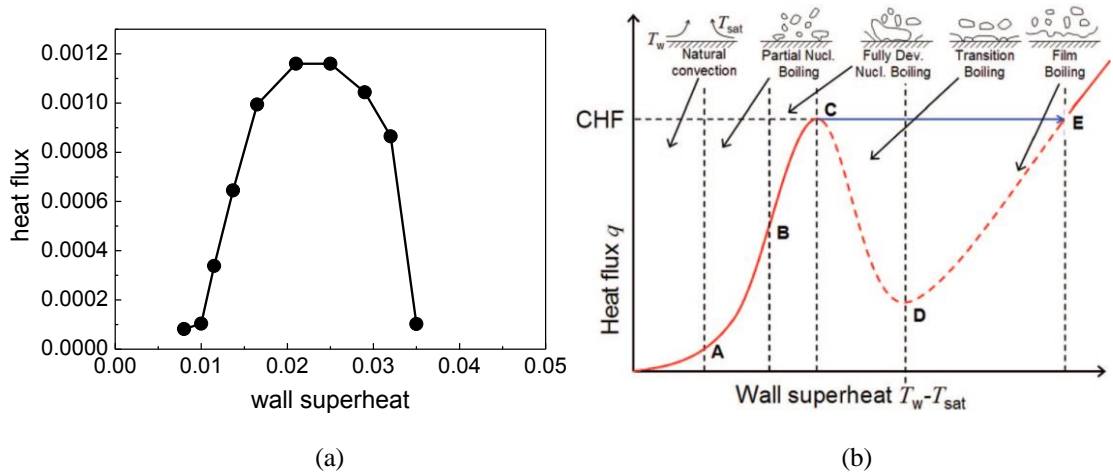

(a)                                                                 (b)

**Fig. 25**. (a) A boiling curve obtained in Li *et al*.'s study with the liquid-phase contact angle being 44.5° [104] (reprinted with permission of Elsevier) and (b) a typical boiling curve of pool boiling, which is reprinted from Chen *et al*. [361] with permission of the American Chemical Society.

Besides the applications in boiling heat transfer, the pseudopotential and the phase-field LB methods have also been applied to simulate droplet evaporation, which is of great importance to many scientific and technical applications, such as liquid-fueled combustion and spray drying. The first attempt was made by Safari *et al*. [110] using their thermal phase-field LB model, who simulated a two-dimensional droplet evaporation process and found that the well-known $D^2$ law was satisfied. The so-called $D^2$ law was firstly found in the study of an evaporating fuel droplet in combustion [362, 363], which predicts that the square of the evaporating droplet diameter varies linearly during the lifetime of the droplet. Later, Safari *et al*. [111] investigated the evaporation process of a droplet in a laminar forced convective air environment. The effects of the Reynolds number and the Schmidt number were studied and compared with an empirical correlation. Recently, Ledesma-Aguilar *et al*. [115] studied the evaporation of a sessile droplet on a flat chemically patterned surface using a binary-fluid phase-field LB model. They showed that the flow pattern of the droplet is reversed when the contact line moves from a hydrophobic ring to another hydrophobic ring. Moreover, Albernaz *et al*. [116] have employed a thermal pseudopotential LB model to simulate droplet evaporation under convective effects. It was shown that the evaporation rate increases when the convection becomes stronger. Furthermore, they found that the evaporating droplet exhibits an oscillatory behavior at higher Reynolds and Peclet numbers.

### 5.4 Energy storage with phase change materials

In recent years, phase change materials have attracted significant attention due to their ability to store thermal energy and have been widely used in thermal energy storage systems [358, 364-368] for heat pumps, solar engineering, and spacecraft thermal control applications. Unlike conventional (sensible) storage materials, phase change materials are *latent heat* storage materials and capable of storing and releasing large amounts of energy at a nearly constant temperature [368]. The solid-liquid phase change is the most used phase change process for phase change materials. The corresponding thermal energy transfer occurs when the material changes from solid to liquid (melting), or liquid to solid (solidification) [358].

In recent years, the LB method has been extensively used to simulate solid-liquid phase change problems with complicated moving boundaries of solid-liquid interface and variable thermophysical properties. Generally, the existing LB models for solid-liquid phase change problems can be classified into two major categories: the solid-liquid phase-field LB method [369-373] and the thermal LB method with an enthalpy-updating scheme [374-383]. Additionally, a couple of models were recently constructed based on interface-tracking methods [384, 385]. In the solid-liquid phase-field LB method, the governing equation of the order parameter, namely the interface-capturing equation [386, 387], is different from that in the phase-field method for liquid-gas two-phase flows. This method was first introduced by Miller *et al*. [369], who have applied the method to simulate the melting of gallium, anisotropic crystal growth [370], dendritic growth of a pure crystal [371], and binary-alloy solidification [372]. In addition, Medvedev and Kassner [373] have employed the solid-liquid phase-field LB method to study the crystal growth in external flows.

The thermal LB method with an enthalpy-updating scheme is a commonly used method in the current LB simulations of solid-liquid phase change. Both the DDF thermal LB approach and the hybrid thermal LB approach have been applied to investigate solid-liquid phase change phenomena. The first work was conducted by Jiaung *et al*. [374], who proposed an extended temperature-based thermal LB model in conjunction with an enthalpy formulation in which the interfacial position of solid-liquid phase change was determined through the liquid-phase fraction. The net enthalpy is defined as $H = c_p T + L_a f_l$, where $c_p T$ is the sensible enthalpy and $L_a f_l$ is the latent enthalpy in which $L_a$ is the latent heat of phase change and $f_l$ is the liquid-phase fraction ($f_l = 1$ and $0$ represent the liquid and solid phases, respectively). A source term was added to the thermal LB equation to account for the latent heat source term $\partial_t (L_a f_l)$. Using the model, Jiaung *et al*. [374] successfully simulated some melting and solidification problems. Nevertheless, iterations were needed in Jiaung *et al*.'s work because the transient term $\partial_t f_l$ was calculated by a forward finite-difference scheme and the liquid-phase fraction at the $t + \delta_t$ level was *a priori* unknown.

Later, following the line of He *et al*.'s internal-energy based DDF LB model, Chatterjee and Chakraborty [201, 375, 376] proposed a series of enthalpy-based DDF LB models for solid-liquid phase change. In their models, a fixed-grid enthalpy-porosity approach was utilized. Moreover, the morphology of the phase change region (i.e., the mushy zone) was treated as an equivalent porous medium that offers a resistance force towards the fluid flow through the phase change region, thus one does not need to impose hydrodynamic or thermal boundary conditions at the interface. Using the models, Chatterjee and Chakraborty have simulated the dendritic growth [375], the melting of gallium [201], and the Bridgman crystal growth [376]. In addition, Chakraborty and Chatterjee [377] have presented a hybrid thermal LB model for studying convection-diffusion phenomena with solid-liquid phase change. The thermofluidic aspects of solid-liquid phase change were also handled with the fixed-grid enthalpy-porosity approach, but the temperature field was solved by a control-volume-based fully implicit finite-difference method.

Using a modified version of Jiaung *et al.*'s model, Huber *et al.* [378] proposed a temperature-based DDF LB model for simulating melting coupled with natural convection. The model was employed to investigate convection melting in a porous medium. Later, Parmigiani *et al.* [379] developed an LB model for simulating multiphase flows with phase change (melting/solidification) of the solid phase in a porous medium by coupling the pseudopotential LB method with Huber *et al.*'s model. The effects of melting on the flow of an invading non-wetting fluid in a buoyancy-driven capillary channel were studied. Figure 26 shows the influence of the Stefan number on the stability of the non-wetting fluid channel. From the left panel of Fig. 26 it can be seen that the non-wetting fluid channel is well-defined and stable at a low Stefan number. However, with the increase of the Stefan number, the channel becomes less stable and the fluid eventually breaks into slugs or bubbles.

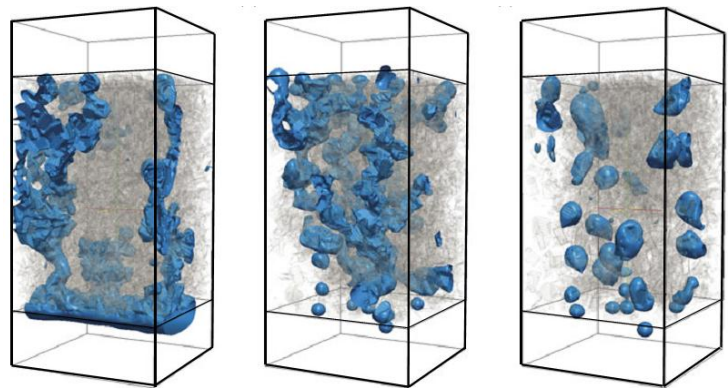

**Fig. 26**. Comparison of the non-wetting phase distribution between three cases with different melting efficiencies. The Stefan number, which represents the melting efficiency, is given by $St = 0.01$, $0.1$, and $1$ (from left to right). Reprinted from Parmigiani *et al.* [379] with permission of Cambridge University Press.

Gao and Chen [380] have developed a temperature-based DDF LB model for melting with convection heat transfer in porous media at the representative elementary (REV) scale on the basis of the generalized non-Darcy model. The effects of the Rayleigh number, Darcy number, and porosity on the melting processes were investigated. In addition, Liu and He [381] have presented an temperature-based DDF LB-MRT model for solid-liquid phase change with natural convection in porous media at the REV scale. However, to deal with the latent heat source term, iterative schemes were used in most of the aforementioned studies, which may greatly increase the computational cost and reduce the efficiency of the model. To overcome this drawback, Eshraghi and Felicelli [382] proposed a modified temperature-based thermal LB model for simulating solid-liquid phase change phenomena, in which the latent heat source term was treated by an implicit approach, but iterations were avoided by solving a group of linear equations. Using the model, Eshraghi and Felicelli [382] successfully simulated binary-alloy solidification and showed that their model is superior over the finite-element method in terms of computational efficiency.

By combining the transient term $\partial_t (c_p T)$ with the latent heat source term $\partial_t (L_a f_l)$, Huang *et al*. [383] developed a novel enthalpy-based DDF LB model for solid-liquid phase change. In their model, the equilibrium distribution function for the energy field was modified so as to recover the target governing equation of enthalpy. The temperature field and the liquid-phase fraction were determined by the enthalpy, which avoids iterations or solving a group of linear equations in numerical simulations. The efficiency of the model was demonstrated by the problems of conduction-induced melting and melting coupled with natural convection.

Furthermore, Huang and Wu [384] have proposed an immersed boundary DDF LB-MRT model for simulating solid-liquid phase change problems. In the model, the usual enthalpy-updating scheme was not used and the solid-liquid interface was viewed as a boundary with no thickness immersed in the fluid and traced by the Lagrangian grid. The velocity and thermal boundary conditions at the solid-liquid interface were handled by the immersed boundary method. The model was successfully applied to simulate convection melting in a circular cylinder with the motion of the solid phase being considered. The effect of the motion of the solid phase on the melting processes at different Fourier numbers (the Fourier number $\mathrm{Fo} = \alpha t / l_c^2$) were investigated, which can be found in Fig. 27. When the solid phase is free, it can be seen that the solid phase is pushed toward the bottom of the circular cylinder as time progresses, while the fluid rises to the top. The numerical results showed that the motion of the solid phase accelerates the melting process [384].

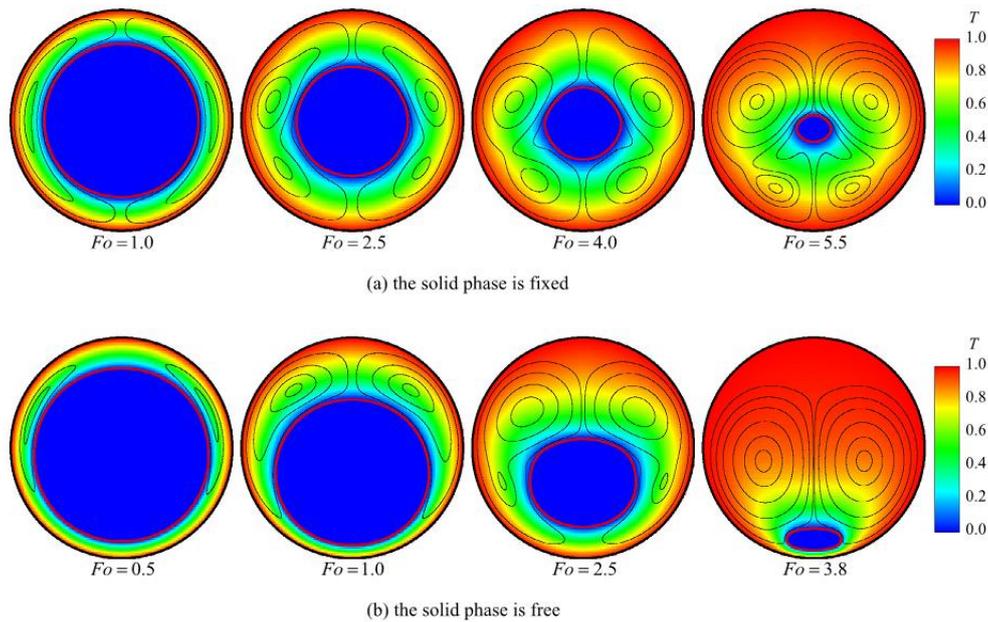

**Fig. 27**. The temperature field, the streamlines (denoted by the black solid lines), and the solid-liquid phase interface (denoted by the red solid line) of melting in a circular cylinder at different Fourier numbers. Reprinted from Huang and Wu [384] with permission of Elsevier.

Recently, Li *et al.* [385] have also developed an LB model for simulating solid-liquid phase change. In their model, the Navier-Stokes equations and the temperature equation were mimicked by a DDF LB model, while the melting front location was tracked by an interface-tracking scheme. The model has been validated by both conduction- and convection-dominated melting problems. In some of the aforementioned studies, the enthalpy, which consists of the sensible enthalpy and the latent enthalpy, was claimed to be "*total*" enthalpy. It should be noted that in thermodynamics the *total* enthalpy is the enthalpy at a stagnation point. For the liquid phase, the total enthalpy contains a term associated with the kinetic energy of the liquid [202].

Using Huang *et al*.'s model, Luo *et al*. [388] have investigated convection melting in a shell-and-tube energy storage system filled with phase change materials. The influences of the Rayleigh and Stefan numbers as well as the arrangements of tubes on the melting dynamics of the shell-and-tube energy storage system were studied in detail. The effect of the number of inner tubes can be seen in Fig. 28. Luo *et al*. [388] showed that the melting time decreases as the number of inner tubes increases, which was found to be attributed to the increase of the surface area and the enhancement of the convection around the tubes.

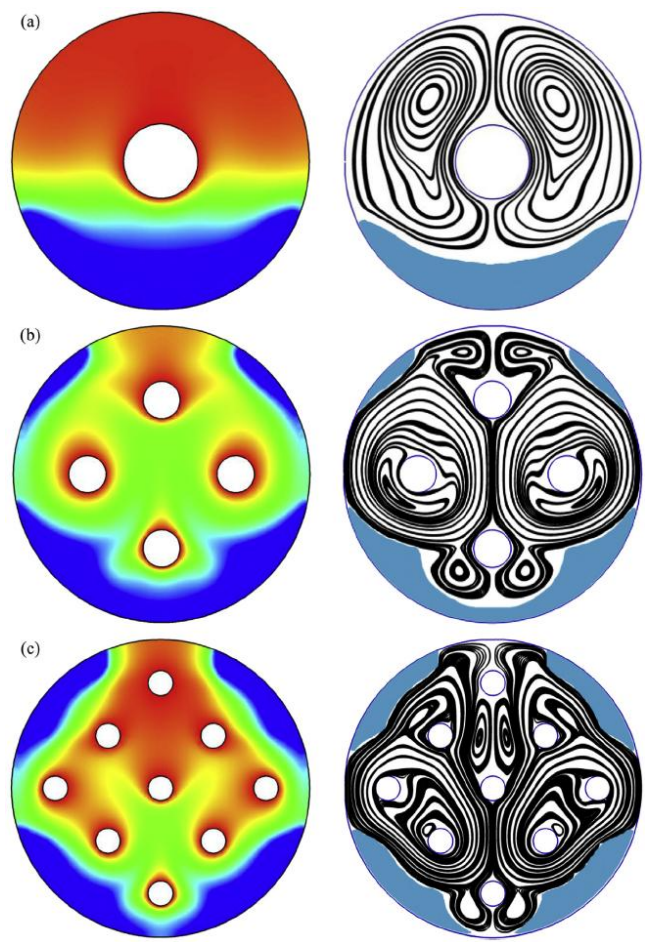

**Fig. 28**. Simulation of the melting process in a shell-and-tube energy storage system. (a) one tube at $\mathrm{Fo}=20$, (b) four tubes at $\mathrm{Fo}=8$, and (c) nine tubes at $\mathrm{Fo}=4$. The temperature distribution (left) and the streamlines (right). Reprinted from Luo *et al*. [388] with permission of Elsevier.

Furthermore, Huo and Rao [389] have employed Huang *et al*.'s model to study the solid-liquid phase change phenomenon of phase change materials under constant heat flux. Three kinds of heat flux distributions were investigated. It was found that increasing the uniform heat flux accelerates the rate of the solid-liquid phase change process. Nevertheless, the ability of the phase change material for maintaining the temperature becomes poorer after raising the heat flux. In addition, Talati and Taghilou [390] have applied Eshraghi and Felicelli's model to simulate the solidification of phase change materials within a rectangular finned container. It was shown that the maximum required time for the solidification process of phase change materials inside the container occurs when the container aspect ratio equals 0.5 and changing the fin's material from aluminum to copper has no significant impact on the freezing history even at low aspect ratios.

## 6. Summary and outlook

As a latecomer to the family of modeling and simulation tools, the LB method has witnessed an astonishing growth in its methodology development and application over the past quarter of a century. It fills a vital gap between the macroscopic continuum approaches such as the Navier-Stokes solvers and the particle-based microscopic approaches such as molecular dynamics. Such a mesoscopic approach has found applications in almost all areas of energy and combustion. The present review has focused on multiphase flow and phase-change heat transfer, in which the LB method has demonstrated distinctive strengths in comparison with both macroscopic and microscopic approaches. On the other hand, the LB method is still an evolving methodology yet to reach its maturity, where both theoretical exploration and new applications are continually appearing.

For multiphase flows, the pseudopotential LB method and the phase-field LB method have been particularly successful and popular. They have been developed and applied to study the dynamics of a wide range of multiphase flows at realistic density ratios, Reynolds numbers and Weber numbers. In the latest development, the dynamics process of droplet collision has been successfully simulated by a cascaded-based pseudopotential LB model with an improved forcing scheme at a density ratio of 1000,

Reynolds number of 6200 and Weber number of 440 [391].

In the pseudopotential multiphase LB method, the interfacial dynamics and phase separation are described by interparticle interactions. As a result, in the pseudopotential LB simulation of multiphase flows, the interface can arise, deform and migrate naturally without resorting to any techniques to track or capture the interface. In recent years, advances have been made in eliminating the thermodynamic inconsistency, reducing the spurious currents, adjusting the surface tension and the interface thickness, etc. These developments have significantly enhanced the capability of the pseudopotential LB method for multiphase flow simulations. It should be noted that the forcing scheme plays a very important role in the pseudopotential LB method because it affects the mechanical stability condition, which determines the coexistence densities given by the pseudopotential LB models. Moreover, to decouple the surface tension from the density ratio, the mechanical stability condition should remain unchanged. Future developments can focus on the enhancement of numerical stability at lower relaxation times and further reducing the spurious currents around the three-phase contact line.

In the phase-field multiphase LB method, the interface is captured with the evolution of an order parameter, which is governed by the Cahn-Hilliard equation or a Cahn-Hilliard-like equation. In this paper, the basic formulations of the phase-field theory have been described and the different definitions of interface thickness have been clarified. Moreover, developments in the phase-field multiphase LB method for isothermal models, elimination of hydrodynamic inconsistency, implementation of contact angles, and thermal models for liquid-vapor phase change have been discussed in detail. With the aid of a mixed difference scheme or a finite-difference solver for the Cahn-Hilliard equation, the phase-field LB method has been successful in modeling dynamic multiphase flows at large density ratios. Nevertheless, further efforts are still required as the mixed difference scheme suffers from the lack of Galilean invariance.

For thermal applications, we have reviewed thermal LB approaches on standard lattices. In particular, a simplified internal-energy-based DDF LB model and the usual temperature-based DDF LB model have been analyzed, showing the error terms in the recovered macroscopic equations (see Eqs. (64), (67), and (68)). It is worth mentioning again that, when employing a multiphase LB model together with a thermal LB model to simulate thermal multiphase flows, special attention should be paid to the correct recovery of the target temperature/energy equation. Some error terms, which may be very small in single-phase flows, are non-negligible in multiphase flows.

Finally, applications of multiphase and/or thermal LB methods have been highlighted for fuel cells, batteries, droplet collisions, boiling, evaporation, and energy storage with phase change materials. For instance, the liquid-gas transport in PEM fuel cells has been extensively investigated with the pseudopotential and the phase-field LB methods, leading to detailed information about the dynamic features of PEM fuel cells under various conditions. Moreover, it is very inspiring that a typical boiling curve of pool boiling and the three boiling stages have been numerically reproduced by the LB method without using the empirical assumptions involved in conventional algorithms. There is no doubt that ever more applications of the LB method as a powerful mesoscopic numerical tool will be found in a variety of fields including energy and combustion. An important future development will be in the hybrid methods, which will significantly expand the capability of the LB method. For example, the combination of the LB method with the immersed boundary method [16] can tackle phase change problems with complex boundaries and/or solid-structure interactions, e.g., boiling flow in rod-bundle geometries, which is a critical issue in boiling water reactors in nuclear power plants. Furthermore, applications of the LB method in micro energy systems and energy system analysis and optimization [392] also have great potential.

**Acknowledgments**

The authors gratefully acknowledge the support from the Los Alamos National Laboratory's Lab Directed Research & Development (LDRD) Program, the National Natural Science Foundation of China (No. 51506227), and the Engineering and Physical Sciences Research Council of the United Kingdom (under the project "UK Consortium on Mesoscale Engineering Sciences (UKCOMES)",

Grant No. EP/L00030X/1). Q. K. also acknowledges the support from a DOE NETL Unconventional Oil & Gas Project.